\documentclass[article]{jfm}

\usepackage{graphicx}
\usepackage{newtxtext}
\usepackage{newtxmath}
\usepackage{natbib}
\usepackage{hyperref}
\usepackage{soul}
\usepackage{marginnote}

\hypersetup{
    colorlinks = true, 
    urlcolor   = blue,   
    citecolor  = black,
}

\newcommand{\RomanNumeralCaps}[1]
\linenumbers

\usepackage{amssymb,amsmath}
\usepackage{array}
\usepackage{latexsym}
\usepackage{bezier}
\usepackage{psfrag}
\usepackage{bm}
\usepackage{float}
\usepackage{multirow}
\usepackage{mathrsfs}
\usepackage{color}
\usepackage{marginnote}
\usepackage[normalem]{ulem}
\usepackage[pdf]{pstricks}
\usepackage{newtxtext}
\usepackage{newtxmath}
\usepackage{enumitem}

\newcommand{\bvec}[1]{\mbox{\bf #1}}

\newcommand{\bsy}[1]{{\boldsymbol{#1}}}

\graphicspath{{./Figs/}}

\shortauthor{}

\title[Emergence of chaos in the basin boundary]{Emergence of chaos and fractality in the basin boundary of subcritical shear flows}
\author[B.\,Wang,  R.\,Ayats, K.\,Deguchi, A.\,Meseguer \and\, F.\,Mellibovsky]{
  B.\,Wang\aff{1},
  R.\,Ayats\aff{2},
  K.\,Deguchi\aff{3} \corresp{\email{kengo.deguchi@monash.edu}},
  A.\,Meseguer\aff{2} \and\,
  F.\,Mellibovsky\aff{2} \corresp{\email{fernando.mellibovsky@upc.edu}}
}

\affiliation{
  \aff{1} Institute of Science and Technology Austria (ISTA), 3400 Klosterneuburg, Austria
    \aff{2} Departament de F{\'\i}sica, Universitat Polit\`ecnica de Catalunya - BarcelonaTech (UPC), 08034 Barcelona, Spain
  \aff{3} School of Mathematics, Monash University, VIC 3800, Australia 

  }

\begin{document}

\maketitle
\thispagestyle{plain}
\pagestyle{headings}

\begin{abstract}
From a dynamical systems perspective of subcritical transition in shear flows, the basin boundary separating the laminar and turbulent attractors, along with
the edge state that governs the long-term dynamics on that boundary, are of fundamental interest. 
As the Reynolds number is increased from small values, a multiplicity of simple exact coherent structures (ECS) appear in phase space, of which one often undertakes the role of the edge state. 
At higher values of the Reynolds number, however, the dynamics on the basin boundary and the edge state are sensitive to initial conditions, as shown by a wealth of numerical and experimental studies. The mechanism behind this transition remains unclear.
To address this,
we examine the subcritical regime of Taylor–Couette flow using a minimal computational box and reveal a generic mechanism whereby the edge state becomes chaotic. The first step in this process involves the formation of a heteroclinic tangle between a travelling-wave-type ECS 
and an independently engendered chaotic saddle. The interaction causes the basin boundary to incorporate the saddle, thus inheriting its fractal structure, while
the edge state itself remains the simple ECS. 
The transition of the edge state to chaos
requires a second step: the formation of a heteroclinic cycle involving the ECS and the chaotic saddle. In consequence, the observation of a simple non-chaotic edge state at given values of the parameters is no guarantee that the same situation will hold at nearby values.

\end{abstract}


\section{Introduction}\label{sec:intro}

Comprehending the physical mechanisms behind subcritical transition is a major unresolved challenge in fluid dynamics that dates back to the celebrated pipe flow experiments of \citet{Re83}. 
A puzzling phenomenon already observed in Reynolds' apparatus, and in many subsequent studies of shear flows \citep[e.g.,][]{DaMu95,ScEc97,HoJuMu03,FaEc04,MeMe06,MeME07}, is that the occurrence of turbulent transition depends on both the shape and amplitude of the initial disturbance. This amplitude dependence evinces the need for a mathematical tool that reaches beyond the limited realm of linear stability analysis. 
Moreover, the strong sensitivity to disturbance shape hinders the study of subcritical transition through brute-force approaches 
relying solely on statistical analysis of data generated by experiments or direct numerical simulation.

\begin{figure}
  \begin{center}
      \includegraphics[height=.4\linewidth]{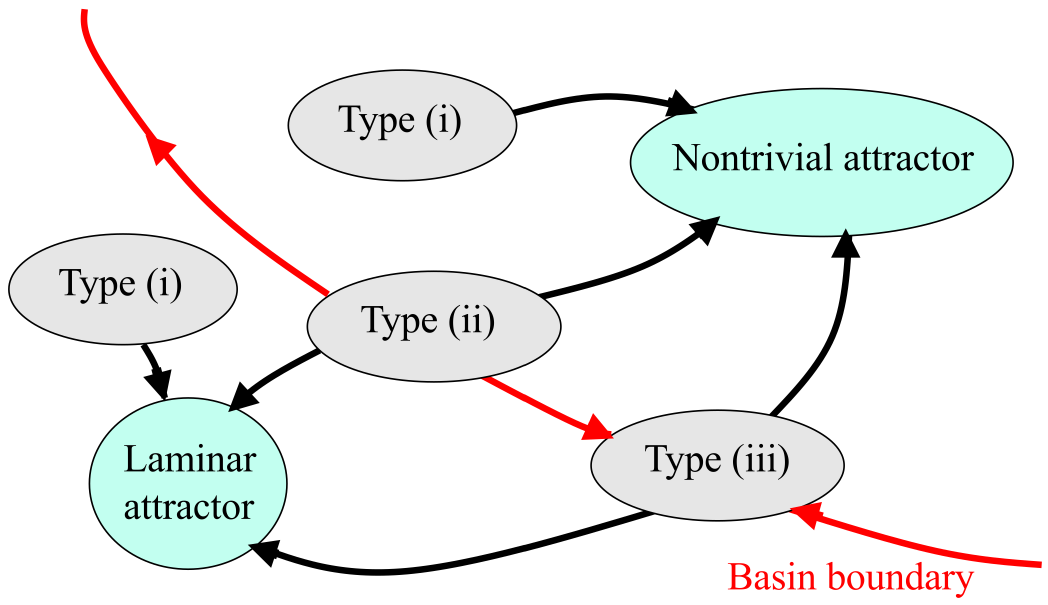}
  \caption{A schematic of the phase space illustrating a typical subcritical transition problem, in which two attractors coexist. 
}
  \label{fig:Sketch} 
  \end{center}
\end{figure}

Over the past two decades, dynamical systems theory has become 
an increasingly {useful tool for addressing the subcritical transition problem} \citep{Ke05,EcSchHoWe07,GraFlo21}.
The {usual} strategy is to analyse the system in phase space to identify the key invariant objects that regulate the dynamics, namely the attractors, saddles and connecting manifolds. 
As shown in figure~\ref{fig:Sketch}, two attractors typically coexist in phase space in the subcritical regime of shear flows: the linearly stable laminar solution and a nontrivial attractor. The latter attractor may be turbulence or, in the early subcritical regime, some precursor to turbulence. 
The manifold that separates the basins of attraction of the attractors (illustrated by the red curve) is known as the \textit{basin boundary}. By definition, initial conditions on one side of this boundary decay to laminar flow, while those on the other side evolve towards the nontrivial attractor. Accordingly, understanding subcritical transition necessarily depends on elucidating the properties of the basin boundary.

When the nontrivial attractor is chaotic, like turbulence, the basin boundary is sometimes referred to as the {\it edge of chaos} \citep{SkYoEc06}. 
The basin boundary is often depicted as a curve in two-dimensional sketches, but it is actually a hypersurface that has one less dimension than the full phase space. Visualising this hypersurface for high-dimensional systems is challenging, but it has been accomplished with some success by \cite{SkYoEc06} for a low dimensional model. 
Their method consisted in systematically varying the initial condition and computing the duration of the transients before laminar decay \citep[also known as lifetimes; see also][]{ScEc97}. If the trajectory did not relaminarise after a pre-determined but sufficiently long integration time, the initial condition was regarded as appertaining to the basin of attraction of the nontrivial {pseudo-}attractor {(i.e. mildly unstable saddle).} Their results showed that the basin boundary hypersurface may develop a large number of folds, thus introducing critical sensitivity to initial conditions.

Lifetime studies have been widely used in the investigation of subcritical shear flows, in particular to characterise the sensitivity to initial conditions and to reveal the apparent fractal structure of the phase space.
For example, \citet{KrEcSch14} demonstrated that similar representations as \cite{SkYoEc06} could be generated for plane Couette flow using appropriate phase map projections. Their findings suggest that the phase space may have a fractal structure, but its self-similar nature is yet to be confirmed.
The parameter dependence of {lifetimes} was also explored by \cite{EcMe99} and \cite{MoFaEc04} for simple model problems. 
The promising results of these preliminary studies soon sparked {numerical} research on 
{pipe flow, such as that by \cite{ScEcYo07}.}
Around the same time, experiments \citep{PeMu06,HoWeSc06} and numerical simulations \citep{FaEc04,WiKe07} showed that the reverse transition -- the natural decay to laminar flow from the transitional chaotic state -- occurs invariably in the early transitional regime. 
{Interestingly, the decay process follows a memoryless (Poisson) probabilistic distribution, providing further evidence of the phase space complexity.} For the latest developments of the transition problem in pipes, readers are referred to the review article by \cite{AvBaHo23}. 


{Generalising the notion of a basin boundary to a pseudo-attractor is convenient in shear flow turbulence, because at moderate Reynolds numbers it may decay after extremely long times. 
Indeed,} for the cases studied in \cite{FaEc04,PeMu06,HoWeSc06,WiKe07}, the laminar state is {very likely} the sole attractor. 
For symmetry restricted flows at sufficiently low values of the Reynolds number, however, the existence of a non-trivial attractor is sometimes relatively easy to ascertain {and a basin boundary can be defined rigorously}
\citep{MeEc11,KrEc12,AvMeRo13,ZaEc15b,RiMeAv16}.
{A widespread method for systematically analysing the basin boundary is 
the computation of \textit{edge states}.} 
A key observation in {mathematically} defining an edge state is that any trajectory starting exactly on the basin boundary will remain on it indefinitely (the basin boundary is an invariant set). Edge states are attractors of the dynamical system that arise when the dynamics are restricted to the basin boundary.
In practice, an edge state emerges as the final state of trajectories when the initial condition is controlled to deliberately inhibit evolution toward either {full-space} attractor. The first implementation of such a control algorithm, known as the edge-tracking technique, in a subcritical shear flow problem was presented by \cite{ItTo01}.  
Edge tracking in small periodic domains, typically of the size of the minimal flow unit \citep[i.e. the smallest periodic domain in which turbulence is sustained, see] []{JiMo91,HaKiWa95}
often converges to simple solutions such as equilibria, travelling waves {or limit cycles}, depending on the shear flow under scrutiny.
This highlights the importance of non-chaotic solutions, known as exact coherent structures (ECS), in understanding the subcritical transition problem \citep{Na90,ClBu92,FaEc03,WeKe04} and explains why they are observed repeatedly, if only briefly, in carefully controlled experiments \citep{HoVaWeNiFaEcWeKeWa04,DeMeAv12}. 


A necessary condition for an ECS to be an edge state is that it has a single unstable eigenvalue, the one that pushes trajectories away from the basin boundary. ECS with such stability properties typically arise from saddle-node bifurcations.
%
When this occurs, the nodal (or upper branch) solution {forms the backbone of}
the non-trivial attractor, {while} 
the saddle (or lower branch) solution has a single unstable eigenvalue and acts as the edge state; {see \cite{WaGiWa07} who studied plane Couette flow in the minimal box, for example.}
Although the lower branch solution tends to retain a single unstable eigenvalue up to increasingly high Reynolds numbers, a behaviour that is consistent with asymptotic analyses \citep{DeHa16}, 
there is no guarantee that the ECS will maintain its role as an edge state indefinitely. 
Indeed, given the sensitivity to initial conditions that is inherent in prototypical subcritical transition, it seems unlikely that a simple ECS should remain the edge state at sufficiently high Reynolds numbers. If the edge state were to become chaotic, which mechanisms could be held responsible? The existence of chaotic edge states has indeed been confirmed in shear flows \citep[see, e.g.,][]{ScEc06}.

All of the complex scenarios of subcritical transition described above involve a \textit{chaotic saddle}, i.e. an unstable chaotic set in phase space. Chaotic saddles can play different roles in phase space, and we classify them comprehensively into {three mutually exclusive} types:
\begin{enumerate}
\item ~~those not on the basin boundary, 
\item ~~those on the basin boundary but not an edge state, and 
\item ~~those that are edge states, 
\end{enumerate}
as illustrated in figure\,\ref{fig:Sketch}. As already noted, when turbulence is only transient, 
a true basin boundary does not exist,
and only type (i) is possible. 
%
Even in this case, edge tracking can converge to {a saddle solution, 
which has} only one unstable direction \citep{SchEc09}. 
Strictly speaking, this is not an edge state, {in the sense that it is not attached to any basin boundary.}
{However,} {since}
a chaotic state that persists beyond feasible time horizons can be considered as an attractor for all practical purposes,
the {outcome} of edge tracking effectively behaves as an {edge state attached to the basin of attraction of that pseudo-attractor} \citep{SkYoEc06,EcSchHoWe07}. {While such generalisations of edge states are now common in shear flow studies, in this paper we adopt the strict definition.}

Type (i) chaotic saddles typically emerge from the collision of a chaotic attractor with its basin boundary in a global bifurcation known as boundary crisis \citep{GrOtYo83,GrOtYo86}. The collision usually involves a saddle that was contained in the basin boundary and {may have even acted as an edge state prior to the crisis}. In plane Couette flow, {a} chaotic attractor arising from a period doubling cascade undergoes a boundary crisis that turns it into a type (i) chaotic saddle \citep{KrEc12,AvMeRo13,ZaEc15b}.
The transient turbulence observed in pipe flow at low Reynolds numbers \citep{PeMu06} is also type (i), and its origin would be related to a boundary crisis {akin to that} identified by \cite{AvMeRo13,RiMeAv16}.
Thus far, the mechanisms by which chaotic saddles of type (ii) or (iii) appear has not been explored. 

It is noteworthy that \cite{LuKaVa19} and \cite{BuDoHo19} recently applied the homoclinic tangency theory to plane Couette and pipe flow, respectively. To fully appreciate the motivation behind these studies, a profound understanding of discrete dynamical systems theory is required. Unlike in continuous dynamical systems, the unstable and stable manifolds of fixed points can intersect transversally in discrete systems.
The presence of such a transversal homoclinic point entails the existence of infinitely many homoclinic points forming a homoclinic tangle. This, in turn, leads to chaos on a certain Cantor set, as described by the Smale-Birkhoff theorem \citep{Bi1950,Sm65}. 
The theory of discrete dynamical systems can be applied to shear flows by defining an appropriate Poincar\'e section in phase space.
In the light of the numerical results of \cite{LuKaVa19}, the tangency they identified likely relates to the emergence of a chaotic saddle of either type (ii) or (iii). However, their analysis did not include a detailed investigation of this aspect.

The aim of this paper is to demonstrate clearly and directly how fractal structures and memoryless processes can emerge at the basin boundary and produce chaotic edge states in subcritical shear flows. With this goal in mind, we select a flow configuration that 
{favours the appearance of a chaotic set that sequentially evolves from type (i) to type (ii) and, finally, to type (iii). Specifically, we focus on the}
subcritical parameter region of Taylor-Couette flow {investigated by} \citet{WaAyDe22}.
{At the root of the analysis is the stabilisation of an ECS \citep[a drifting rotating wave originally found by][]{DeMeMe14}, when considered in the parallelogram-annular domain of \cite{DeAl13}.}

{The paper is structured as follows.} In \S\ref{sec:bifdiagram}, we set out to review relevant known results of the system and present the full bifurcation diagram. The {various types of} temporal dynamics observed at selected values of the Reynolds number are presented in order to motivate the analysis that follows. Return map analysis is then {employed} in \S\ref{sec:emergence} to explore how a type (i) chaotic saddle emerges and acquires a fractal structure in phase space. 
Then \S\ref{sec:bridge} uncovers how the chaotic saddle merges with the basin boundary and {becomes} type (ii). {Its eventual transition into a type (iii) saddle, i.e. a chaotic edge state, is addressed in} \S\ref{sec:Bc}.
{Finally, we summarise the key findings and draw conclusions} in \S\ref{sec:conclusions}.

\section{Problem statement, bifurcation diagram and typical time series}\label{sec:bifdiagram}

\begin{figure}
  \begin{center}
  \begin{tabular}{cc}
  \raisebox{13.0em}{ (a)} &
  \begin{tabular}{cc}
      \raisebox
      {-12.5em}{\includegraphics[width=.38\linewidth]{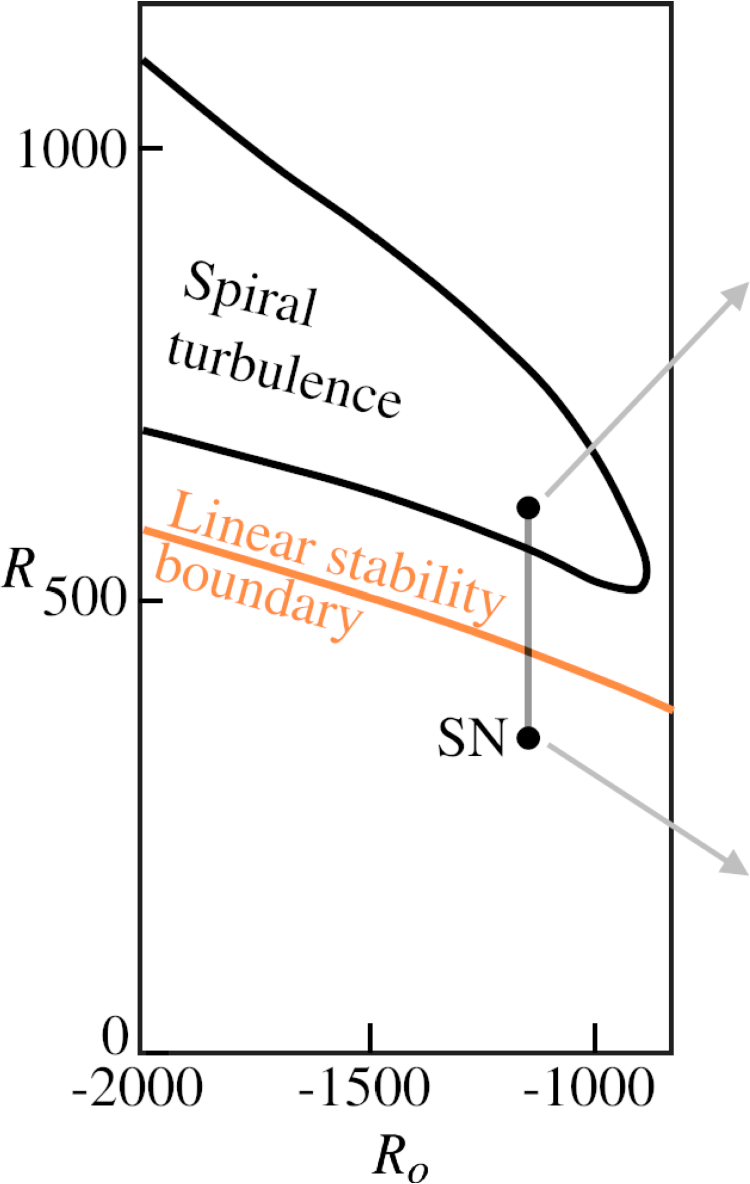}}&
      \begin{tabular}{cc}
       \raisebox{12.5em}{(b)} & \includegraphics[width=.5\linewidth]{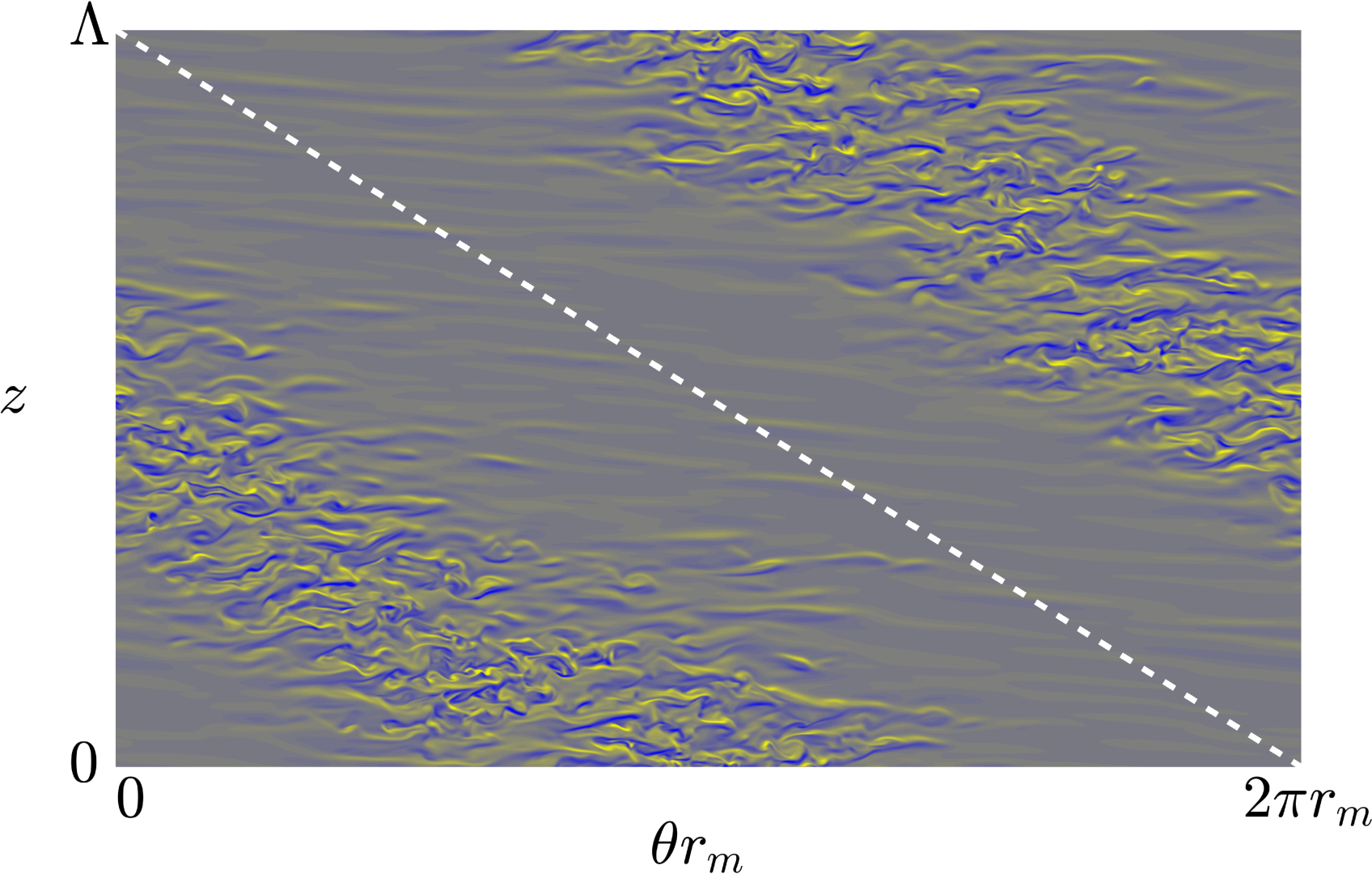} \\
       \raisebox{12.5em}{(c)} &\includegraphics[width=.5\linewidth]{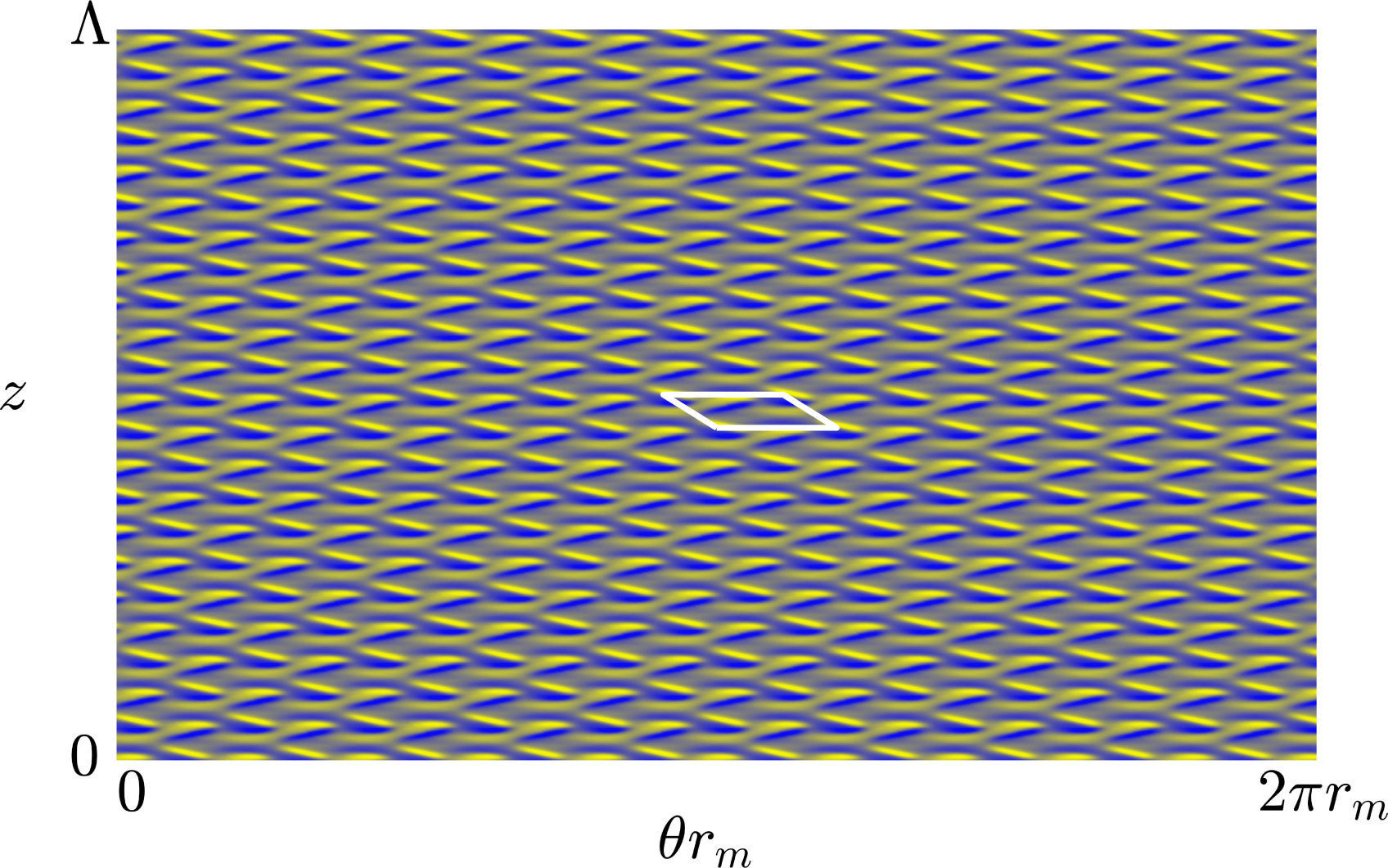} \\
      \end{tabular} \\
   \end{tabular} \\
   \raisebox{15em}{(d)} &
   \includegraphics[width=.7\linewidth]{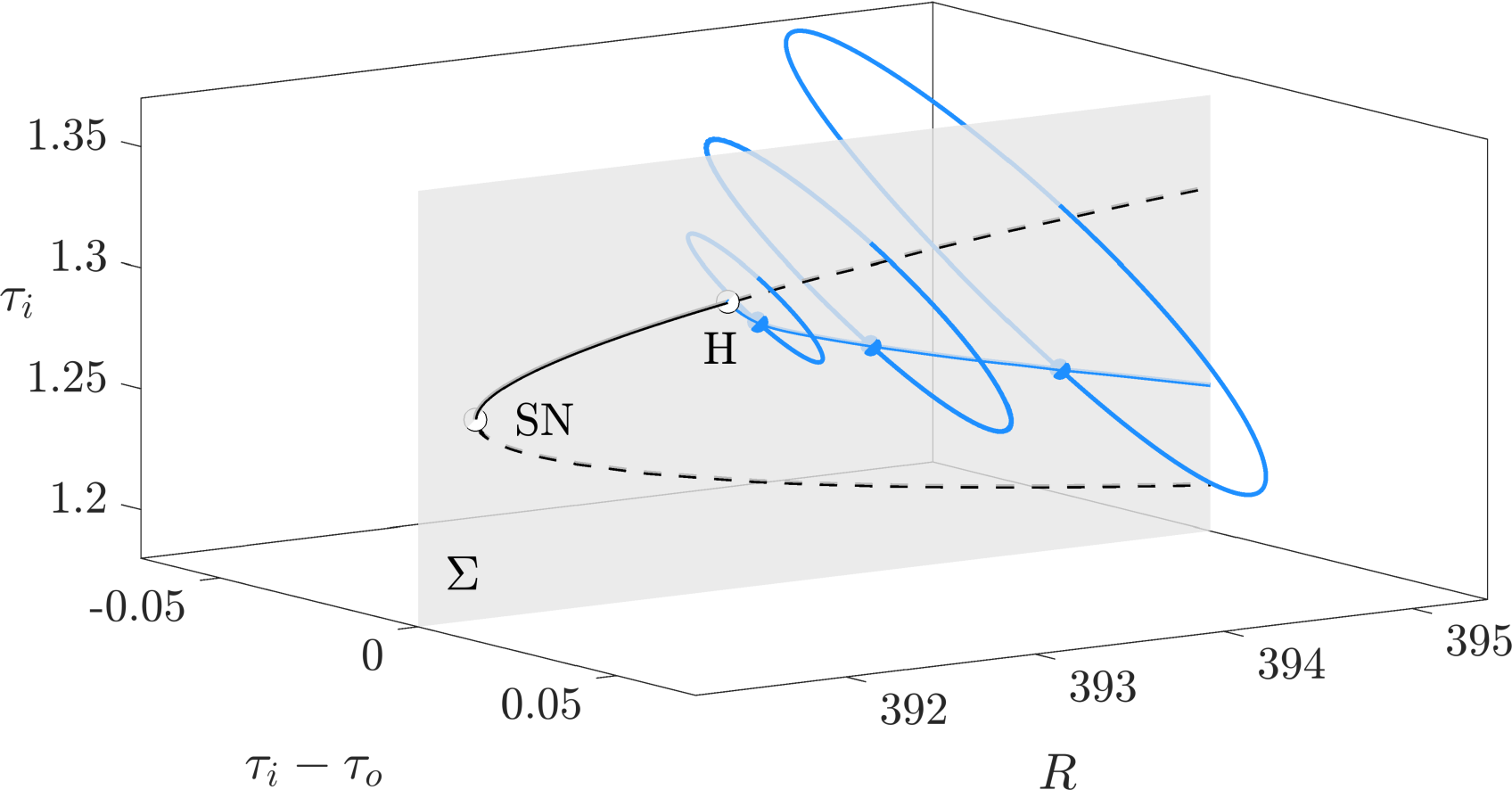} \\
   \end{tabular}
 \end{center}  
\caption{
Summary of results from \cite{WaAyDe22}. (a) Transition diagram for {counter-rotating} Taylor–Couette flow with radius ratio $\eta=0.883$. The bullets on the $R_o=-1200$ line indicate the parameter values used for flow field visualisations in panels (b) and (c). (b) Spiral turbulence at $R=600$ visualised by radial vorticity $\omega_r\in[-4000,4000]$ at $r=r_m = (r_i+r_o)/2 \approx 8.05$. The {dimensionless height of the domain}
is $\Lambda=31.4$.
(c) {DRW at {SN ($R=391.5$)}}
{visualised in the same format as panel (b), but} with $\omega_r\in[-1200,1200]$. The white parallelogram shows {the computational domain.}
(d) Bifurcation diagram at $R_o=-1200$. SN and H denote saddle-node and Hopf bifurcation points, respectively. The black curve represents DRW (solid for the stable portion, dashed when unstable). The P$_1$ relative periodic orbit is represented by the blue closed curves at selected values of $R$. Its representation on the Poincaré section $\Sigma$ is given by the intersection of the loops with the grey plane.}
\label{fig:summary}          
\end{figure} 

We consider the Taylor-Couette flow setup, taking the radial gap $d=r_o^*-r_i^*$ and $d^2/\nu$ as units for length and time, respectively. Here, $r_i^*$ and $r_o^*$ are the dimensional inner and outer cylinder radii, respectively, and $\nu$ is the kinematic viscosity. The cylinders rotate with angular velocities $\Omega_i$ and $\Omega_o$.
The flow obeys the incompressible Navier–Stokes equations, which we express in cylindrical coordinates $(r,\theta,z)$. 
{A zero axial mass flux condition is imposed to mimic the experimental conditions in a numerical setup of infinite axial extent.}
We fix the radius ratio $\eta=r_i^*/r_o^*=0.883$ and the outer cylinder Reynolds number $R_o=d{r_o^*}\Omega_o/\nu=-1200$ following previous studies \citep{MeMeAv09b,MeMeAv09a,DeMeMe14,WaAyDe22,WaAyDeMeMe25a,WaAyDeMeMe25b}. The only parameter we vary throughout the study is the inner cylinder Reynolds number $R\equiv R_i=d{r_i^*}\Omega_i/\nu$. 

{The total velocity field $\mathbf{v}$ is decomposed into a base flow $\bvec{v}_{\rm b}$ and a perturbation $\bvec{u}=u_r\hat{\bsy r}+u_{\theta}\hat{\bsy \theta}+u_z\hat{\bsy z}$.}
The {base state of the system is} the \textit{circular Couette flow} (CCF) solution, {which has a well-known closed-form expression}
\begin{equation}\label{defccf}
  \bvec{v}_{\rm b} = V_{\rm b} \hat{\bsy \theta} = \displaystyle \left(Ar+\frac{B}{r}\right)\,\hat{\bsy \theta}, 
\end{equation}
with $ A=(R_o-\eta R_i)/(1+\eta)$ and $B=\eta(R_i-\eta R_o)/\left[(1-\eta)(1-\eta^2)\right]$. 

The numerical methods used to solve the {governing} equations were thoroughly presented in \cite{WaAyDe22}, to which we refer the reader for details. The computational domain is of parallelogram-annular shape and periodic in the parallelogram directions $(\xi,\zeta)$. 
The spatial discretisation for the direct numerical simulations (DNS) employs Chebyshev--Fourier--Fourier spectral methods, and the spatial resolution {of $[0,50]\times[-8,8]\times[-8,8]$ modes} deployed by \cite{WaAyDeMeMe25a} has been confirmed as sufficient and kept. To compute and track unstable solutions in parameter space, we use a Poincar\'e-Newton-Krylov 
method, duly coupled with the method of slices \citep{FrCv2012,BuCviDaRu15,WaMeAy23} to cope with drift. 

We characterise flow states by the \textit{normalised kinetic energy} $\kappa$ of the perturbation velocity 
and by the corresponding inner and outer cylinders \textit{normalised torque}, $\tau_{\rm{i}}$ and $\tau_{\rm{o}}$,
\begin{eqnarray}
   \kappa = \frac{E(\bvec{u})}{E(\bvec{v}_b)},\qquad \tau_{i, o}= \left. 1+\frac{\partial_r(r^{-1}\langle u_{\theta}\rangle_{\xi\zeta})}{\partial_r(r^{-1}V_{b})}\right|_{r=r_i,r_o},
\label{defnormtorque}
\end{eqnarray}
where
\begin{equation}
  E(\bvec{v}) = \frac{1}{2\mathcal{V}} \iiint_\mathcal{V}{\bvec{v}\cdot\bvec{v}\,d\mathcal{V}} =
  \frac{1}{2\mathcal{V}}\int_{0}^{2\pi}\int_{0}^{2\pi}\int_{r_i}^{r_o} \bvec{v}\cdot\bvec{v}\,r\,\mathrm{d}r\mathrm{d}\xi\mathrm{d}\zeta =
  \frac{1-\eta}{1+\eta}\int_{r_i}^{r_o}{\langle\bvec{v}\cdot\bvec{v}\rangle_{\xi\zeta}\,r\,\mathrm{d}r}
\end{equation}
is the volume-averaged kinetic energy of some velocity field $\bvec{v}$. The volume of the computational domain is $\mathcal{V}=2\pi^2(r_{o}^2-r_{i}^2)=2\pi^2(1-\eta)/(1+\eta)$, and $\langle~\rangle_{\xi\zeta}$ implies averaging in both parallelogram directions. With these definitions, $\kappa=0$ and $\tau_{i}=\tau_{o}=1$ for CCF.

We start by summarising in \S\ref{subsec:previous} what is known regarding the onset of chaos in counter-rotating Taylor-Couette flow. We then extend the bifurcation diagram of \cite{WaAyDeMeMe25a} to include new solution branches and investigate their role in the transition process and the morphing of phase space with varying $R$. In \S\ref{subsec:emergP3} we present a collection of time series at selected values of $R$ to illustrate the most relevant dynamical phenomena encountered along the complex bifurcation scenario.
{The key critical values of the Reynolds number {for which} the qualitative changes in the time series {occur} are {briefly commented}} in \S\ref{subsec:globbif}. 



\subsection{
{The onset of chaos as reported in previous studies}
}\label{subsec:previous}

Figure~\ref{fig:summary}a shows a schematic representation of the transition diagram in $R_o$--$R$ parameter space, with the {grey} vertical line indicating an $R$-sweep at constant $R_o=-1200$ for $\eta=0.883$, which is the focus of our study. 
CCF is linearly unstable above the linear stability boundary (orange line), which is crossed at $R\approx 447.35$ in our exploration.
The solid black curve encapsulates a region of spatio-temporal intermittency, characterised by a flow consisting of a turbulent helical band coherently sustained over a laminar background flow, the spiral turbulence regime (figure~\ref{fig:summary}b). \citet{WaAyDe22} found a finite-amplitude drifting rotating wave (DRW) that exists for $R\geq R_{SN}\approx 391.5$. 
This solution, computed in the small parallelogram domain shown in figure \ref{fig:summary}c, persists in the intermittency region 
and appears to play an important role in the formation and sustainment of spiral turbulence.
The use of this specific parallelogram domain, with two of its sides aligned with the tilt of the turbulent spiral, has been instrumental in our analysis. As evident from the bifurcation diagram in figure \ref{fig:summary}d, the upper branch of DRW (the one with higher torque) is stable at onset ($R=R_{SN}$) in this particular domain. This stability is {briefly} preserved up to the Hopf bifurcation point H ($R=R_{H}\approx 395.48$), whence a family of stable relative periodic orbits (P$_1$, blue loops) is issued supercritically.
To better convey the time periodic nature of P$_1$, we have included a third axis, $\Delta\tau = \tau_i-\tau_o$, in the bifurcation diagram, such that cross-sections at constant $R$ correspond to phase map projections on the $\tau_i$--$\Delta \tau$ plane. 
Since relative equilibria must satisfy $\tau_i=\tau_o$, the DRW solution branch lies within the $\Delta\tau=0$ plane (grey plane). To simplify the ensuing analysis, we introduce a Poincar\'e section 
\begin{equation}\label{eq:PoincSec}
  \Sigma=\left\{\tilde{{\bf v}}\in \mathbb{X}\left |\;\tau_i(\tilde{{\bf v}})=\tau_o({\tilde{\bf v}}),\;\dfrac{d\tau_i}{dt}>\dfrac{d\tau_o}{dt}\right. \right\},
\end{equation}
where $\mathbb{X}$ is the set of all possible {$\tilde{\bf v}$, representing velocity fields modulo spatial drifts.}
Relative periodic orbits are represented by discrete points on $\Sigma$. In particular, the P$_1$ solution branch is represented by the blue curve within the grey plane.

\begin{figure}                                                             
  \begin{center}
    \begin{tabular}{c}
        \includegraphics[width=.9\linewidth]{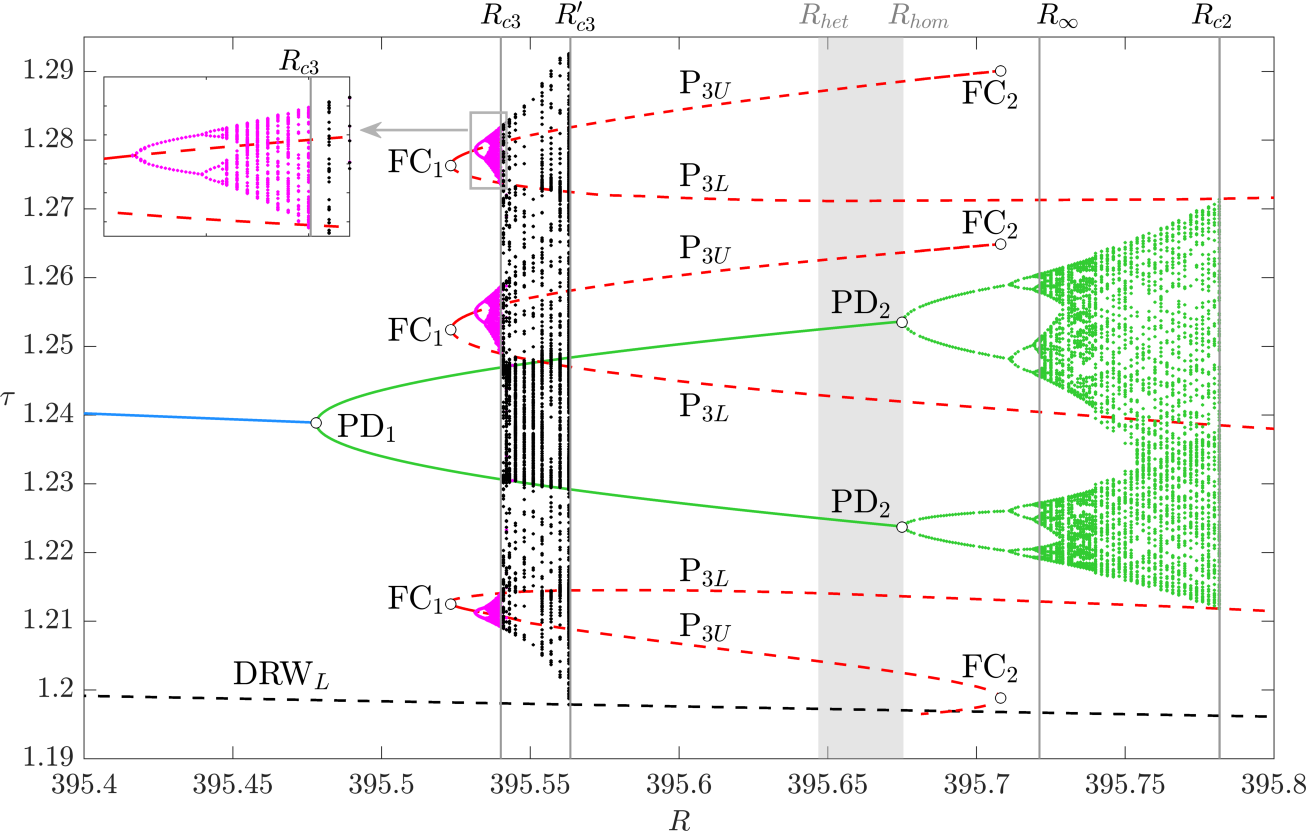}
    \end{tabular}
 \end{center}  
  \caption{Bifurcation diagram constructed from points on $\Sigma$. 
  {At PD$_1$, the P$_1$ orbit (blue line) undergoes a period-doubling bifurcation producing P$_2$, which subsequently initiates a cascade leading to the chaotic set C$_2$ beyond the accumulation point $R=R_{\infty}$ (P$_2$, the cascade, and C$_2$ are {all} shown in green).}
  {The P$_3$ orbit (dashed red line) emanates from a cyclic fold FC$_1$.} Its upper branch (P$_{3U}$),  exhibits another period-doubling cascade (magenta) generating the chaotic set C$_3$. 
  The black dots represent transient chaotic data {{initiated} from P$_{3L}$, 
  } 
  collected before {{the} departure towards P$_2$} (see figure \ref{figE}b). The black dashed curve represents the lower branch of the drifting rotating wave (DRW$_L$). The {grey} vertical lines at $R=R_{c3}$, $R_{c3}'$ and $R_{c2}$ indicate global bifurcation points. {The shaded region indicates the range of $R$ between the {first} heteroclinic bifurcation point {($R_{het}$)} and the homoclinic bifurcation point {($R_{hom}$)} shown in figure \ref{fig:Rhom}a.}
}
\label{fig:biffig}          
\end{figure} 
Figure~\ref{fig:biffig} depicts a bifurcation diagram summarising all solutions {involved} in the transition process, {as viewed on} $\Sigma$.
Since $\tau_i=\tau_o$ on $\Sigma$, we dispense with the subscripts in the torque (i.e. $\tau=\tau_i=\tau_o$).
The P$_1$ solution (blue) {of figure~\ref{fig:summary}d} undergoes a Feigenbaum cascade (green), as found by \cite{WaAyDeMeMe25a}. 
$\mathrm{PD}_1$ indicates the first period-doubling bifurcation as $R$ is increased, where P$_1$ loses stability and a new periodic solution $\mathrm{P}_2$ of about twice the period of P$_1$ emerges. The new stable solution crosses the Poincar\'e section twice every period, hence the subscript {and its representation as two branches in the figure}. At somewhat larger $R$, $\mathrm{P}_2$ exhibits another period doubling bifurcation, $\mathrm{PD}_2$, leading to $\mathrm{P}_4$, which crosses $\Sigma$ at four distinct points. A period doubling cascade follows as $R$ is increased further, with periodic orbits of ever increasing period accumulating at $R_\infty\approx395.721266624$ \citep{WaAyDeMeMe25a}, in accordance with Feigenbaum's self-similar scenario \citep{Fe78,Fe79,Fe80,Fe82}. From this point onwards long term dynamics become unpredictable, following the emergence of a chaotic attractor (cloud of green dots {beyond $R_\infty$} in figure~\ref{fig:biffig}). 
{We shall denote the chaotic set originating from P$_2$ as {C$_2$}. Strictly speaking, the cascade {starts from} P$_1$, but we adopt the subscript 2 {because} P$_2$ plays a more significant role in the {$R$-range of interest.}}
\cite{WaAyDeMeMe25a} and \cite{WaAyDeMeMe25b} demonstrated that the chaotic attractor is embedded in a nearly {one-dimensional} manifold in $\Sigma$. This particularly low dimensionality {is} {precisely} {the key enabling feature }
of our analysis.
{The unexplained features in the figure are findings specific to the present study and will be discussed later.}

\begin{figure}                                                            
 \begin{center}
  \begin{tabular}{l}
  (a) \\
  \includegraphics[width=.95\linewidth]{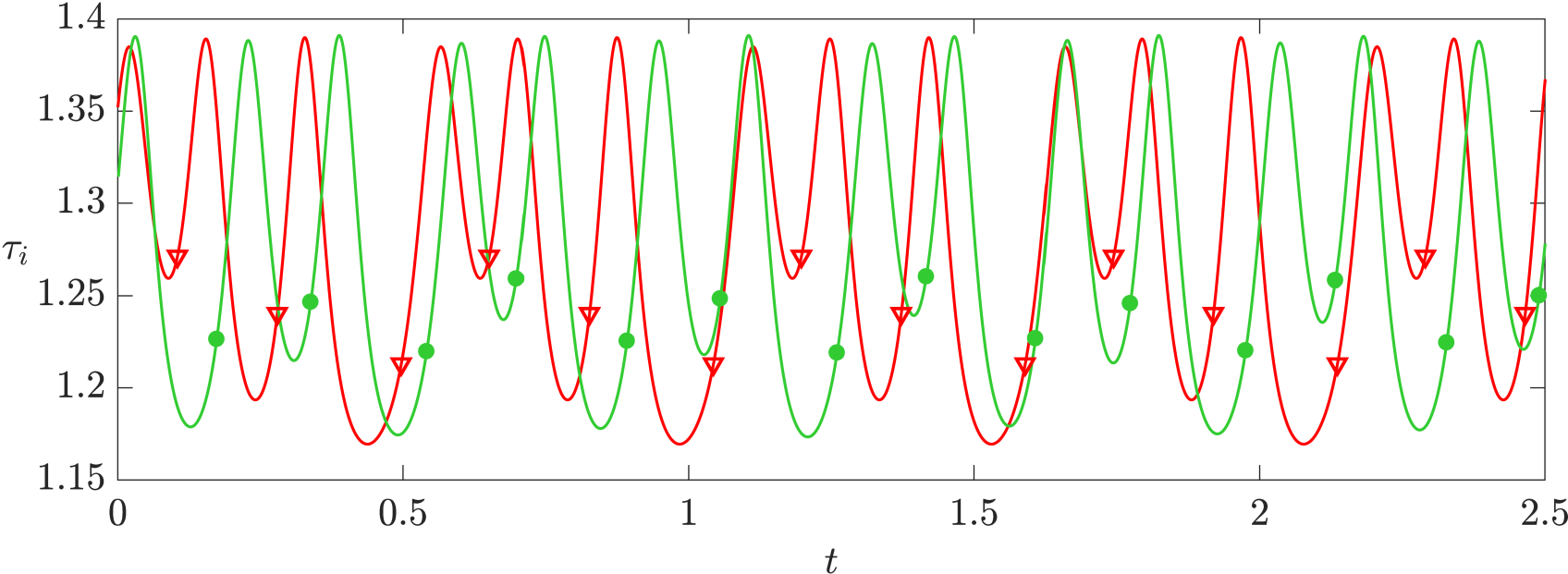} \\
  \begin{tabular}{ll}
    (b) & (c) \\
      \includegraphics[height=.43\linewidth]{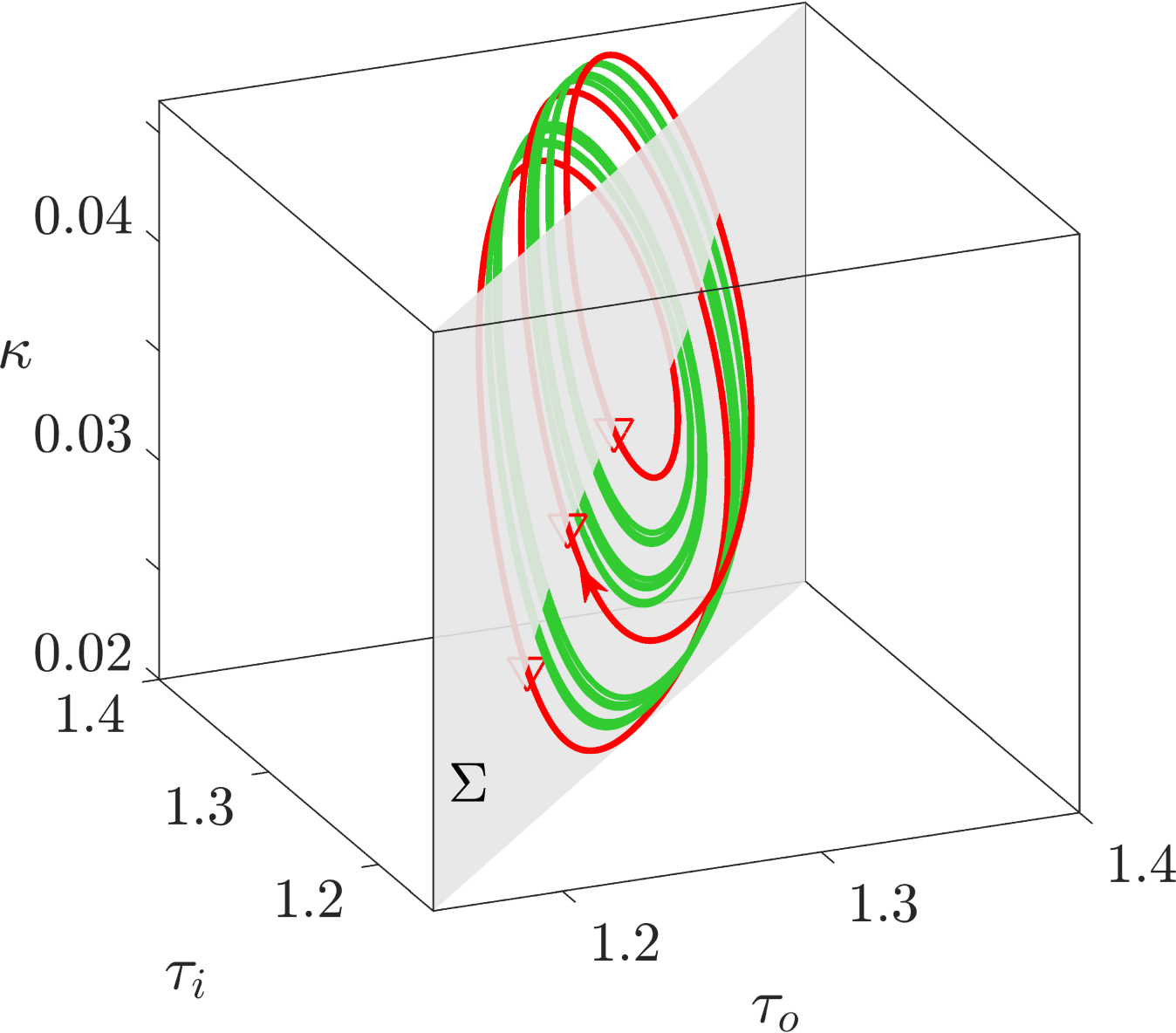} &
      \includegraphics[height=.43\linewidth]{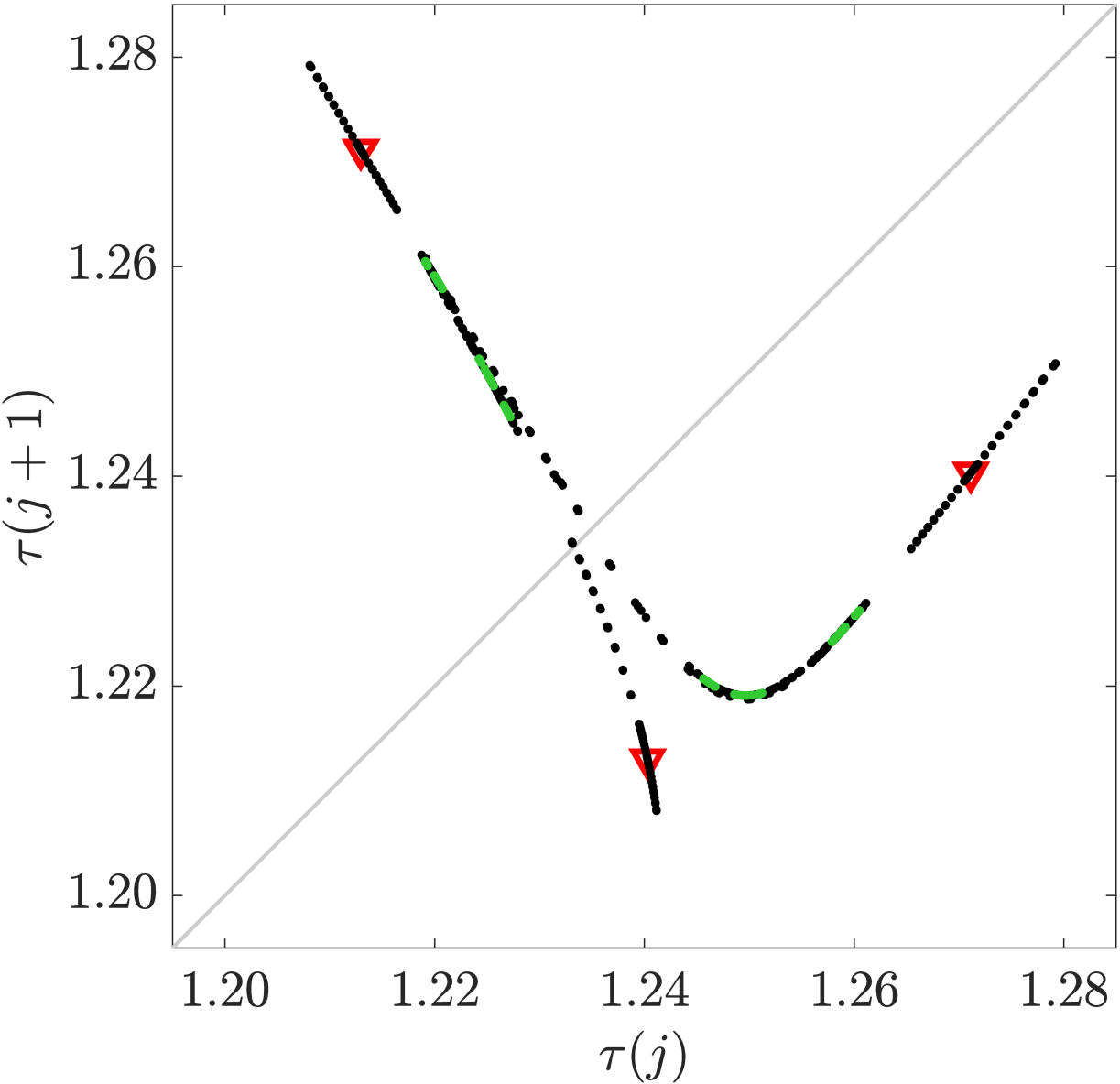} \\
   \end{tabular} 
   \end{tabular}
 \end{center}  
  \caption{Dynamics at the accumulation point $R=R_{\infty}$.
  (a) Time series of $\tau_i$ for the chaotic attractor obtained with DNS (green) and for $\mathrm{P}_3$ (red). Symbols indicate points on $\Sigma$.
The $\mathrm{P}_3$ solution here corresponds to P$_{3L}$ in figure \ref{fig:biffig}, which is unstable and has been computed with the Poincar\'e Newton-Krylov method.
  (b) Projection of the phase space onto the $(\tau_i,\tau_o,\kappa)$ coordinates. The grey plane indicates $\Sigma$. (c) Iteration map constructed from the torque sequence recorded on $\Sigma$. The black dots represent transient dynamics obtained by slightly perturbing $\mathrm{P}_{3L}$.
 }
\label{fig:Rinf}          
\end{figure} 
In the following, we will examine the dynamics employing various types of diagrams, all of which are exemplified in figure \ref{fig:Rinf} at the accumulation point {of the cascade} $R_\infty$. The green curve in figure \ref{fig:Rinf}a shows the $\tau_i$ time series of the chaotic attractor as obtained with DNS. Points on $\Sigma$ are represented with green bullets. While time signals are valuable for their simplicity, projecting both the {phase space} trajectory and $\Sigma$ onto the three-dimensional space spanned by $(\tau_i, \tau_o, \kappa)$, as {done} in figure~\ref{fig:Rinf}b, reveals their relationship in a more clear way. Collecting the torque at consecutive $\Sigma$ crossings in an ordered sequence ${\{\tau(n)\} = \{} \tau(1), \tau(2),\dots{\}}$, and plotting each value as a function of the one preceding it, yields the green points in the return map of figure \ref{fig:Rinf}c. 
The map appears quasi-one-dimensional, as revealed by the {transients} (black points) obtained {from perturbations of} the `period-3 orbit' identified by \citet{WaAyDeMeMe25b} (red triangles).
We refer to this periodic orbit as P$_{3}$, and represent it with a red line in figures~\ref{fig:Rinf}a and \ref{fig:Rinf}b, its three $\Sigma$ crossings shown as red triangles. While the details of P$_{3}$ will be discussed in the next section, the rationale behind its naming should already be clear from figure \ref{fig:Rinf}b. We shall shortly see that P$_{3}$ has upper and lower branches, the one depicted in figure \ref{fig:Rinf} corresponding to the lower branch P$_{3L}$.


\subsection{Crisis bifurcations and the {origin} of the P$_3$ solution family}\label{subsec:emergP3}

Now let us briefly outline some of the new findings already apparent in the bifurcation diagram of figure~\ref{fig:biffig}. The chaotic attractor C$_2$ generated by the Feigenbaum cascade loses stability at the critical Reynolds number $R_{c2}\approx 395.782$. At this point, the chaotic attractor (green points) collides with the P$_3$ (red dashed curve) {in} a {global bifurcation known as} boundary crisis in dynamical systems theory 
\citep{GrOtYo83,GrOtYo86}. The phenomenon is clearly observed in the bifurcation diagram thanks to the appropriate choice of $\Sigma$. 

To find the origin of P$_3$, we have tracked the branch {to} lower values of $R$. The branch has a turning point at FC$_1$, where a fold bifurcation {of limit cycles} occurs. We refer to the solutions before and after this turning point as P$_{3L}$ (lower branch) and P$_{3U}$ (upper branch), respectively, the former being the solution responsible for the boundary crisis at $R_{c2}$.
{Remarkably}, P$_{3U}$ is stable in the vicinity of FC$_1$ and triggers {its own} period doubling cascade as $R$ is increased (see the inset of figure~\ref{fig:biffig}). The {resulting} chaotic attractor (C$_{3}$, magenta points) {eventually} collides with P$_{3L}$ at $R_{c3}\approx 395.540$, also in a boundary crisis. When $R$ exceeds this value, the 
chaotic state can no longer be sustained 
indefinitely 
and trajectories are 
ultimately attracted to P$_2${, which is still stable in that region of parameter space}.
A representation of the resulting transient chaos 
can be obtained by initialising DNS simulations with slight perturbations of P$_{3L}$ and collecting the initial transients
(black dots in the bifurcation diagram, corresponding to multiple runs with varying initial perturbations of P$_{3L}$; note that they do not form an invariant set and are therefore not a chaotic saddle as defined in dynamical systems theory). 
{As $R$ is increased, the transient dynamics encompass an increasingly larger portion of the phase space until colliding with DRW$_L$ (dashed black line in figure~\ref{fig:biffig}) at $R_{c3}'\approx 395.563$.}
The {transient chaos does not disappear at this point, but its} qualitative properties change, as will be clarified later in \S\ref{subsec:globbif}.

\begin{figure}                        
  \begin{center}
   \begin{tabular}{ll}
     (a) & (b) \\
       \includegraphics[height=.37\linewidth]{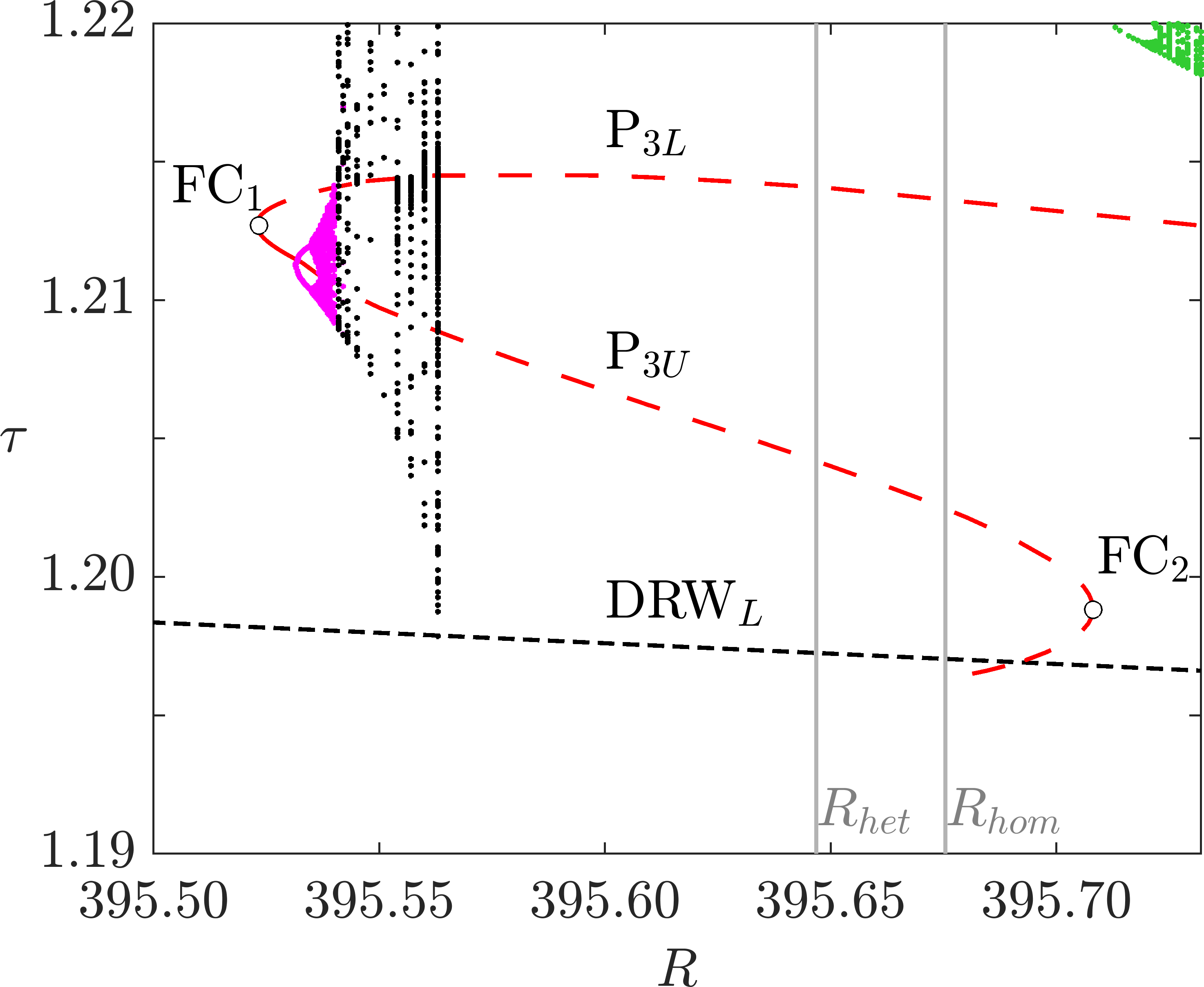} &
       \includegraphics[height=.365\linewidth]{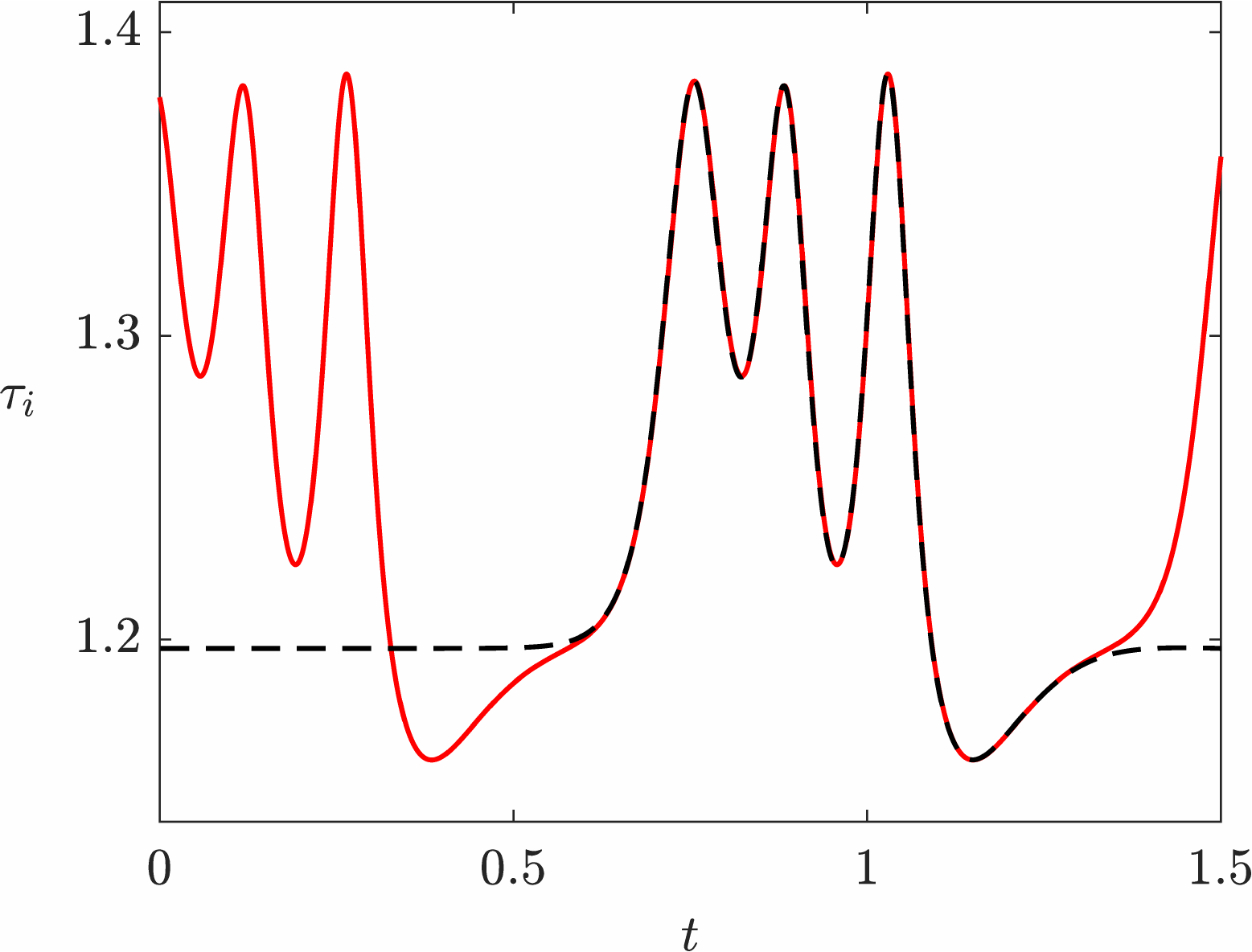} \\
    \end{tabular} 
 \end{center}  
  \caption{Disappearance of P$_3$ in a homoclinic bifurcation.
  (a) Enlarged view of the bifurcation diagram (figure \ref{fig:biffig}) near the homoclinic bifurcation point $R_{
  {hom}}$. 
  (b) Time series at $R=395.675 \approx R_{{hom}}$ of the unstable manifold of DRW (dashed black) and the P$_3$ solution (solid red) at the closest attainable {$R = 395.679$}.
  %
    }
\label{fig:Rhom}          
\end{figure} 
At much larger $R$, the P$_{3U}$ branch in figure~\ref{fig:biffig} undergoes another cyclic fold bifurcation, FC$_2$, at $R\approx 395.708$.
Shortly after the turn, the P$_3$ branch collides with DRW$_{L}$ and disappears. This critical point is denoted as $R_{{hom}}$ in the enlarged view of the bifurcation diagram shown in figure~\ref{fig:Rhom}a, and corresponds to the formation of a 
{homoclinic orbit} from DRW to itself. 
Figure~\ref{fig:Rhom}b compares {the} {$\tau_i$ time series} {along} the unstable manifold of DRW$_{L}$ at $R = 395.675 \approx R_{{hom}}$ (dashed black line), i.e. the approximate homoclinic orbit, and the P$_3$ solution (solid red) calculated at the closest attainable value of the Reynolds number. The excellent match between the time series provides strong evidence for the occurrence of a homoclinic bifurcation involving P$_3$ and DRW. Note that \cite{WaAyDe22} confirmed that DRW$_L$ {has only} one real unstable eigenvalue and, therefore, {its} unstable manifold can be easily approximated using DNS. 




\subsection{Changes in the dynamics due to global bifurcations}\label{subsec:globbif}
In general, global bifurcations {such as boundary} crises {induce} a significant {transformation} {of} the phase space topology. Among the critical Reynolds numbers discussed so far, $R_{c2}$ produces the most obvious alteration, 
as the dynamics on C$_{2}$ cease to be permanent
and all initial 
perturbations 
eventually decay to CCF, the sole remaining attractor.
As a consequence, C$_{2}$ can be straightforwardly classified as a type (i) chaotic saddle, similar to that found by \cite{BuDoHo19}. Relaminarision events of saddle C$_2$, particularly in the vicinity of $R_{c2}$, are often preceded by dynamics that are reminiscent of the $P_3$ orbit, an observation that motivated its discovery in the first place \citep{WaAyDeMeMe25b}. However, a connection between P$_3$ and CCF is not immediately apparent from the bifurcation diagram of figure~\ref{fig:biffig}, {which shows} no {direct branching} between the two solutions. The relation, of a dynamical nature, will become clear in {section 5.} 

{The bifurcation diagram of figure~\ref{fig:biffig} shows all stable states along with the most relevant unstable solution branches, i.e. DRW$_L$ (dashed black), the unstable portion of P$_3$ (dashed red) and the {chaotic transients generated by the} C$_3$ chaotic saddle (cloud of black dots).}
{In the parameter region of interest,}
{from {PD$_1$} to $R_{c2}$, there are at least two attractors in the phase space. One is CCF, which exists {and is stable} all along, and the other is either P$_2$ (up to PD$_2$) or the cascade that follows.}
{In the interval between FC$_1$ and $R_{c3}$, however, the attractor count rises to three. The third attractor is initially P$_{3U}$, then the cascade it spawns.}
In order to gain a qualitative understanding of the phase space changes, it is useful to monitor trajectories starting from the close neighbourhoods of P$_{3L}$ {and DRW$_L$}. From such trajectories, computed with DNS, the following observations can be readily made:

\begin{enumerate}[label=(\alph*)]
\setlength{\itemsep}{0.2cm}


\item ~{For $R\lesssim R_{c3}$, trajectories starting from close to P$_{3L}$ either lead to $P_2$ {(grey circles in figure~\ref{figE}a)} or evolve towards the chaotic attractor C$_3$ (black crosses).
\begin{figure}                                                                 
  \begin{center}
   \begin{tabular}{cc}
    \raisebox{16.5em}{ (a)} & 
\includegraphics[width=.7\linewidth]{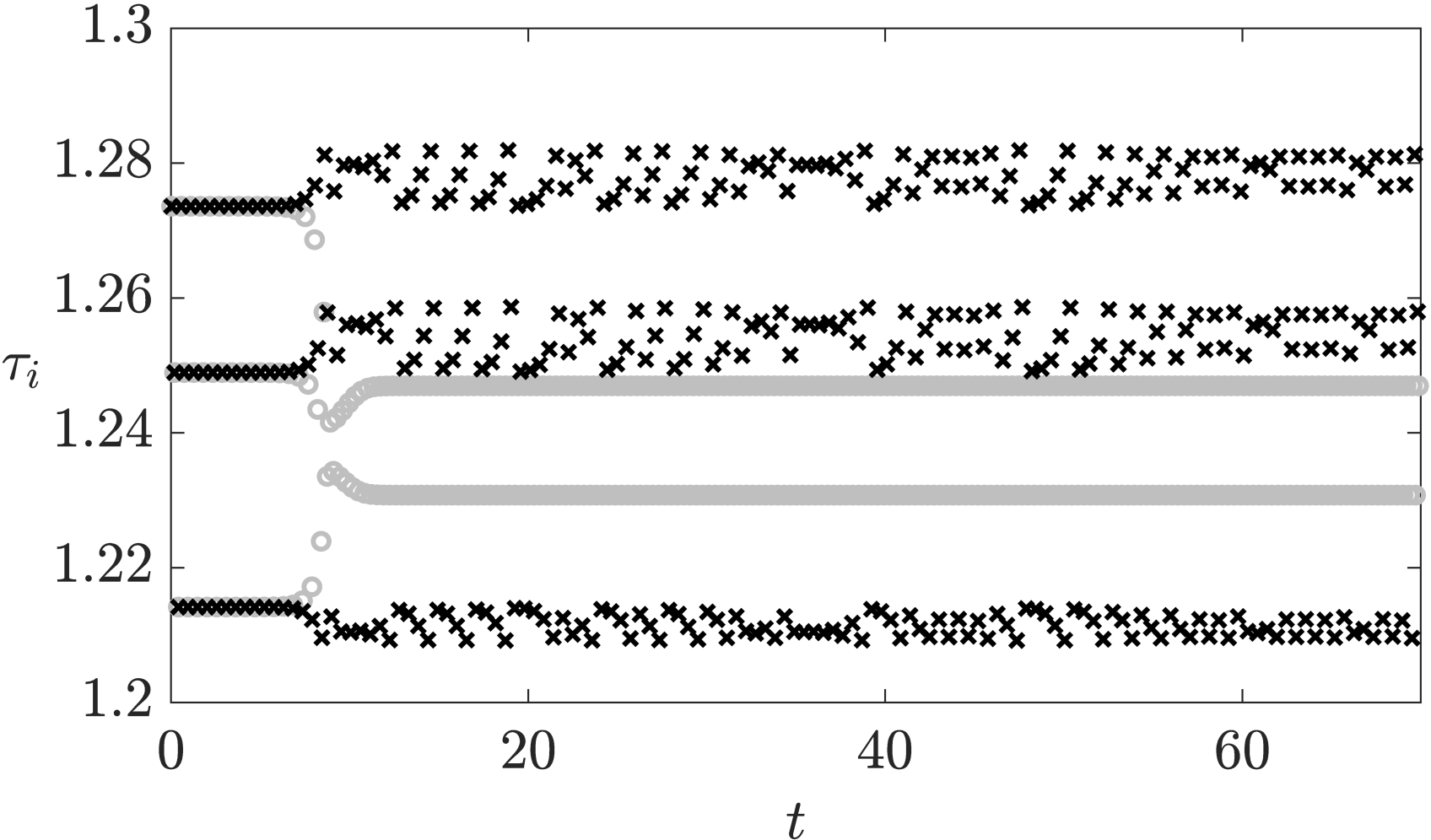} 
\put(-30,135){\vector(2,-1){10}}
\put(-40,138){C$_3$}
\put(-30,55){\vector(2,1){10}}
\put(-40,50){P$_2$}
\put(-220,74){\vector(-1,2){6.4}}
\put(-225,65){P$_{3L}$}
\put(-225,140){$R\lesssim R_{c3}$}
\\
      \raisebox{16.5em}{(b)} & 
       \includegraphics[width=.7\linewidth]{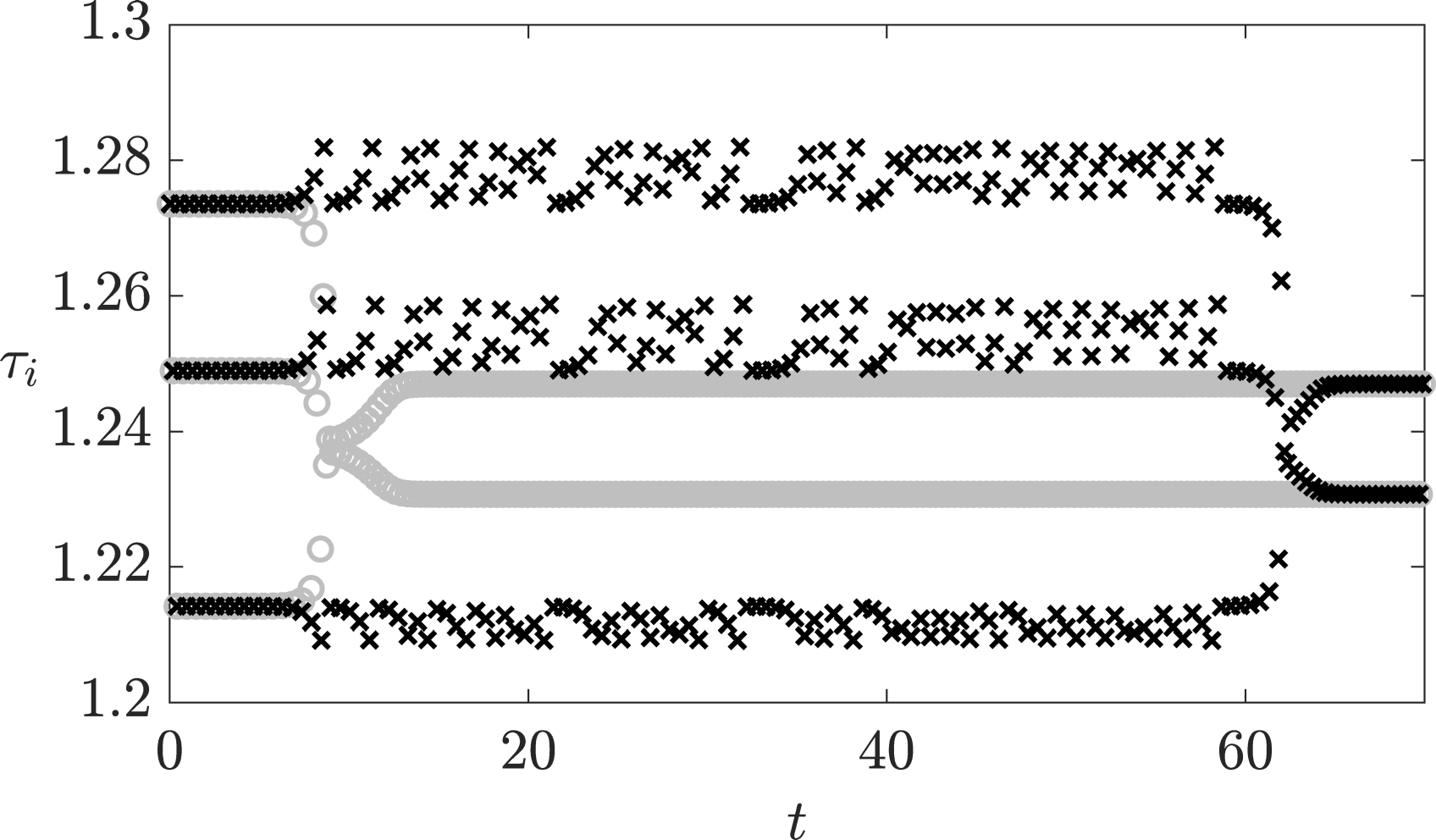}
\put(-110,138){\vector(2,-1){10}}
\put(-120,141){C$_3$}
\put(-110,55){\vector(2,1){10}}
\put(-120,50){P$_2$}
\put(-220,75){\vector(-1,1){10}}
\put(-225,65){P$_{3L}$}
\put(-225,140){$R\gtrsim R_{c3}$}
\\
    \raisebox{16.5em}{ (c)} & 
       \includegraphics[width=.7\linewidth]{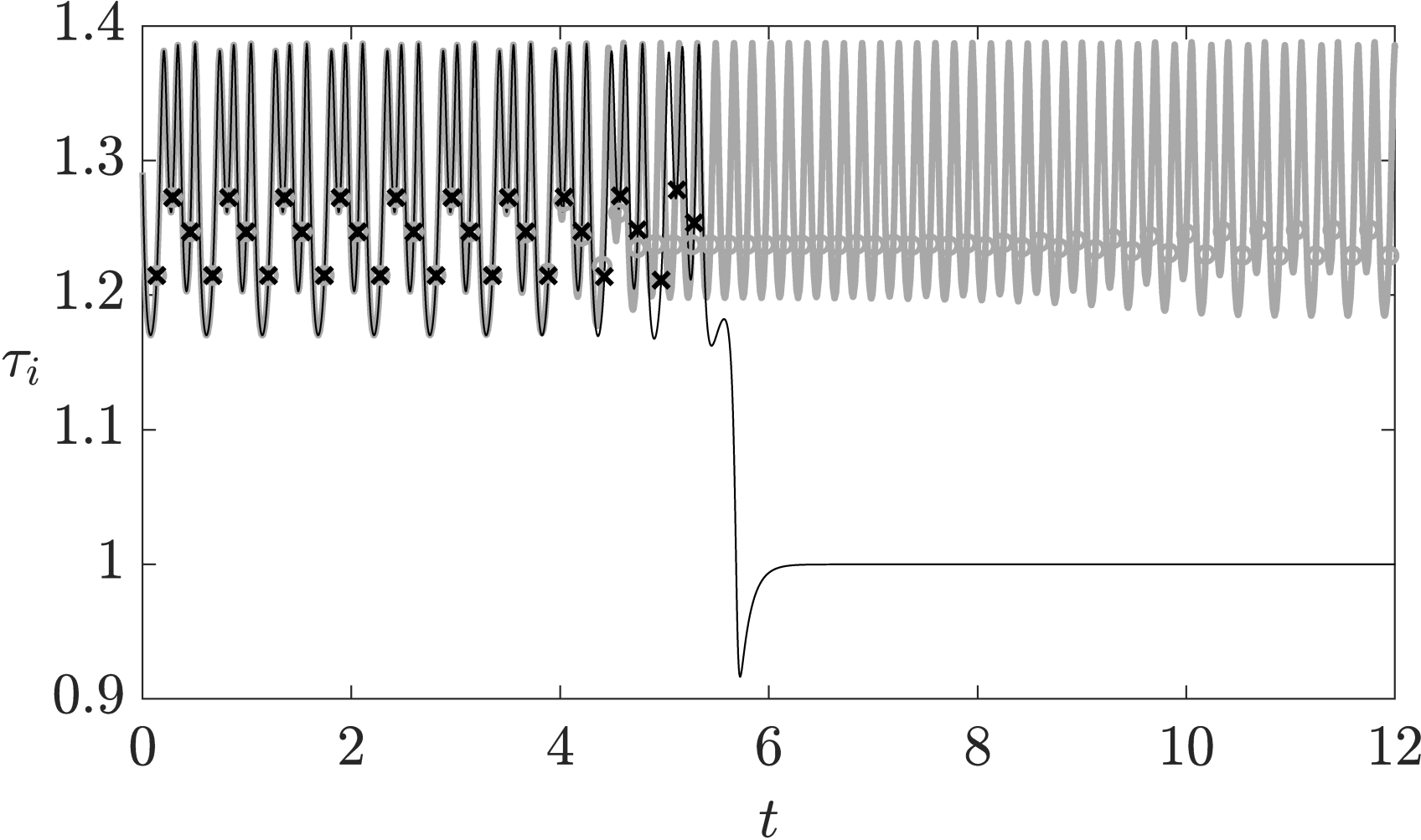} 
\put(-30,90){\vector(2,1){10}}
\put(-40,85){P$_2$}
\put(-225,85){\vector(-1,1){10}}
\put(-225,75){P$_{3L}$}
\put(-30,60){\vector(2,-1){10}}
\put(-40,65){CCF}
\put(-225,40){$R\gtrsim R_{c3}'$}
       \\
    \end{tabular} 
 \end{center}  
  \caption{
Time series of $\tau_i$ corresponding to DNS started from small random perturbations of P$_{3L}$ at (a) $R=395.5400\lesssim R_{c3}$, 
 (b) $R=395.5401\gtrsim R_{c3}$ and (c) $R=395.564\gtrsim R_{c3}'$. 
Only two time series of the many run are shown in each panel to illustrate the two qualitatively different behaviours observed at each of the values of $R$. Panels (a) and (b) only depict the values of $\tau_i$ on $\Sigma$, while (c) superposes also the continuous time series to capture the decay towards CCF. 
{Perturbation amplitudes are of order $O(10^{-8})$ relative to P$_{3L}$.}
 }
\label{figE}          
\end{figure}
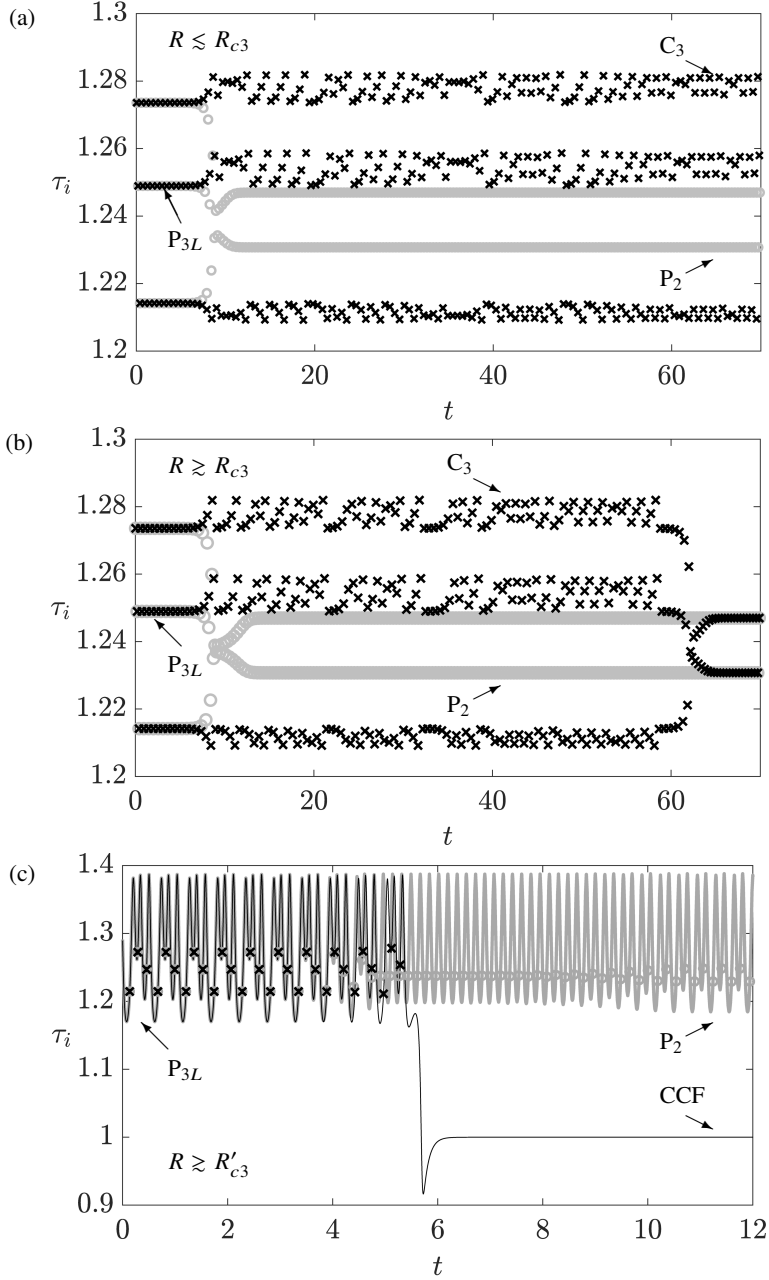
{On the other hand,} starting from the vicinity of DRW$_L$, the choice is between CCF (black line in figure~\ref{fig:DRWDNS}a) and P$_{2}$ {(grey line)}.
\begin{figure}                        
  \begin{center}
  \begin{tabular}{ll}
  (a) & (b) \\
     \includegraphics[width=.45\linewidth]{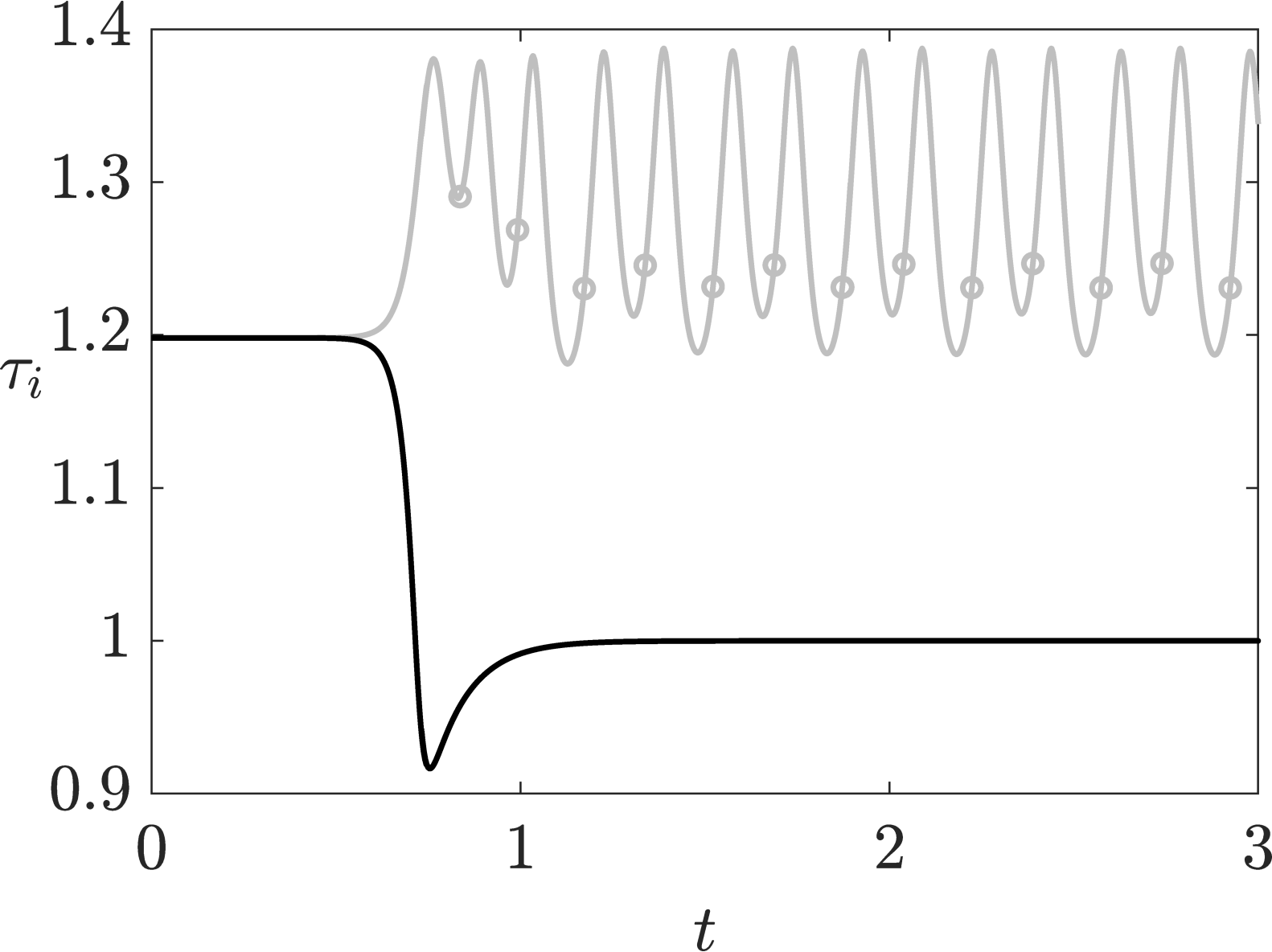}  &
     \includegraphics[width=.45\linewidth]{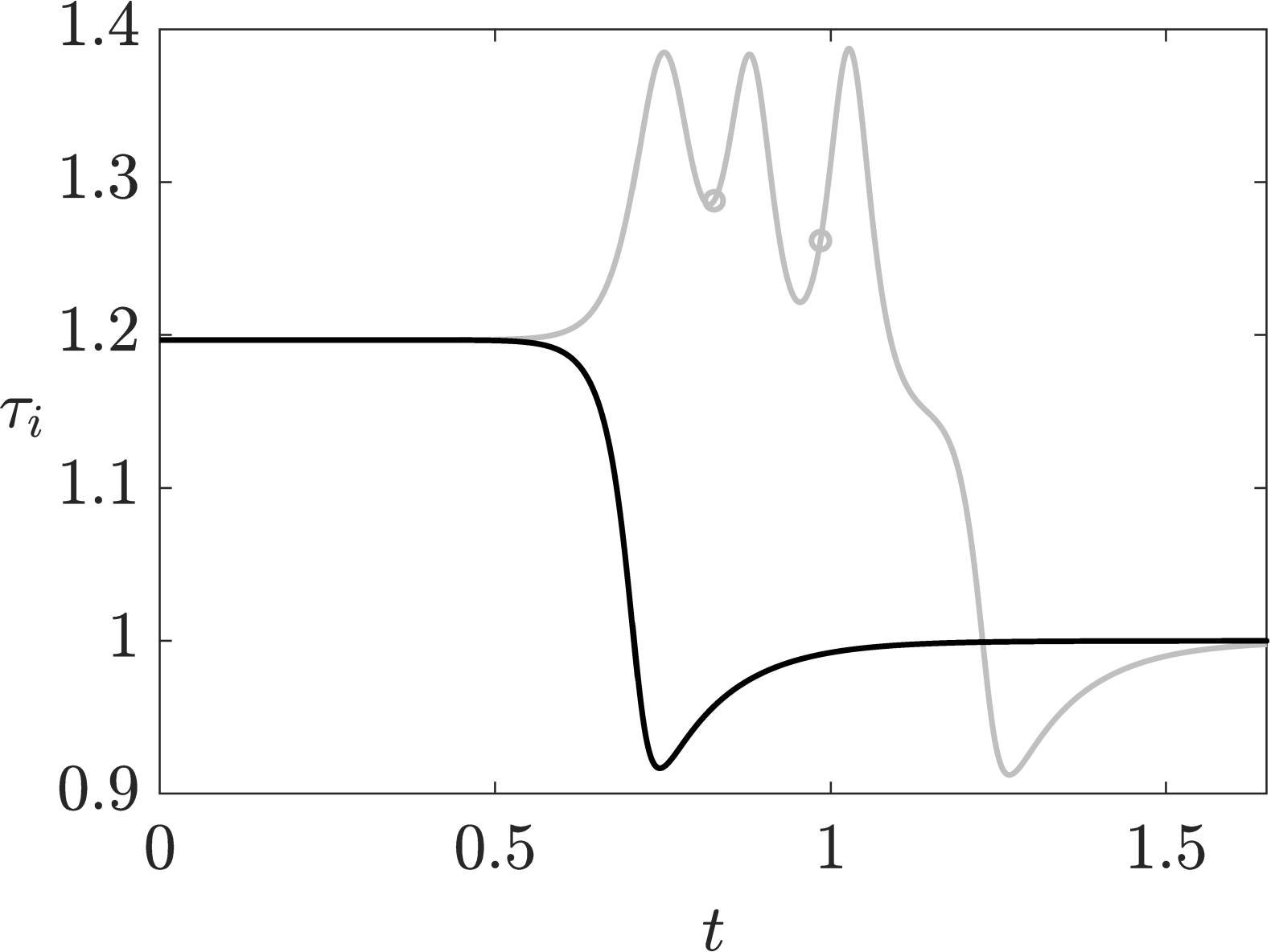} \\
    \end{tabular}
 \end{center}   
  \caption{
  Unstable manifold of DRW$_L$ at (a) $R=395.5400\lesssim R_{c3}$ and (b) $R=R_{\infty}>R_{{hom}}$.  
  {The manifolds are closely approximated with DNS started from very slight perturbations of DRW$_L$ to either side in the direction of its only unstable eigenfunction. Perturbation amplitudes are of order $O(10^{-10})$ relative to DRW$_L$.}
    }
\label{fig:DRWDNS}          
\end{figure} 
The edge tracking algorithm confirms that both P$_{3L}$ and DRW$_L$ act as edge states {at this stage}, as will be argued later.
}


\item ~{For $R\gtrsim R_{c3}$, C$_3$ {has become} a chaotic saddle and the attractor count is reduced to two. As illustrated by figure~\ref{figE}b, trajectories starting from close to P$_{3L}$ can only reach P$_2$, but two different paths may be observed. The departure towards P$_2$ can either be uneventful (grey) or transiently visit C$_3$ prior to relaxing on P$_2$ (black). Since the transient dynamics of C$_3$ lead always to P$_2$, the chaotic saddle is type (i). The clouds of black points in figures~\ref{fig:biffig} and \ref{fig:Rhom}a depict transient visits to C$_3$ obtained from multiple DNS simulations started from perturbations of P$_{3L}$. 
The duration of these transients is highly sensitive to initial conditions; {we shall see later that this is a consequence of} the fractal structure of C$_3$ in phase space.
As before, starting from close to DRW$_L$ may still lead to CCF or to P$_2$, but now DRW$_L$ is the only solution acting as an edge state.}

\item ~For $R\gtrsim R_{c3}'$, CCF has become within reach of trajectories starting from the vicinity of P$_{3L}$. The dynamics can still {end on P$_2$, either directly (grey line and circles in figure~\ref{figE}c) or after transiently approaching C$_3$ (not shown), }
or can alternatively decay to CCF {after short visits to C$_3$} ({black}). 
We will see in section 3 that the chaotic saddle C$_3$ is now in the boundary separating the basins of attraction of CCF and P$_2$, 
but edge tracking invariably converges on DRW$_L$, which remains the sole edge state. 
{Thus, C$_3$ is not an edge state and is therefore classified as type (ii).}

\item ~For sufficiently large $R$, the route from DRW$_L$ to P$_2$ has been severed altogether. 
Figure~\ref{fig:DRWDNS}b shows typical trajectories starting in the vicinity of DRW$_L$ at the accumulation point {$R_\infty$}. 
Perturbing in one or the other direction of its one-dimensional unstable manifold always results in laminar decay, sometimes uneventfully (black line), sometimes after some transients (grey). Despite still having a single unstable eigenvalue, DRW$_L$ is no longer an edge state. 


\end{enumerate}

A natural question arises as to when {and how} the qualitative change in the unstable manifold of DRW$_L$, from figure~\ref{fig:DRWDNS}a to figure~\ref{fig:DRWDNS}b, occurs. We will show later that this happens at $R=R_{het}$ indicated in figure~\ref{fig:Rhom}a. For $R\gtrsim R_{het}$, the edge state becomes chaotic, which, according to our classification, corresponds to a type (iii) chaotic saddle. 

\vspace{2mm}

\section{The emergence of the chaotic saddle C$_3$}
\label{sec:emergence}


\begin{figure}                                                                 
  \begin{center}
   \begin{tabular}{ll}
     (a) & (b) \\
       \includegraphics[height=.4\linewidth]{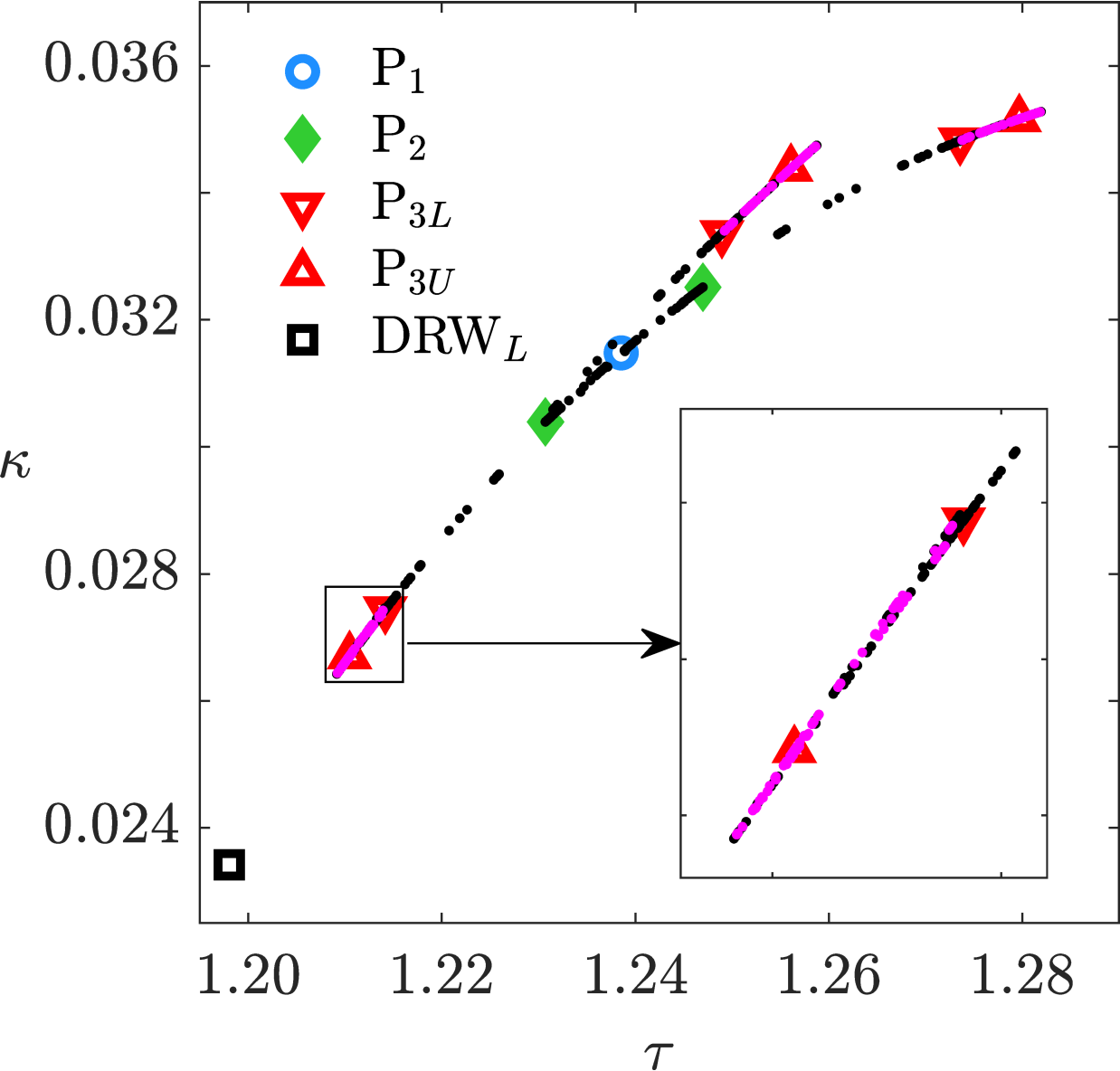} &
       \includegraphics[height=.4\linewidth]{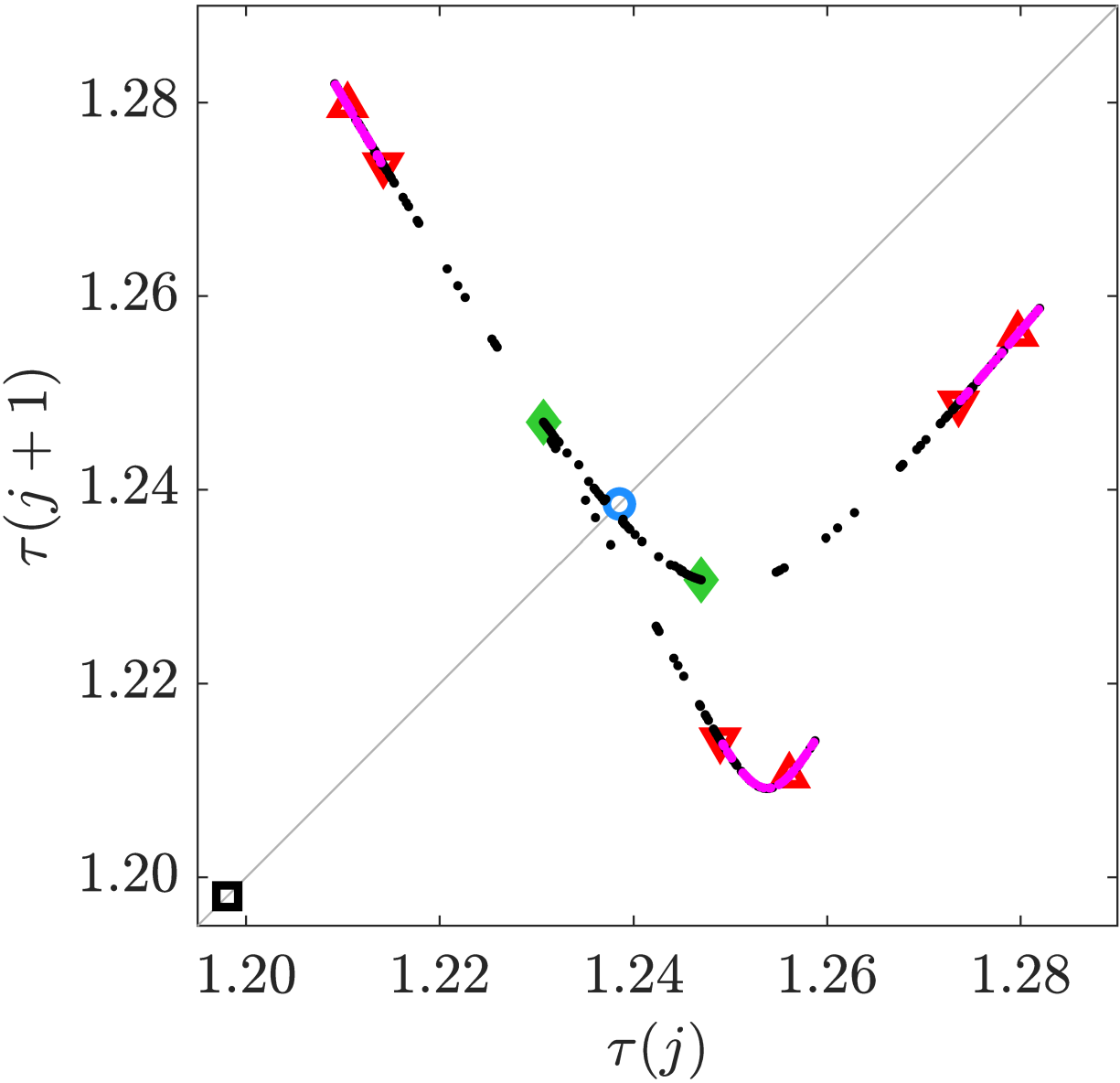} \\
           (c) & (d) \\
       \includegraphics[height=.4\linewidth]{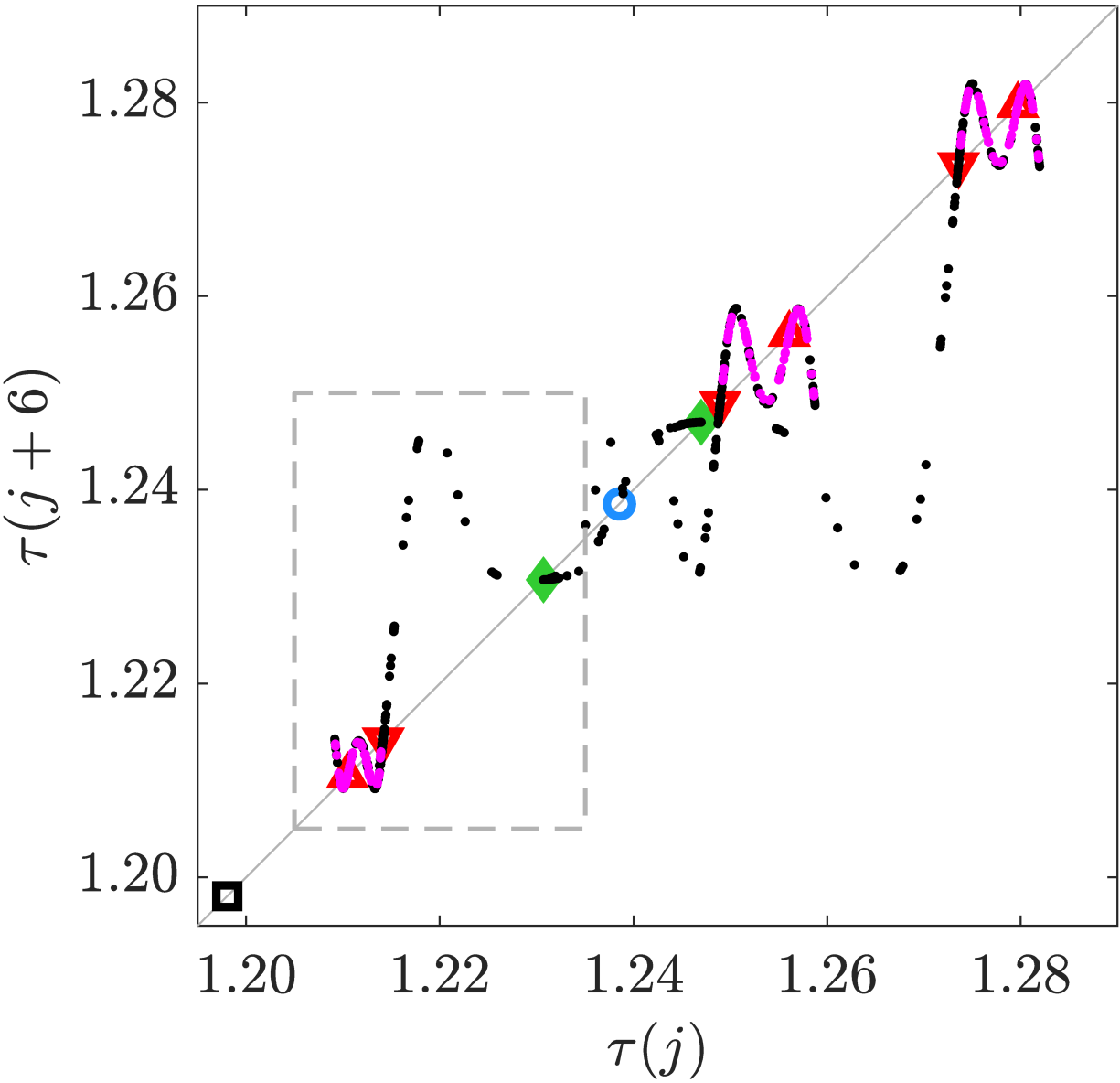} &
       \includegraphics[height=.41\linewidth]{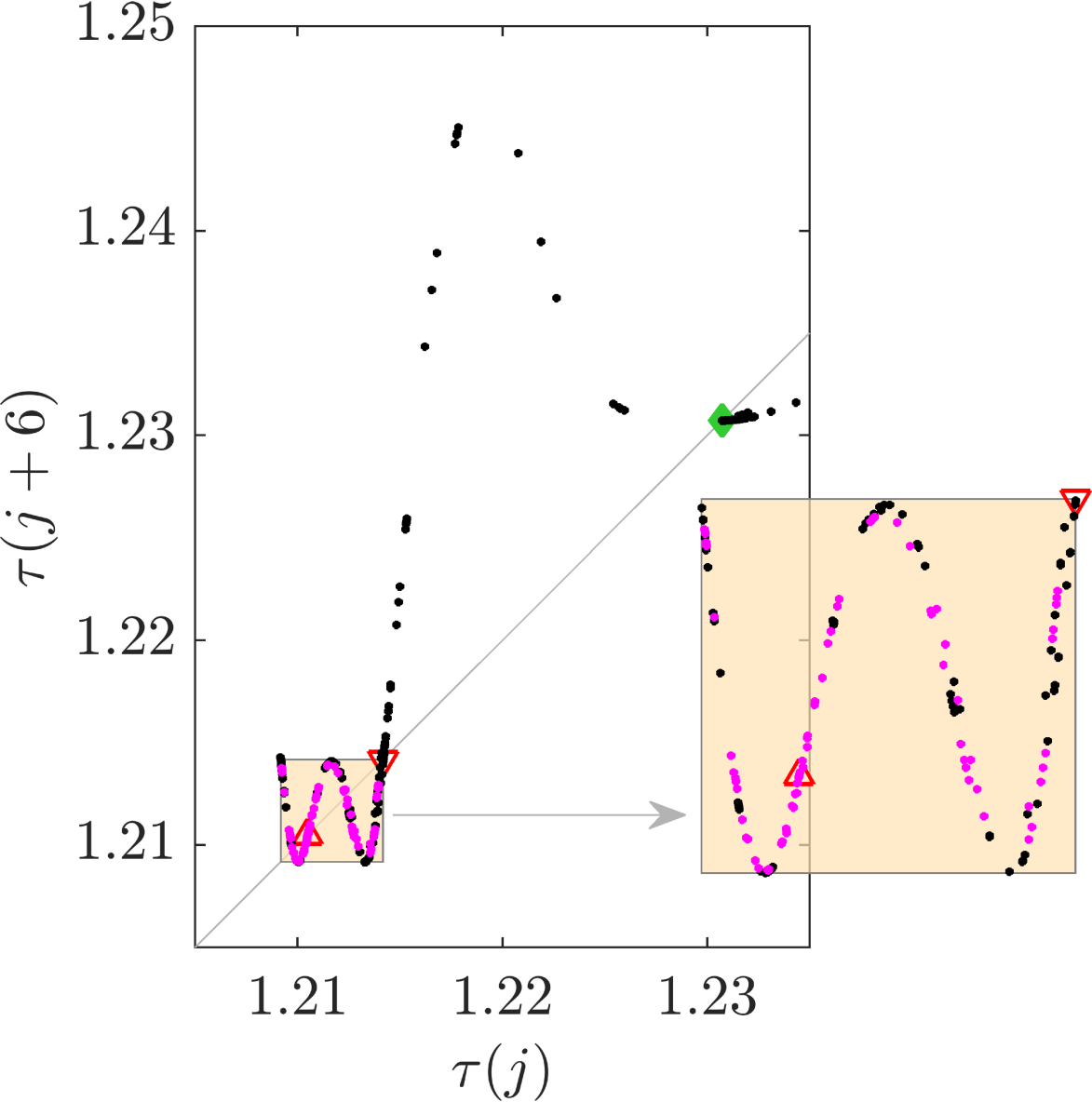} \\
    \end{tabular} 
 \end{center}
  \caption{{The transition of C$_3$ from attractor to saddle {around} $R=R_{c3}$.} All data is recorded on $\Sigma$. 
  The magenta points represent the C$_{3}$ attractor obtained from DNS at $R = 395.5400 \lesssim R_{c3}$, while the symbols shown in the legend correspond to Poincar\'e-Newton-Krylov 
  solutions.
  The black dots show the transient dynamics obtained by perturbing P$_{3L}$ at $R=395.5401 \gtrsim R_{c3}$. (a) Projection of the points onto the $\tau$--$\kappa$ plane. (b) First return map of the torque sequence. (c) Sixth return map. The region enclosed by the dashed box is magnified in panel (d).
  %
}
\label{fig:mapRc1}          
\end{figure}

As noted in the previous section, the first appearance of transient chaos occurs at $R=R_{c3}$. The main aim of this section is to examine the structure of the underlying chaotic saddle and to elucidate its emergence.
The starting point is figure~\ref{fig:mapRc1}a, which shows the {representation on} $\Sigma$ of exact solutions (see the legend) and both permanent (magenta dots) and transient (black dots) chaotic trajectories {at $R\simeq R_{c3}$}. The magenta points represent the chaotic attractor C$_3$ at $R=395.4000\lesssim R_{c3}$. These points have been obtained by evolving the black trajectory in figure~\ref{figE}a for a long time and discarding the initial transients. Other key invariant objects are indicated with symbols in figure~\ref{fig:mapRc1}a. {In particular,} P$_{3L}$ {(down-pointing red triangle)} is located on {one} edge of the chaotic attractor, while P$_{3U}$ {(up-pointing red triangle) is at its core}.
The black dots in figure~\ref{fig:mapRc1}a denote transient chaos at $R=395.4001\gtrsim R_{c3}$.
They have been obtained {from many DNS runs initialised with small random perturbations of} P$_{3L}$, one example of which is shown in figure~\ref{figE}b. Roughly speaking, the chaotic saddle C$_3$ at $R=395.4001$ exists in the region where the black and magenta points overlap (see {inset} of figure~\ref{fig:mapRc1}a).
Note that the black and grey trajectories of figures~{\ref{figE}a and} \ref{figE}b correspond, respectively, to perturbations of P$_{3L}$ toward {and away from} the chaotic set {C$_3$.}

Figure~\ref{fig:mapRc1}a reveals that the discrete dynamics on $\Sigma$ evolve on a nearly one-dimensional manifold of negligible thickness. 
This motivates the construction of the return map shown in figure~\ref{fig:mapRc1}b. The function $f$ relating consecutive torques on $\Sigma$, defined by $\tau(j+1) = f(\tau(j))$, is multi-valued, but a straightforward analysis reveals that the branch selection rules are univocal. In the following, we first dissect in \S\ref{subsec:3.1} the branch selection rules of the return map using the period-doubling bifurcation starting from P$_2$ as a guide, before revisiting the global bifurcations that turn C$_3$ into a saddle in \S\ref{subsec:3.2} and \S\ref{subsec:3.3}.


\subsection{Structure of the return map and period-doubling originated from P$_2$}\label{subsec:3.1}

\begin{figure}
  \begin{center}
   \begin{tabular}{lclc}
     (a) & & (b) & \\
       & \includegraphics[height=.35\linewidth]{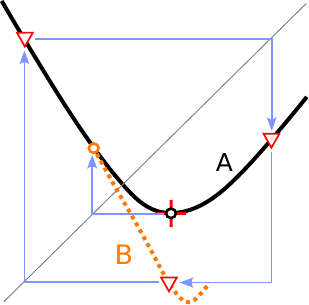} &
       & \includegraphics[height=.35\linewidth]{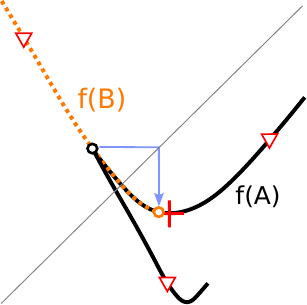} \\
           (c) & & (d) & \\
       & \includegraphics[height=.35\linewidth]{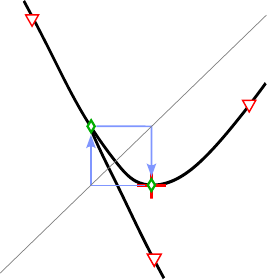} &
       & \includegraphics[height=.35\linewidth]{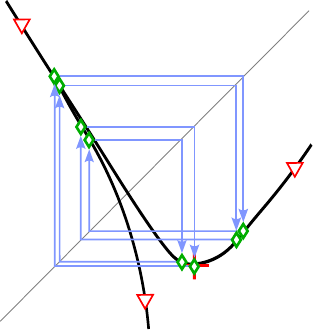} \\
    \end{tabular} 
 \end{center}  
  \caption{
Sketch of the iteration maps. 
Panels (a) and (b) are based on the data from figure \ref{fig:mapRc1}b. The branches A and B labeled in panel (a) are mapped under $f$ as illustrated by panel (b). Downward triangles represent P$_{3L}$, and the arrows indicate the action of $f$. Panels (c) and (d) show the return maps when P$_2$ and P$_8$ are superstable, respectively.
  }
\label{fig:skitmap}          
\end{figure} 
Figure~\ref{fig:skitmap}a shows a sketch of the map $f$, with the two distinct branches labelled as A and B. The blue arrows indicate how $f$ acts on points of the map. Points corresponding to a periodic orbit form a closed loop as illustrated for P$_{3_L}$ (triangles).
Meanwhile, $f$ maps the local minimum of branch A (black circle on top of the red plus sign) onto the point where branch B bifurcates from branch A (orange circle). In general, application of $f$ to the branching point (black circle) does not map back onto the minimum of branch A (red plus sign), as exemplified, at the same $R$, in figure~\ref{fig:skitmap}b. The figure also shows the action of $f$ on the two branches as a whole. Branch B (dashed orange) maps onto a portion of branch A (solid black), while branch A maps onto the union of branch B and the portion of itself to the right of the branching point.
The immediate implication of this, is that predicting precisely which branch the trajectory will be on after $n$ iterations is troublesome when $n$ is large. 

An exception to this occurs when the minimum of the map (red plus sign) is a point of a periodic orbit, in which case the orbit is said to be superstable (the multiplier is zero). Figure~\ref{fig:skitmap}c shows an example for a superstable P$_2$.
In this case, the map can be partitioned at the periodic points of P$_2$, so that all the points on a partition are mapped by $f$ onto a partition or a combination of partitions. The map is said to {\it admit} a Markov partition. 
Figure~\ref{fig:skitmap}d is a sketch of the map when P$_8$ is superstable, this being another case for which a Markov partition is possible. {The cobweb of successive superstable orbits helps} better understand the action of $f$, as well as how the period-doubling bifurcations accumulate toward the limiting case shown in figure \ref{fig:Rinf}c. The figures also confirm that the multivalued nature of $f$ does not lead to any peculiar behaviour in the period-doubling cascade, as the map consistently behaves like a quadratic function near the local minimum of branch A.

\subsection{Return map analysis around $R=R_{c3}$}\label{subsec:3.2}

Let us now examine the flow dynamics around $R=R_{c3}$. As clear from figure~\ref{figE}b, 
the P$_2$ and P$_{3L}$ orbits play key roles in the dynamics.
This observation motivates the depiction of the 6th return map in figure~\ref{fig:mapRc1}c, making use of the fact that the least common multiple of 2 and 3 is 6 and, therefore, P$_2$, P$_{3L}$ and P$_{3U}$ are all fixed points of $f^6$.

Figure~\ref{fig:mapRc1}d is a close-up of the region enclosed by the dashed box in figure~\ref{fig:mapRc1}c. Since $f^6$ is single-valued in this region, the usual cobweb analysis can be readily applied. The W-shaped magenta structure corresponds to the C$_{3}$ chaotic attractor that exists at $R=395.5400 \lesssim R_{c3}$.  The shape of the map is similar to the second iteration of the logistic map, 
{which suggests} that a local unimodal structure must have appeared in $f^3${, as is the case indeed}. Returning to the original map $f$ of figure~\ref{fig:mapRc1}b, the unimodal structure originates from 
{the lowermost cluster} of magenta points.
That is, the {period doubling} cascade generating C$_{3}$ can be traced back to the local minimum of the branch labeled B in figure \ref{fig:skitmap}a.




The shaded square enclosing C$_{3}$ {in figure~\ref{fig:mapRc1}d} has been defined 
using the furthermost point that $f^6$ maps onto P$_{3L}$.
Cobweb analysis shows that, for $R=395.5400\lesssim R_{c3}$, any trajectory started from within the square remains in the square. In other words, no trajectory can escape the shaded region.
P$_{3U}$ is embedded in the chaotic attractor, whereas small perturbations to P$_{3L}$, located at a corner of the square, can lead to either the chaotic attractor C$_{3}$ or P$_2$, depending on whether they fall within the square or outside. This is consistent with observation (a) in \S\ref{subsec:globbif}.

The critical point $R_{c3}$ {corresponds to} the {situation in which the} local extrema of $f^6$ touch the {shaded} square. {For $R<R_{c3}$, the W-shaped segment of the map is fully contained within the square.}
When the local extrema of the W-shaped map extend beyond the square region for $R>R_{c3}$, an escape pathway toward P$_2$ is formed, as can be seen from the cobweb analysis. The black dots in figure \ref{fig:mapRc1} depict a collection of trajectories escaping from the square at $R=395.5401\gtrsim R_{c3}$. Notice, however, that these trajectories {are not \emph{on} the chaotic saddle itself but} represent {instead} transient dynamics {\emph{in} its vicinity that eventually lead to P$_2$}, as will be clarified in \S\ref{subsec:3.3}.


\subsection{Fractality of the chaotic saddle}\label{subsec:3.3}

\begin{figure}
  \begin{center}
   \begin{tabular}{ll}
     (a) & (b) \\
       \includegraphics[height=.4\linewidth]{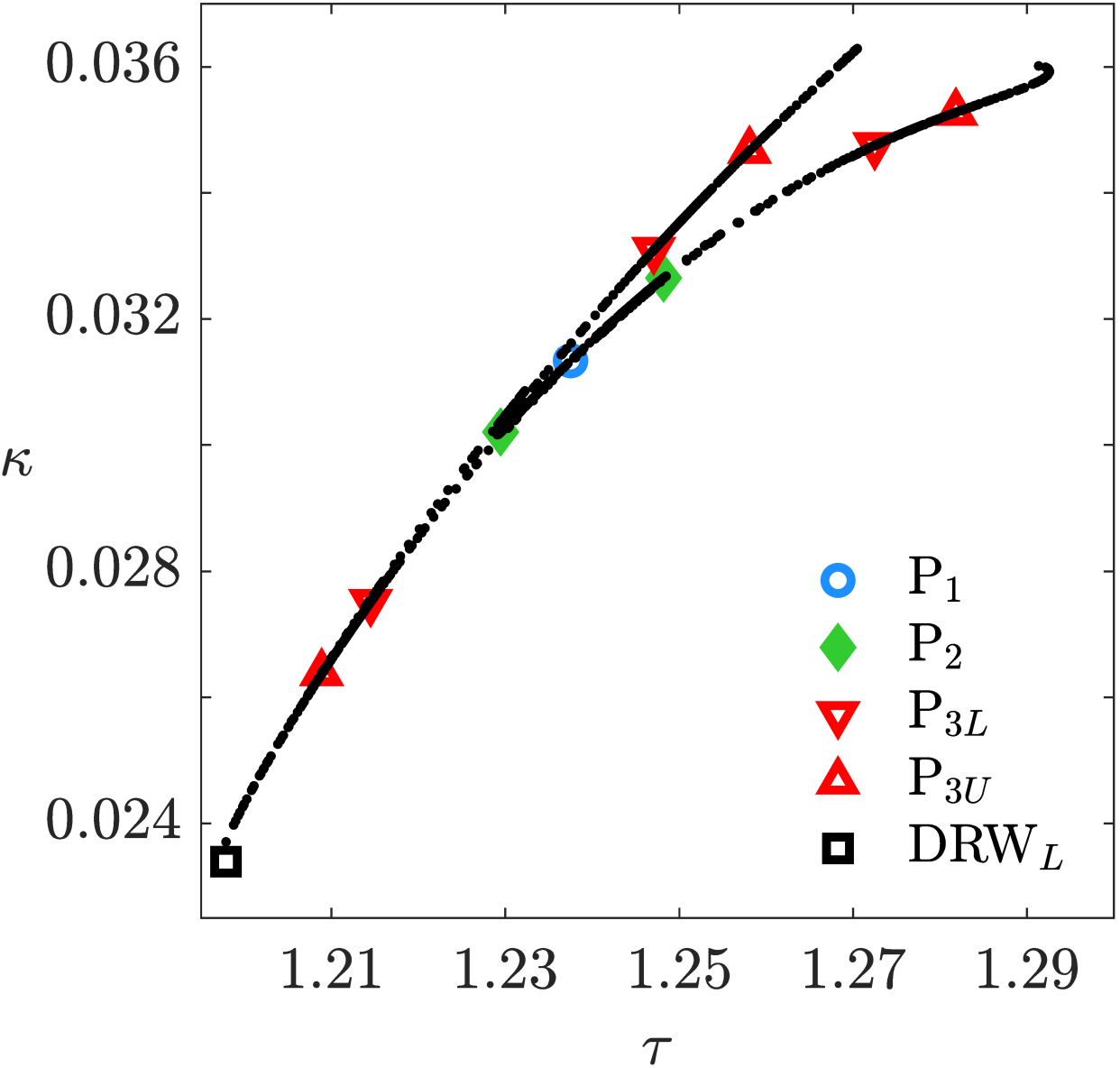} &
       \includegraphics[height=.4\linewidth]{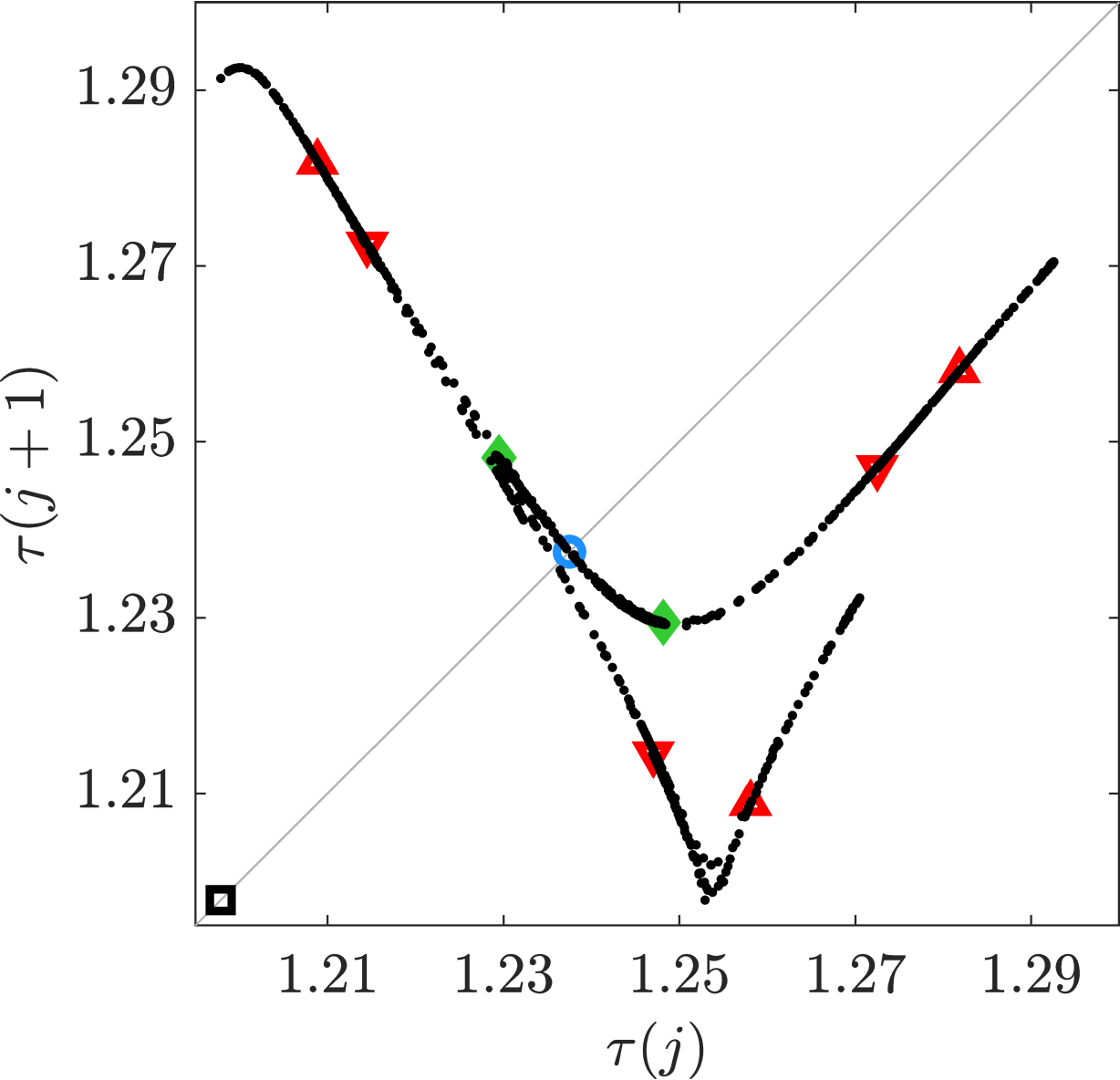} \\
           (c) & (d) \\
       \includegraphics[height=.4\linewidth]{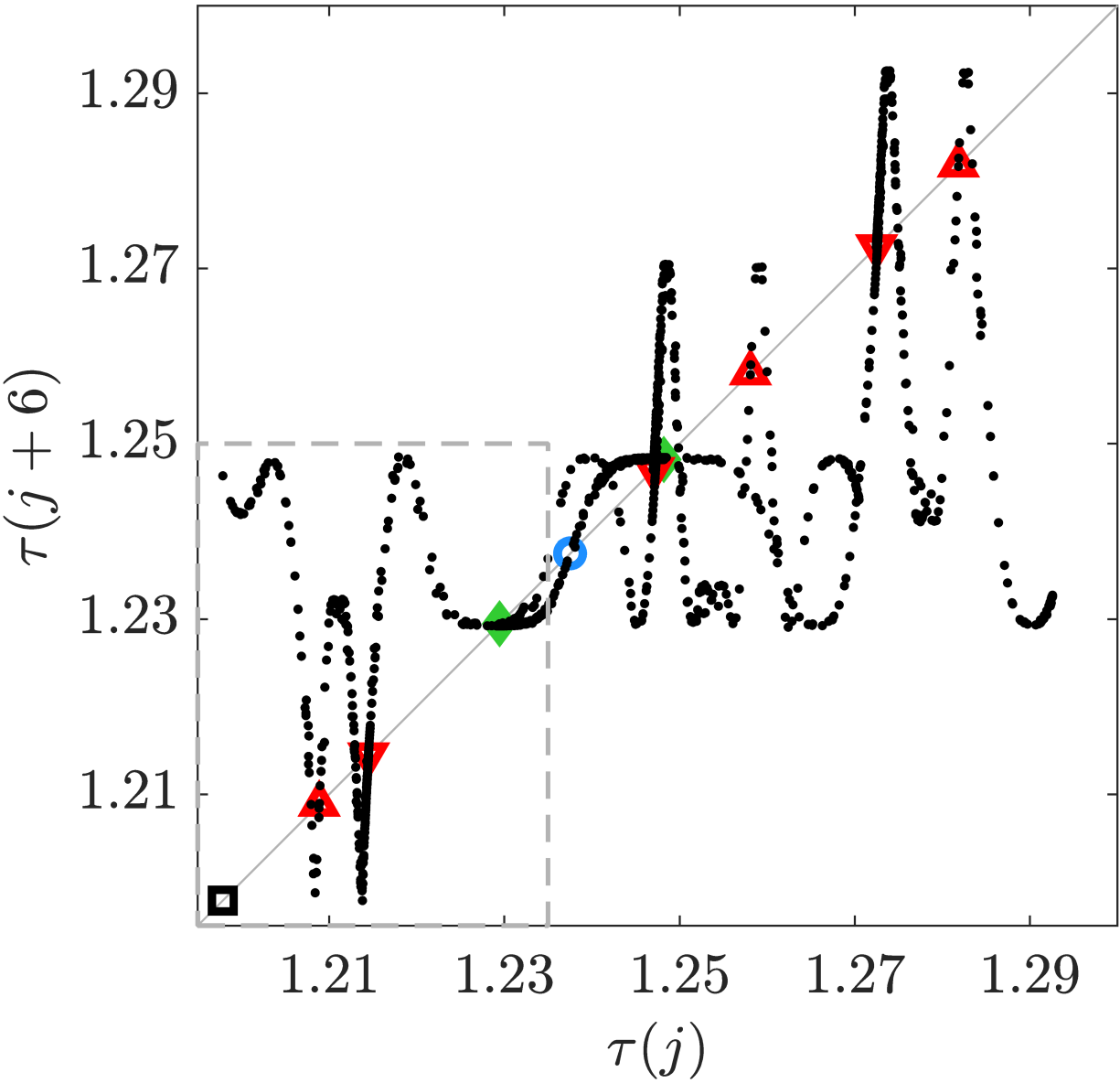} &
       \includegraphics[height=.41\linewidth]{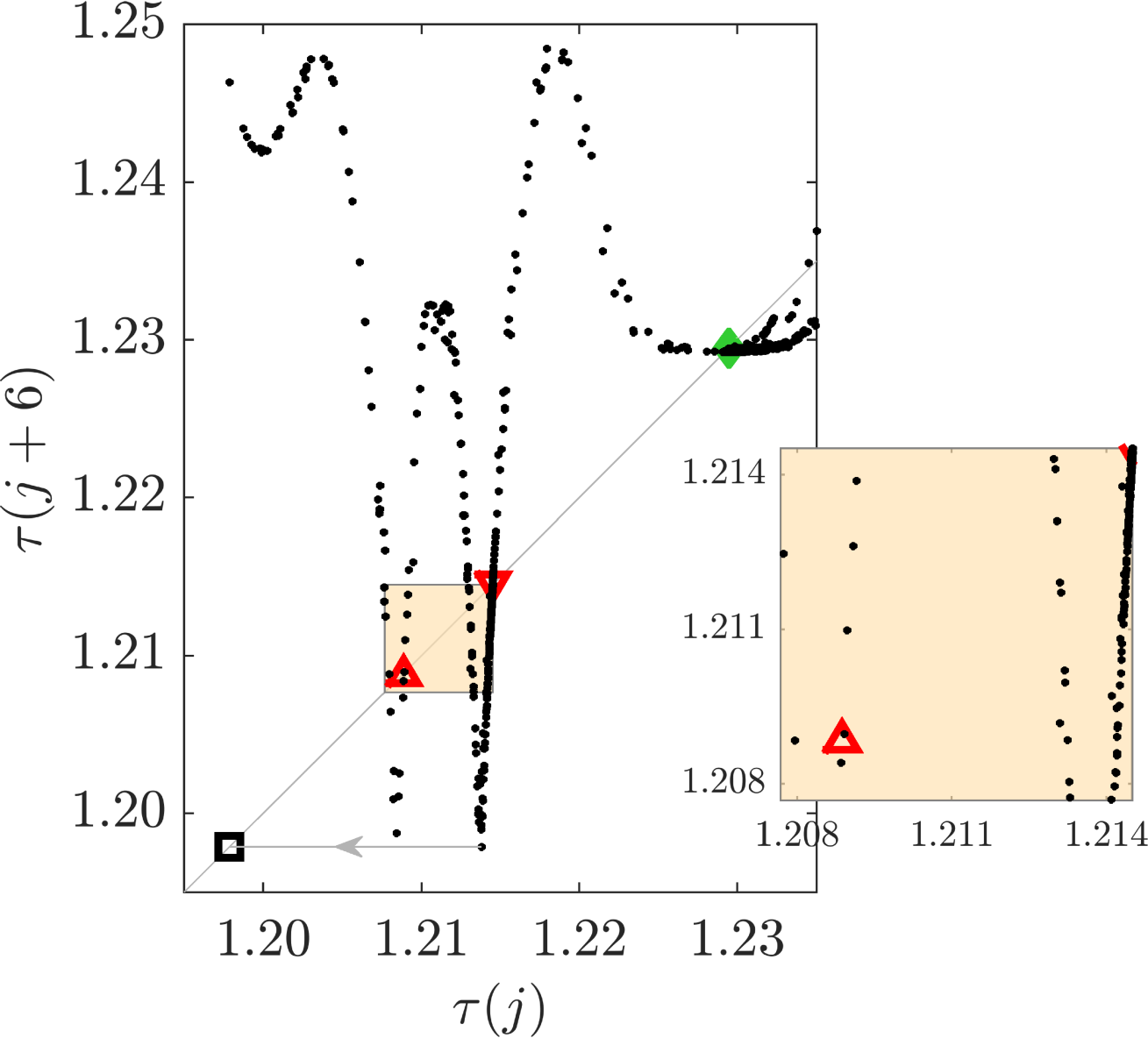} \\
    \end{tabular} 
 \end{center}  
  \caption{
  The phase space at $R=395.564$. Panels, styles and colours as for
    figure \ref{fig:mapRc1}.
}
\label{fig:mapRc2}          
\end{figure}

Figure~\ref{fig:mapRc2} shows the same diagrams as figure~\ref{fig:mapRc1} but at the higher Reynolds number $R= 395.564$ (the same value used in figure \ref{figE}c). 
Figures~\ref{fig:mapRc2}a and \ref{fig:mapRc2}b are qualitatively similar to figures~\ref{fig:mapRc1}a and \ref{fig:mapRc1}b. However, as {evident from} figure~\ref{fig:mapRc2}d, {trajectories can escape the square region more easily (a larger proportion of points maps outside)} at this parameter value than {at $R=395.5401$ depicted} in figure~\ref{fig:mapRc1}d. This makes the higher value of $R$ more convenient in analysing the properties and nature of the chaotic saddle. 

\begin{figure}
  \begin{center}
     \begin{tabular}{ll}
       (a)  &
       (b)  \\
       \multicolumn{1}{r}{\raisebox{-13em}{\includegraphics[width=.44\linewidth]{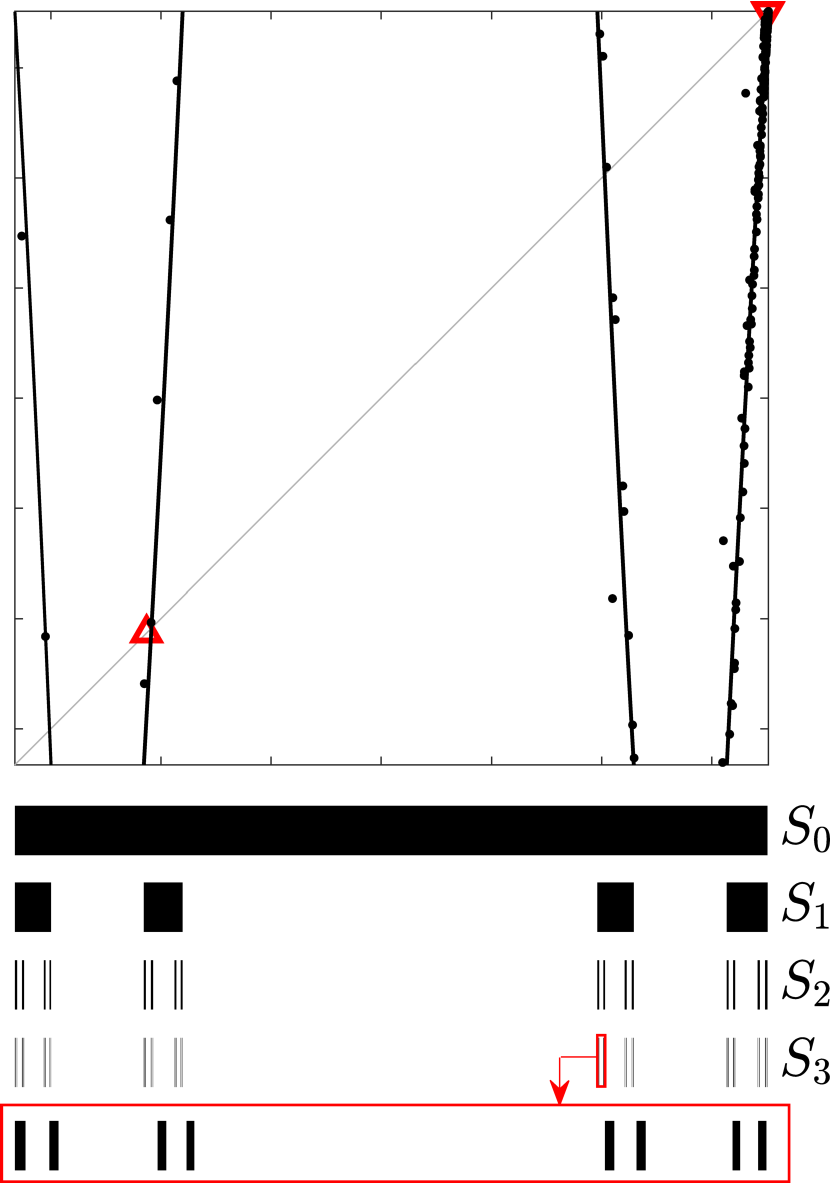}}} &
       \begin{tabular}{l}
         \includegraphics[width=.45\linewidth]{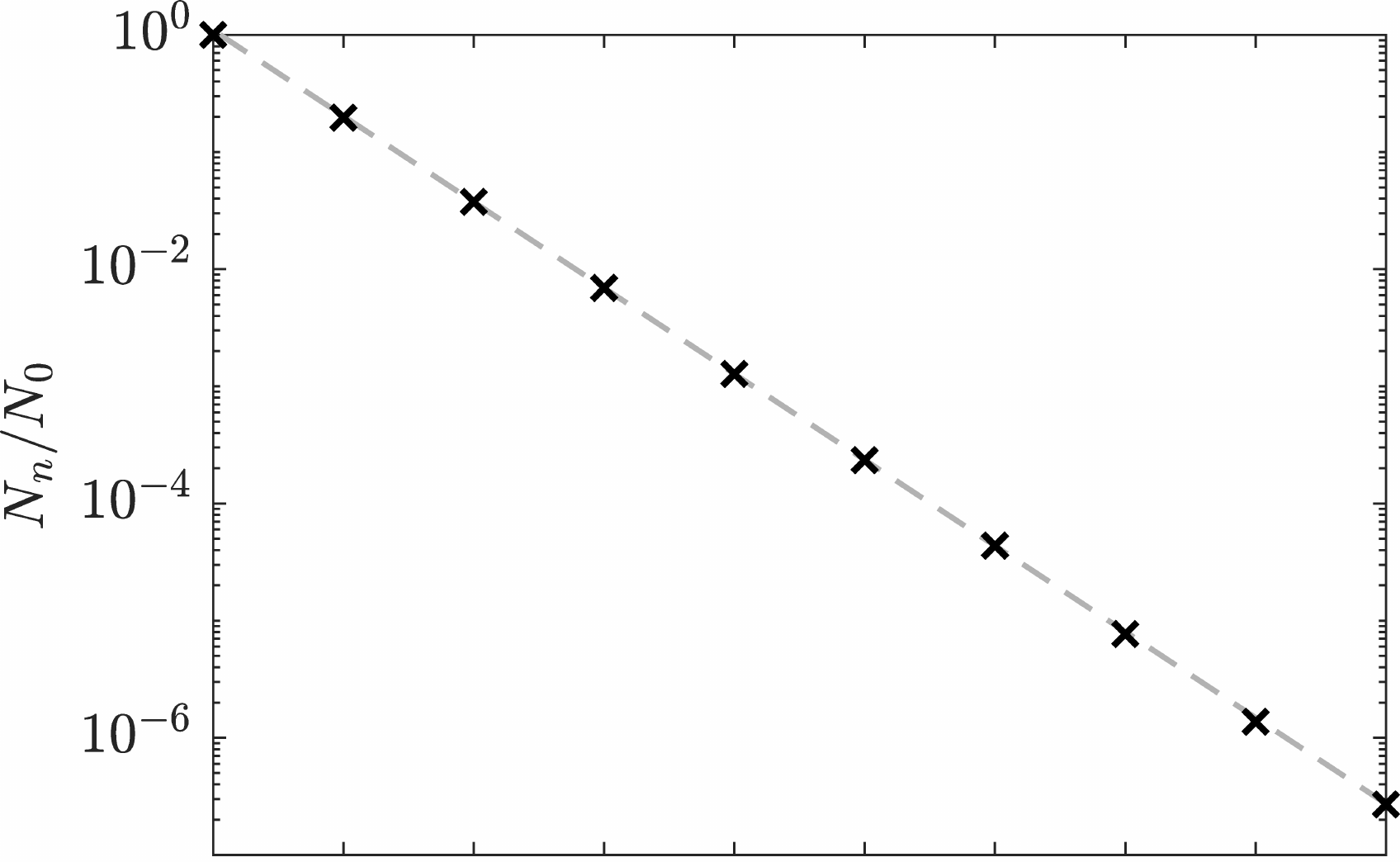} \\
         (c) \\
         \multicolumn{1}{r}{\includegraphics[width=.445\linewidth]{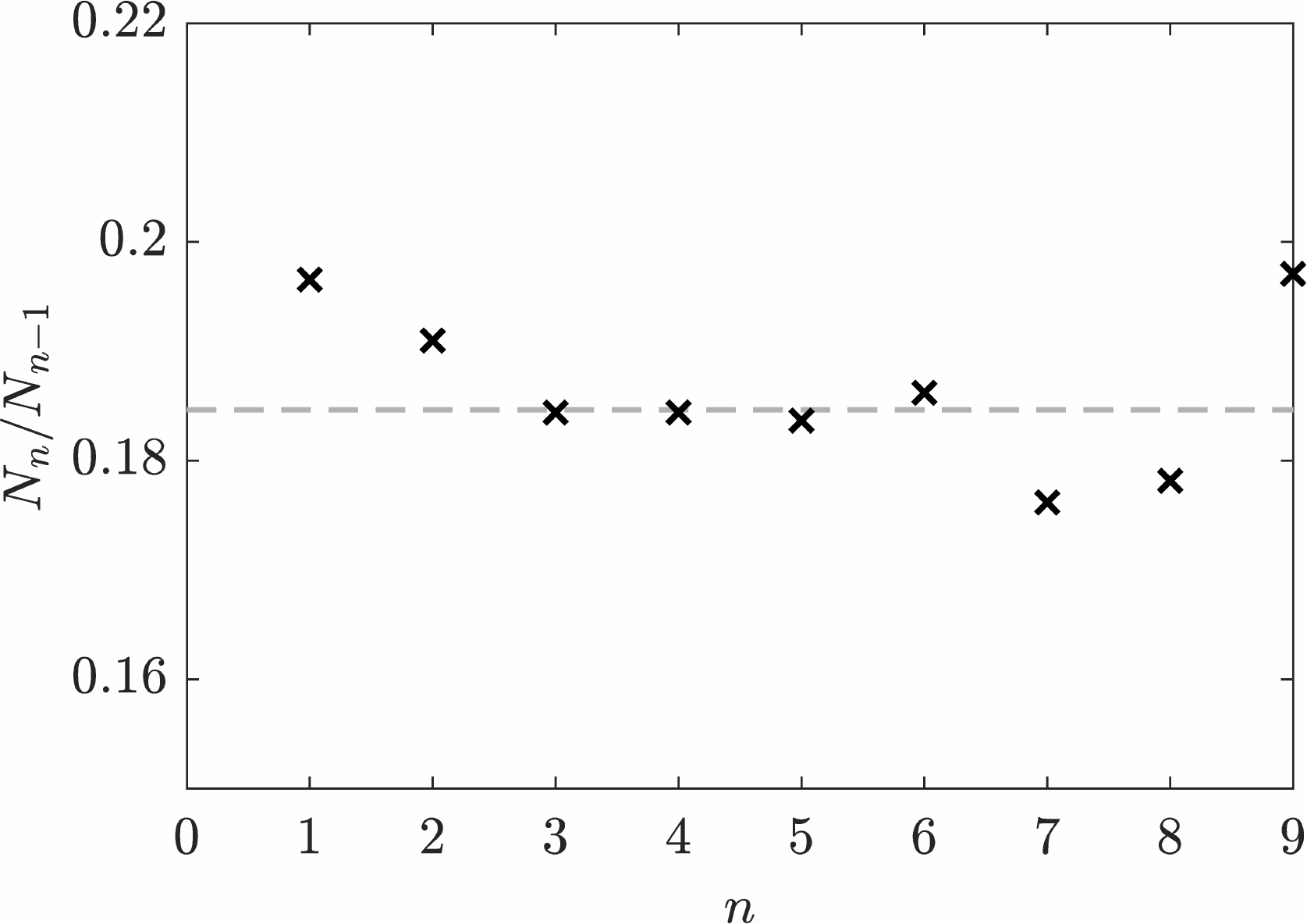}}
       \end{tabular}
     \end{tabular}
  \end{center}
    \caption{
Analysis of the chaotic saddle at $R = 395.564$. (a) Magnified view of the shaded square region in figure \ref{fig:mapRc2}d. The DNS data (black dots) has been interpolated with splines (function $F$, black solid lines). The up- and down-pointing triangles denote P$_{3U}$ and P$_{3L}$, respectively. The intervals below the map correspond to $S_n=F^{-n}(S_0)$, where $S_0$ is the full range {of $F$.} (b) Survival probability $N_{n}/N_0$ of the $N_0=10^8$ points (uniformly distributed in $S_0$) as a function of the number of iterates ${n}$ taken of $F$. (c) Survival rate $N_{n}/N_{n-1}$ of points from one iterate to the next. 
}
\label{fig:lifetime}          
\end{figure}

Figure~\ref{fig:lifetime}a provides an enlarged view of the square region in figure~\ref{fig:mapRc2}d. The DNS data (black dots) has been interpolated with splines (black solid lines) for clarity and the resulting one-dimensional function named $F$. The spline representation evinces that our DNS
computations embody a textbook example of a one-dimensional chaotic saddle, as described in standard dynamical systems literature.
Let us denote by $S_0$ the full range of torque values encompassed by the square region (see bottom of figure \ref{fig:lifetime}a). Points in the set 
$S_1=F^{-1}(S_0)\subset S_0$ remain within the square for at least one iteration of $F$. In other words, $S_0$ and $S_1$ correspond to the range and domain of $F$, respectively. Naturally, $S_1$ consists of four intervals because $F$ consists of four disjoint segments. 
Initial points chosen in $S_1$ may escape the square under $F^2$, since points in $F(S_1)=S_0$ may escape under $F$.
To ensure that the trajectory stays inside the square for two iterations of the map, the initial value must be taken within $S_2=F^{-1}(S_1)=F^{-2}(S_0)$. The set $S_2$ consists of $4^2=16$ subintervals. 
Similarly, to ensure that a point remains in the square under $F^3$, the initial value must be chosen from $S_3=F^{-3}(S_0)$. In the figure, each of the 16 {apparent} bands consists {in actuality} of a set of 4 sub-bands, resulting in a total of 64 subintervals 
(see the enlarged view in the red box).
The chaotic saddle is the {\emph {invariant}} set of {all} points that remain indefinitely within the square, 
and can therefore be formally expressed as $S_{\infty} = \lim_{n \to \infty}S_n$, where $S_n=F^{-n}(S_0)$. Each set $S_n$ consists of ${4}^n$ subintervals and provides an increasingly accurate approximation of the chaotic saddle as $n$ increases. As evident from {its} construction, the chaotic saddle is a Cantor set. To the best of the authors’ knowledge, this is the first direct demonstration of the fractal nature of a chaotic saddle in a shear flow.

We can now estimate the escape rate away from the saddle by taking a large value
$N_0$ of initial conditions in $S$ (\emph{e.g.} uniformly distributed), and then counting the amount of points $N_{{n}}$ that remain in $S$ after the ${n}$th iterative application of $F$.
The logarithm of $N_{{n}}/N_0$ drops {linearly} (figure \ref{fig:lifetime}b), pointing at a nearly constant escape rate from one iteration to the next 
(figure \ref{fig:lifetime}c), which is a hallmark of a Poisson distribution.


%


It should also be noted that, according to dynamical systems theory, a one-dimensional chaotic map such as that in figure~\ref{fig:lifetime}a exhibits \textit{topological transitivity} \citep{Devaney22}. In plain words, this means that a trajectory starting in an arbitrarily small neighbourhood of any point in the chaotic set {will} eventually reach an arbitrarily small neighbourhood of any other point in the set.
Topological transitivity in our map implies that from the neighbourhood of P$_{3L}$, one can reach any neighbourhood of C$_{3}$. For this reason, {we were justified in choosing} P$_{3L}$ as the starting point for the investigation of the chaotic saddle (as done, e.g., to produce figure~\ref{figE}). 


\section{The emergence of {chaos in the} basin boundary}\label{sec:bridge}

\begin{figure}
  \begin{center}
   \begin{tabular}{cc}
    \raisebox{14.5em}{(a)} &  \includegraphics[width=.6\linewidth]{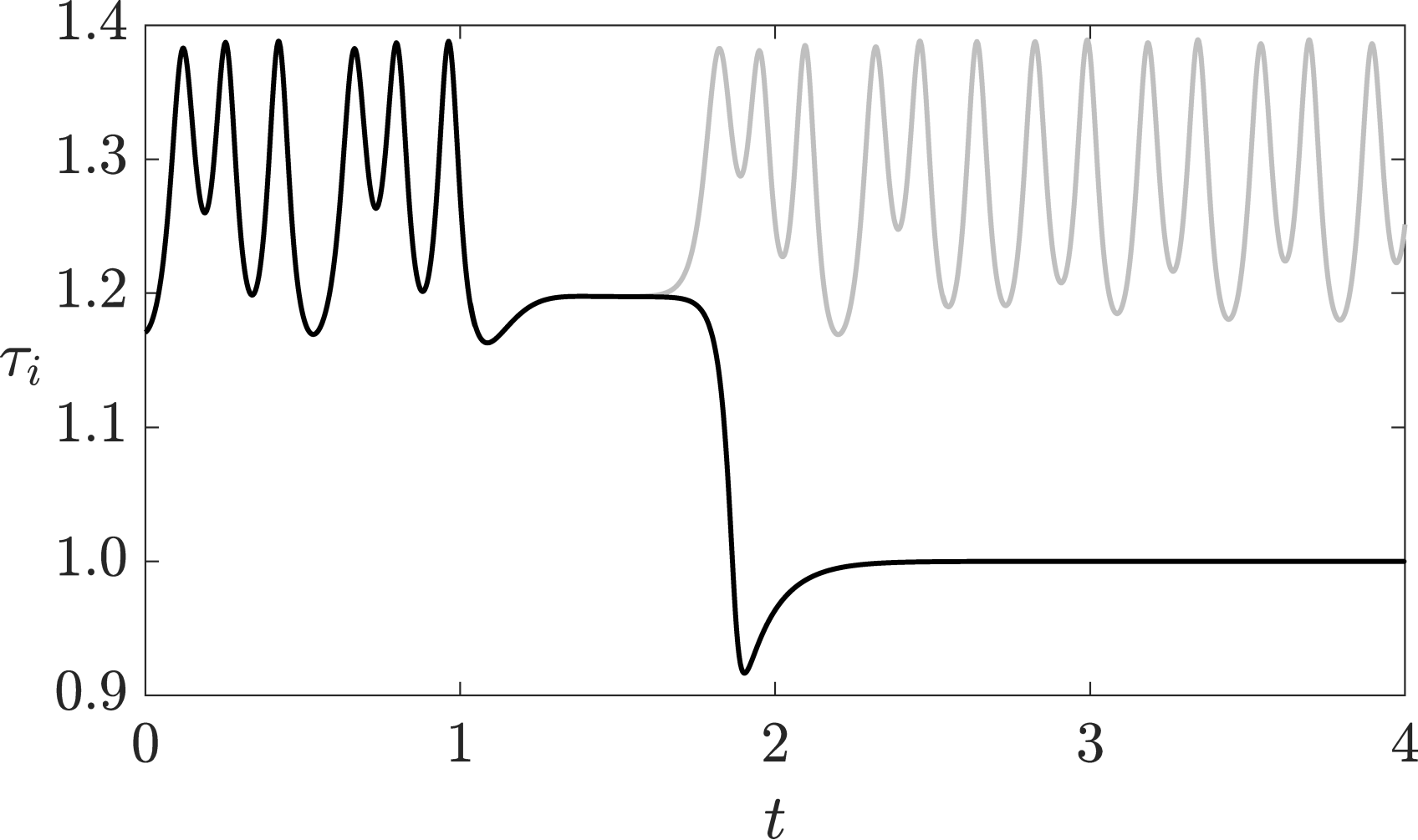} \\
     \raisebox{16em}{(b)} & \includegraphics[width=.6\linewidth]{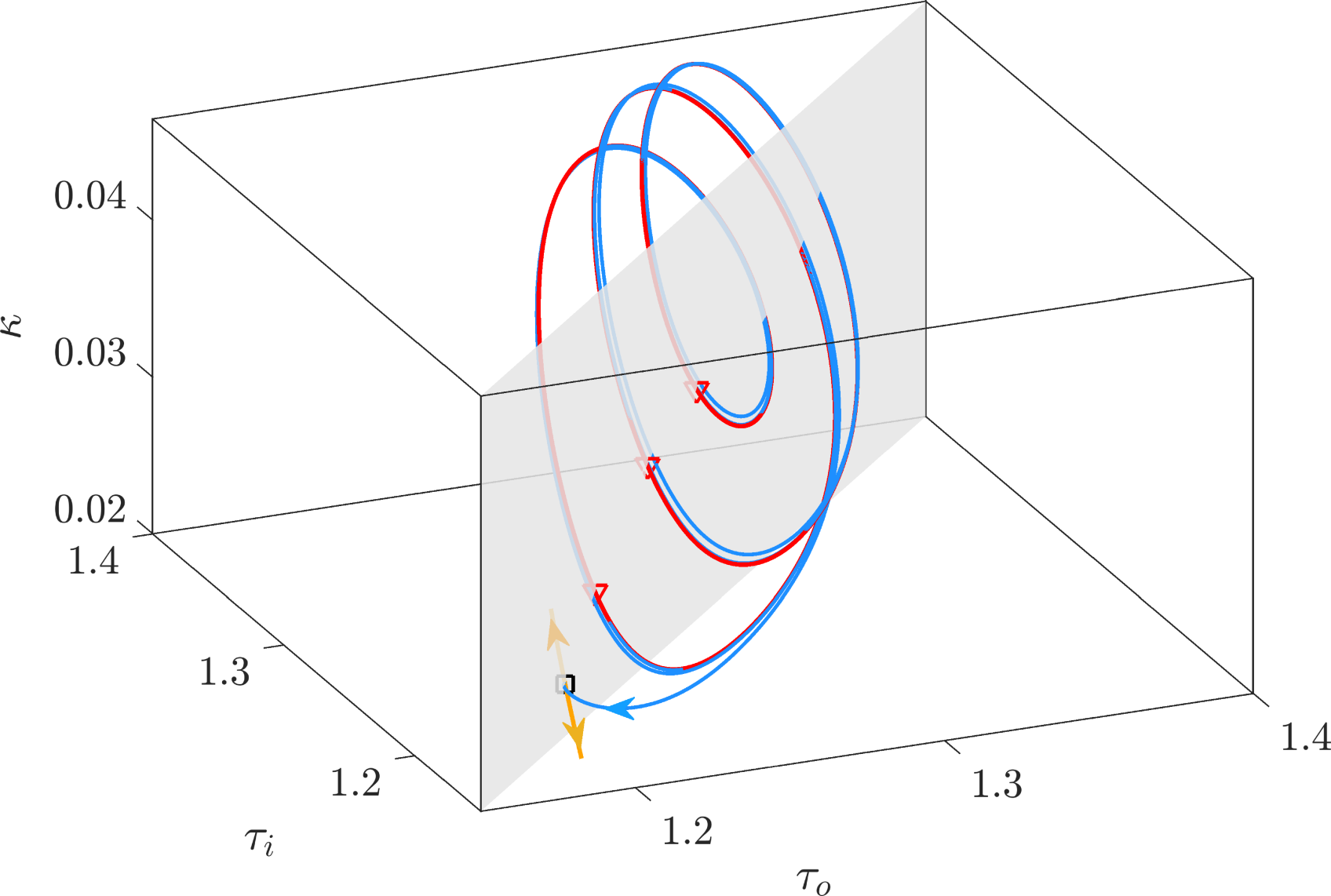} \\
     \raisebox{16em}{(c)} &  \includegraphics[width=.6\linewidth]{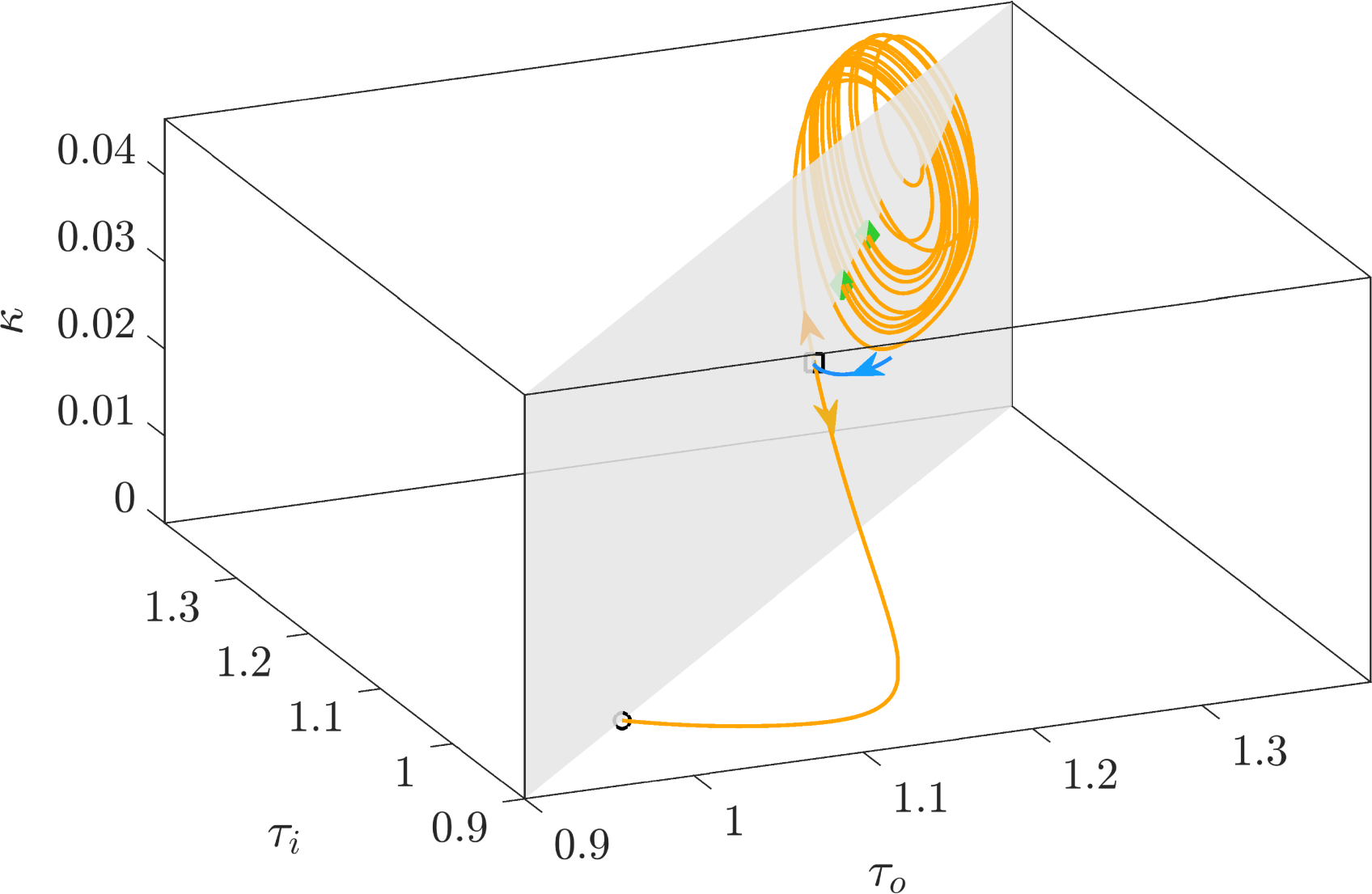} \\
    \end{tabular} 
 \end{center}  
  \caption{Edge state analysis at $R = 395.63$. (a) Two converged bounding trajectories starting from the neighbourhood of P$_{3L}$ and eventually leaving towards CCF ({black}) and P$_2$ ({grey}). The edge tracking algorithm is enforced up to $t\simeq1.3$ to guarantee convergence onto the edge state, DRW$_L$ (plateau), and then turned off.
  (b) Three dimensional projection of the trajectories on $(\tau_i, \tau_o, \kappa)$ space while edge tracking is on (blue), up to $t \leq 1.3$). Also shown are P$_{3L}$ (red loop) and DRW$_L$ (square).
  (c) Projection of the trajectories once edge tracking is turned off for $t> 1.3$ (yellow lines). P$_{2L}$ (green loop) and CCF (black circle) are also shown for reference. 
  }
\label{fig:edgeR395p63}          
\end{figure} 

We now turn our attention to the transition of the chaotic saddle C$_3$ from type (i) to type (ii) at the critical parameter $R_{c3}'$.
The basin boundary becomes chaotic at this point but, as we will see in \S\ref{subsec:4.1}, DRW$_L$ remains the edge state.

To address the advent of chaos in the basin boundary and the later emergence of a chaotic edge state,
we will exploit the theory of homoclinic and heteroclinic tangencies, which occur when manifolds in phase space come into contact. This theory has the advantage of {greater generality than the simpler} one-dimensional discrete map approximation, though the ensuing analysis tends to be more intricate.
As an example, the period-doubling bifurcations leading to C$_{2}$ (see figure~\ref{fig:skitmap}) can also be interpreted as arising from homoclinic tangencies between the stable and unstable manifolds of P$_1$. These tangencies ultimately result in chaos according to the Smale-Birkhoff theorem. In dynamical systems theory, the relationship between one-dimensional dynamics and {manifold} tangencies is often illustrated using the logistic and H\'enon maps \citep{strogatz2015nonlinear, Devaney22}. The phenomena we observe here are analogous
to those reported for
these classical examples.



As we will see in \S\ref{subsec:4.2}, the dynamics in the neighbourhood of $R_{c3}'\approx 395.5401$ 
cannot be fully grasped with the one-dimensional discrete map approximation, hence the need 
for tangency theory.

\subsection{Edge tracking at $R=395.63>R_{c3'}$}\label{subsec:4.1}

We will now show that, although having been subsumed within the basin boundary, C$_3$ is still not an edge state immediately after $R_{c3}'$. We will take advantage of its topological transitivity, as noted in \S\ref{subsec:3.3},
to edge-track between trajectories originated in a vicinity of P$_{3L}$.


Figure~\ref{fig:edgeR395p63}a illustrates the result of edge tracking at $R=395.63$, well above $R_{c3}'$ (see figures \ref{fig:biffig} and \ref{fig:Rhom}a), starting from a close neighbourhood of P$_{3L}$. The algorithm has been initiated by selecting two initial conditions in this neighbourhood, one decaying onto CCF (black) and the other departing towards P$_2$ (grey). 
{Then, the standard process of alternating trajectory bisection and time-advance has been implemented to compute the dynamics on the basin boundary within prescribed tolerance.}
The two trajectories evolve together for as long as the edge tracking algorithm is kept on and, as expected, diverge thereafter. It is clear, however, that keeping the algorithm on for sufficiently long has the dynamics settle on a plateau ($t\in[1.2,1.6]$ in figure~\ref{fig:edgeR395p63}a). The converged solution is DRW$_L$, which remains the edge state at this value of $R$ despite the significant reconfiguration of the basin boundary.

%

 Figure~\ref{fig:edgeR395p63}b shows the same edge tracking up to $t<1.3$ (blue line), projected onto the three-dimensional space $(\tau_i, \tau_o, \kappa)$. The two bounding trajectories start very close to P$_{3L}$ (red closed loop) and approach DRW$_L$ (black square) before parting (yellow arrows) for $t>1.3$. According to the definition in dynamical systems theory, the edge tracking trajectory is contained in the stable manifold of DRW$_L$, $W^s$(DRW$_L$), with the final approach approximating the least stable of its eigenmodes.
%


The yellow lines in figure~\ref{fig:edgeR395p63}c correspond to the same phase map representation but for the diverging portion of the trajectories in figure~\ref{fig:edgeR395p63}a, which result from turning off the edge tracking algorithm for $t>1.3$. These trajectories {start as} close approximations of the one-dimensional unstable manifold of DRW$_L$, $W^u$(DRW$_L$).



Note that, despite having only one unstable eigenvalue, P$_{3L}$ is not an edge state according to the definition given in \S\ref{sec:intro}. As noted there, states possessing a single unstable eigenvalue can, in most cases, be computed through edge tracking by exploiting the different transient behaviour of trajectories issued on either side of their unstable manifold as the separation criterion. {The periodic orbit} P$_{3L}$, however, provides an example that edge tracking offers no general guarantee of convergence for all solutions featuring one-dimensional unstable manifolds.



\subsection{Heteroclinic tangency at $R=R_{c3}'$}\label{subsec:4.2}

\begin{figure}                                                                 
  \begin{center}
  \begin{tabular}{cc}
    (a) &  \includegraphics[width=.6\linewidth]{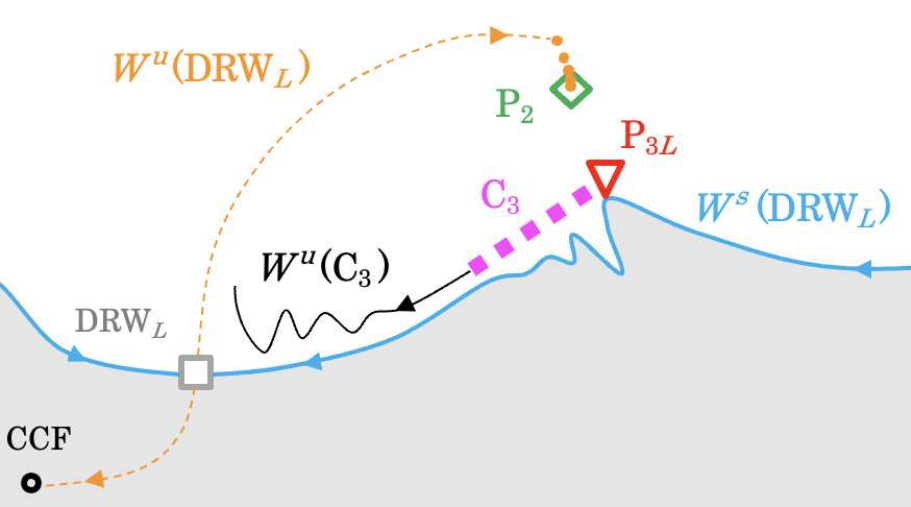} \\
    (b) & \includegraphics[width=.6\linewidth]{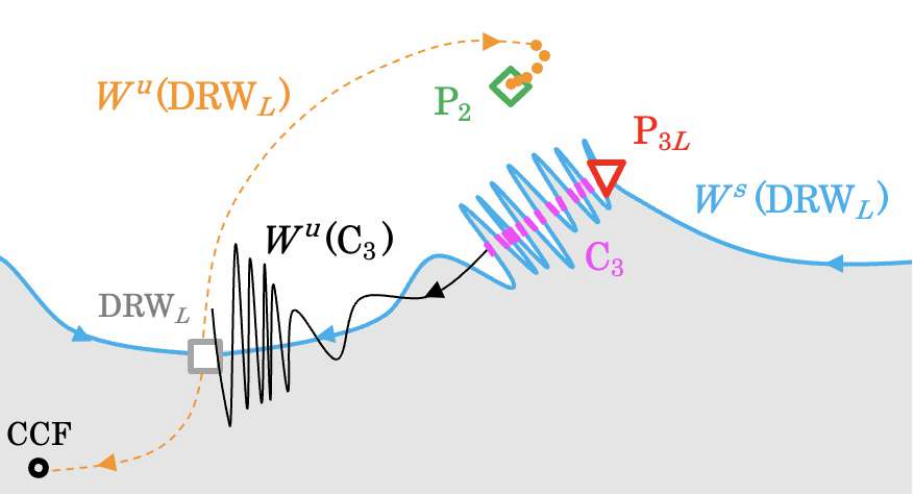} \\
   \end{tabular} 
 \end{center}  
  \caption{
  Sketch of the phase space near the critical Reynolds number $R_{c3}'$. Each periodic orbit is represented by its intersection with $\Sigma$. The unstable manifold $W^u$(DRW$_L$), which does not lie in $\Sigma$, is shown using dashed curves. (a) Before the tangency ($R\lesssim R_{c3}'$). (b) After the tangency ($R\gtrsim R_{c3}'$). 
  }
\label{fig:skbasin}          
\end{figure} 

Let us revisit the return maps in figure~\ref{fig:mapRc2}{b-d}, plotted at $R=395.564\gtrsim R_{c3}'$. 
Recall that the black dots mark points on the unstable manifold of C$_{3}$, i.e., $W^u$(C$_{3}$). 
The global bifurcation at $R=R_{c3}'$ can be ascribed to a heteroclinic tangency between $W^u$(C$_{3}$) and $W^s$(DRW$_L$).
Indeed, the analysis of figure~\ref{fig:mapRc2}d provides compelling evidence for a manifold tangency: 
the cob-web trajectories linking the local minima of the map to their image lands in close proximity of DRW$_L$,
as indicated by the arrow.

Figure~\ref{fig:skbasin} presents sketches of the phase space on $\Sigma$ near the critical parameter {$R_{c3}'$}, constructed from the results obtained so far.
The white and grey regions represent the basins of attraction of P$_2$ and CCF, respectively. They are separated by $W^s$(DRW$_L$), with DRW$_L$ acting as the edge state. As noted earlier, DRW$_L$ possesses only one unstable eigenvalue, so $W^u$(DRW$_L$) forms a one-dimensional manifold in the phase space $\mathbb{X}$, whose two branches are directed toward CCF and P$_2$ at either side of $W^s$(DRW$_L$) (see figure~\ref{fig:edgeR395p63}c).
Although DRW$_L$ belongs in $\Sigma$, its unstable manifold (orange line) has no representation on it, hence its depiction as dashed in figure~\ref{fig:skbasin}. It immediately departs from $\Sigma$ and does not intersect it again until after reaching {the neighbourhood of} P$_2$ 
on one side and converging on CCF on the other.

Figures~\ref{fig:skbasin}a and \ref{fig:skbasin}b illustrate the phase space for $R \lesssim R_{c3}'$ and $R \gtrsim  R_{c3}'$, respectively. In the former case, trajectories originating from {a neighbourhood of} C$_{3}$ always {depart towards} P$_2$, implying that $W^u$(C$_{3}$) must lie entirely within the {basin of attraction of P$_2$}. 
At the critical parameter, a heteroclinic tangency occurs between $W^u$(C$_{3}$) and $W^s$(DRW$_L$), which is followed by a manifold tangle (see figure~\ref{fig:skbasin}b). Note that the occurrence of an intersection between the manifolds implies the existence of infinitely many intersections (heteroclinic points), which is why the manifolds must appear tangled \citep{Ku04, wiggins2010introduction}.
The effect of this tangency extends throughout the entire chaotic saddle C$_{3}$ due to its topological transitivity.
It is evident that after the tangency, the basin boundary becomes extremely intricate, exhibiting fractal geometry. Transitions from this region therefore show strong sensitivity to initial conditions.
%




\subsection{Comparison to plane Couette flow}\label{subsec:4.3}

{The above results up to $R_{c3}'$ share certain features with the transition scenario in plane Couette flow reported by \cite{LuKaVa19}.
This suggests {some} degree of genericity 
in the transition process of shear flows.}
{For example,} figure 4 in \cite{LuKaVa19} {also} shows two tangencies occurring at the critical Reynolds numbers $Re_{cr1}$ and $Re_{cr2}$. At the first tangency, the upper branch of a solution called P2 gives rise to a chaotic attractor that makes contact with the lower branch, producing a chaotic saddle. This chaotic saddle is type (i) by our definition, and the tangency at $Re_{cr1}$ is {akin} to {our observation} at $R_{c3}$. At the second tangency, the chaotic saddle {seems} to collide with the lower branch of {another pre-existing solution,} P1, which is the edge state known as the gentle periodic orbit (GPO). This scenario closely resembles the tangency {we observe} at $R_{c3}'$. The presence of these shared global bifurcations suggests {some} degree of genericity 
in the transition process of shear flows.



{{We must, however, highlight} a significant difference between the results by \cite{LuKaVa19} and ours: while the edge state {was} a periodic orbit ({the} GPO) {there}, it is a travelling wave (DRW$_L$) {here}.}
{Recall that we {specifically} designed the Poincar\'e section $\Sigma$ to contain travelling waves such as DRW$_L$. However, travelling waves are} singular points on $\Sigma$ where the Poincar\'e map cannot be properly defined.
{This renders the} theoretical analysis harder {in our case, and for many other subcritical shear flows featuring steady or travelling-wave edge states (see \S\ref{sec:intro}).} 


{Another important} difference is that \cite{LuKaVa19} reported both homoclinic and heteroclinic tangencies near $Re_{cr2}$, while we {have} only {found {a}} heteroclinic tangency at $R_{c3}'$. {As sketched in} figure \ref{fig:skbasin}, a homoclinic orbit {from} DRW$_L$ {to itself} is clearly impossible, since there is no pathway from the edge state to the chaotic saddle. This {route}  emerges only at higher $Re$, as {we will show} in \S\ref{sec:Bc}. It is likely that, in \cite{LuKaVa19}{, the opening of the pathway precedes}
$Re_{cr2}$.



\section{The emergence of a chaotic edge state}\label{sec:Bc}

As {evident from} figure \ref{fig:edgeR395p63}a, DRW$_L$ remains 
an edge state at least up to  $R=395.63$, with its stable manifold separating the basins of attraction of P$_2$ and CCF. At {somewhat} larger $R$, however, all trajectories starting near DRW$_L$ are {eventually captured by} CCF (see figure \ref{fig:DRWDNS}b).
In \S\ref{subsec:5.1} we shall show that this qualitative change in dynamics occurs at $R=R_{{het}}$ ({see} figure \ref{fig:Rhom}a). {The replacement edge state is then analysed in} \S\ref{subsec:5.2} at $R$ {just} above this critical point.

\begin{figure}                                                                 
  \begin{center}

   \begin{tabular}{ll}
     (a) & (b) \\
 \includegraphics[height=.35\linewidth]{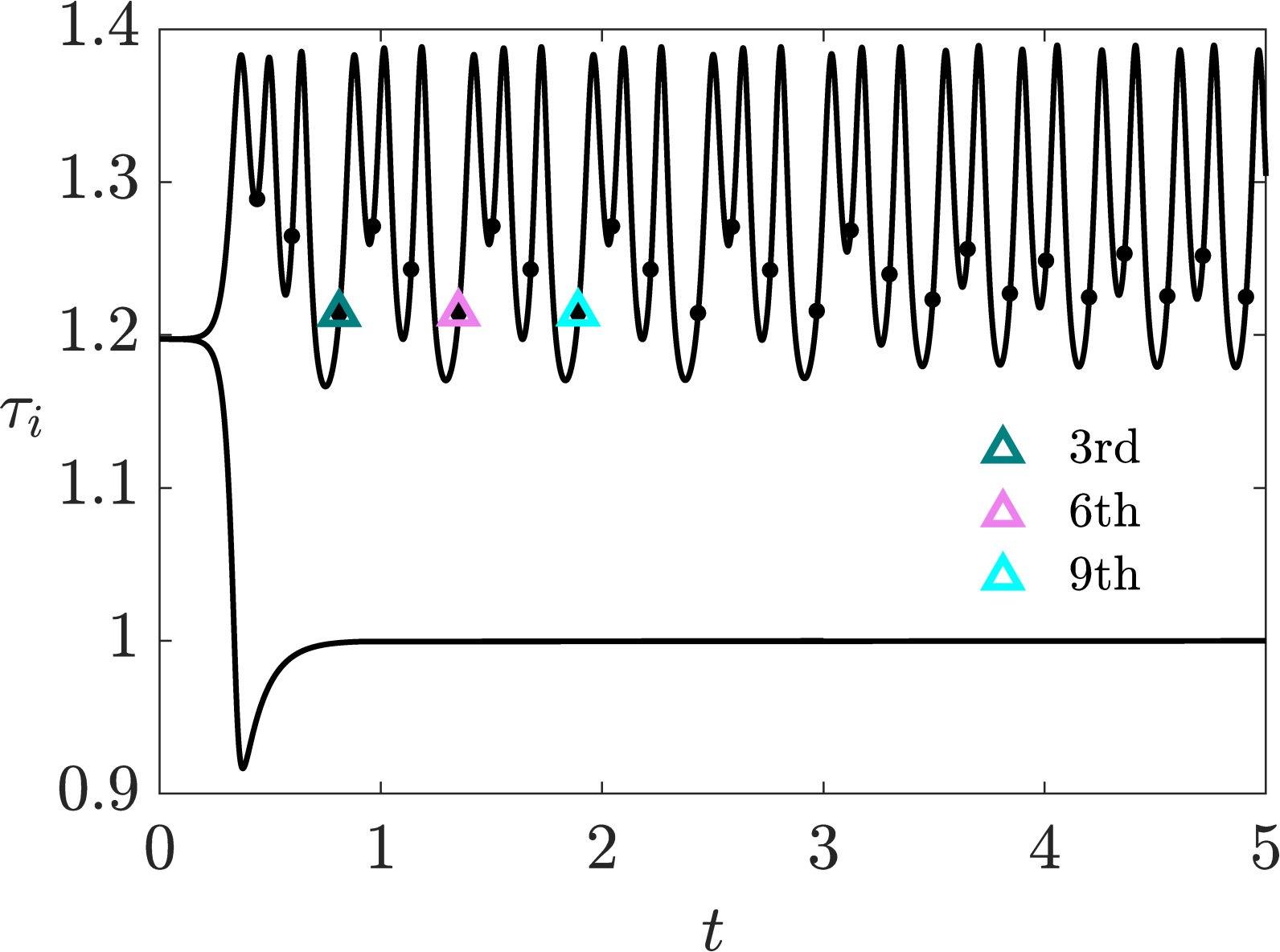} &\includegraphics[height=.35\linewidth]{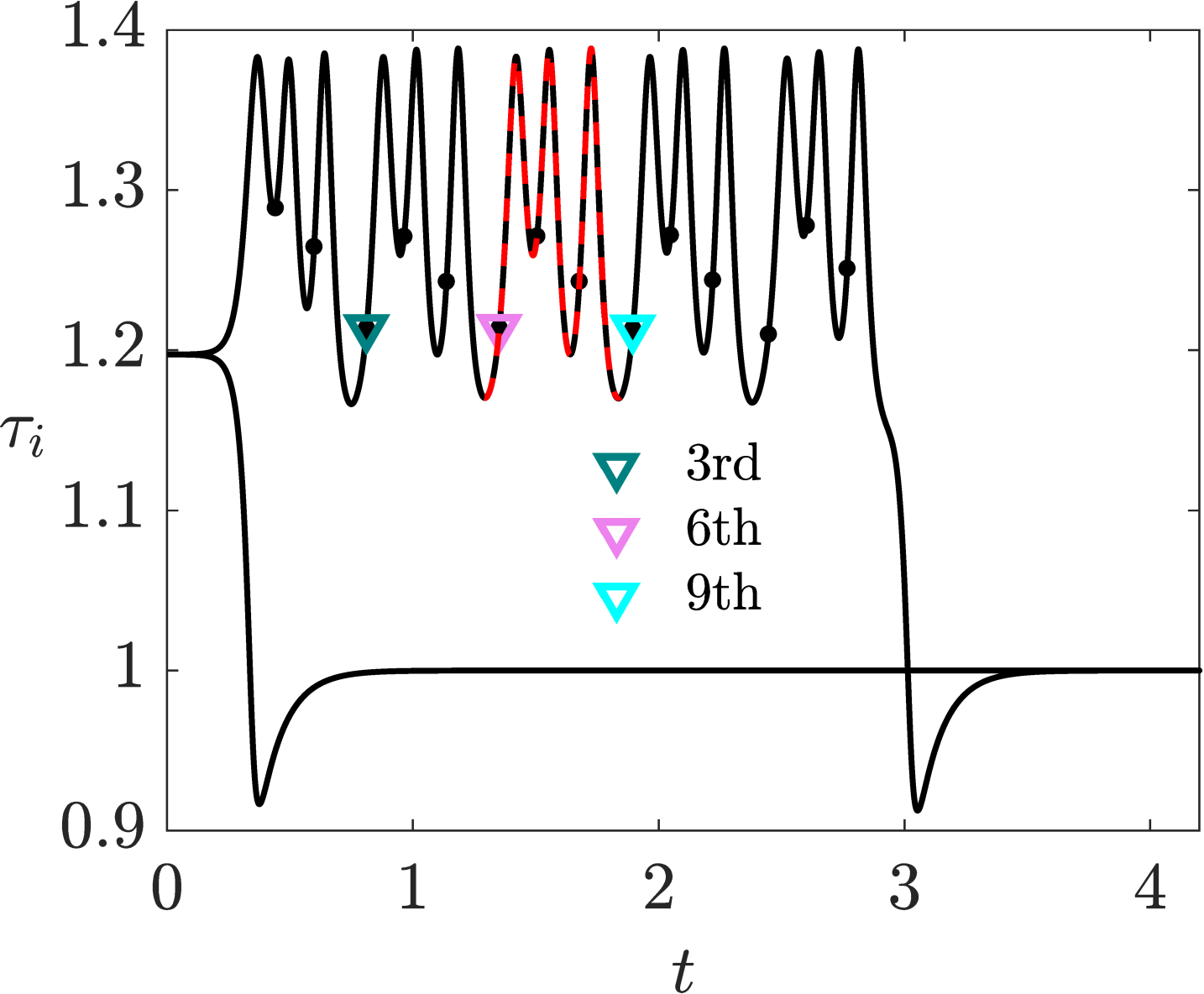} \\
 (c) & (d) \\
 \includegraphics[width=.45\linewidth]{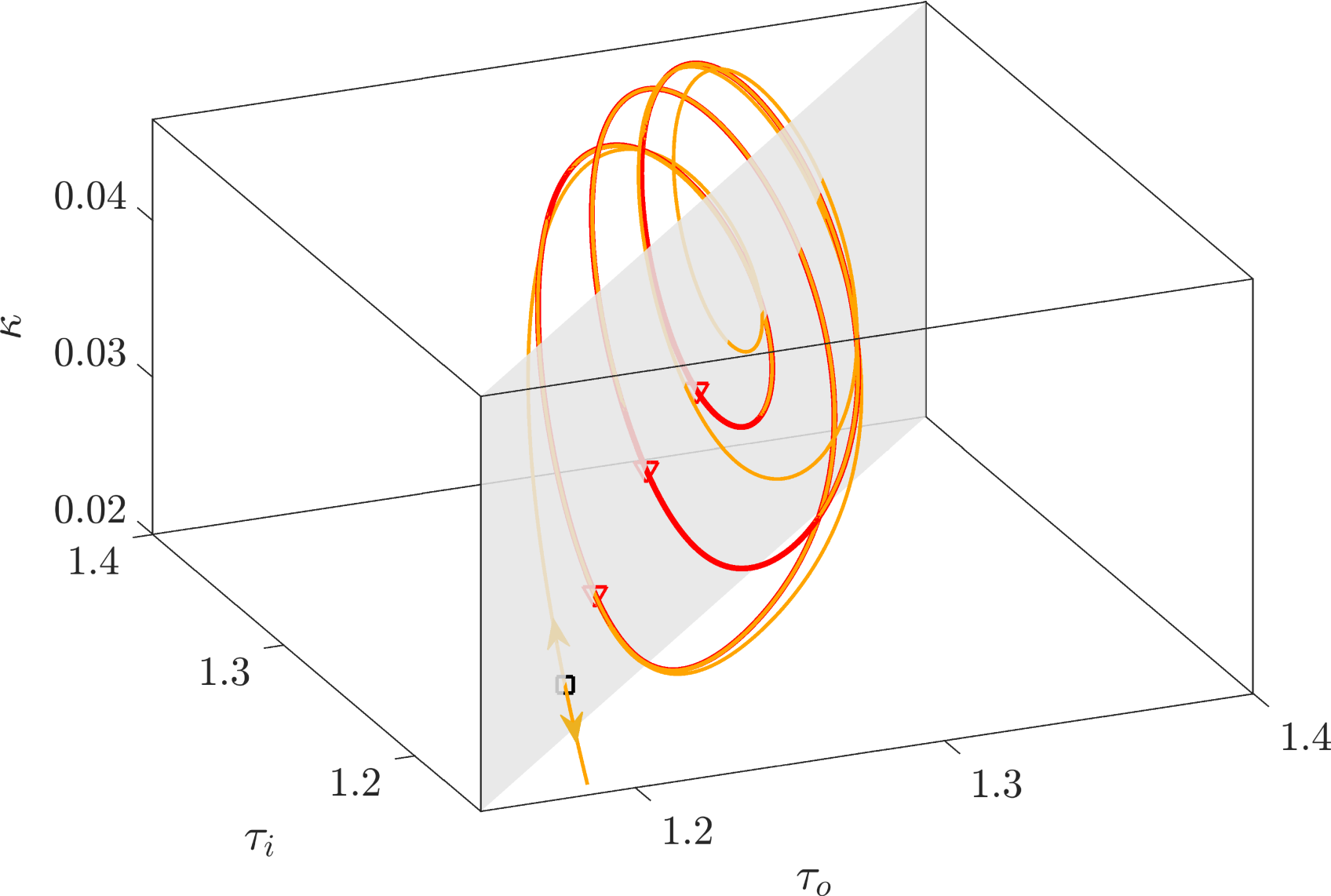} &\includegraphics[width=.4\linewidth]{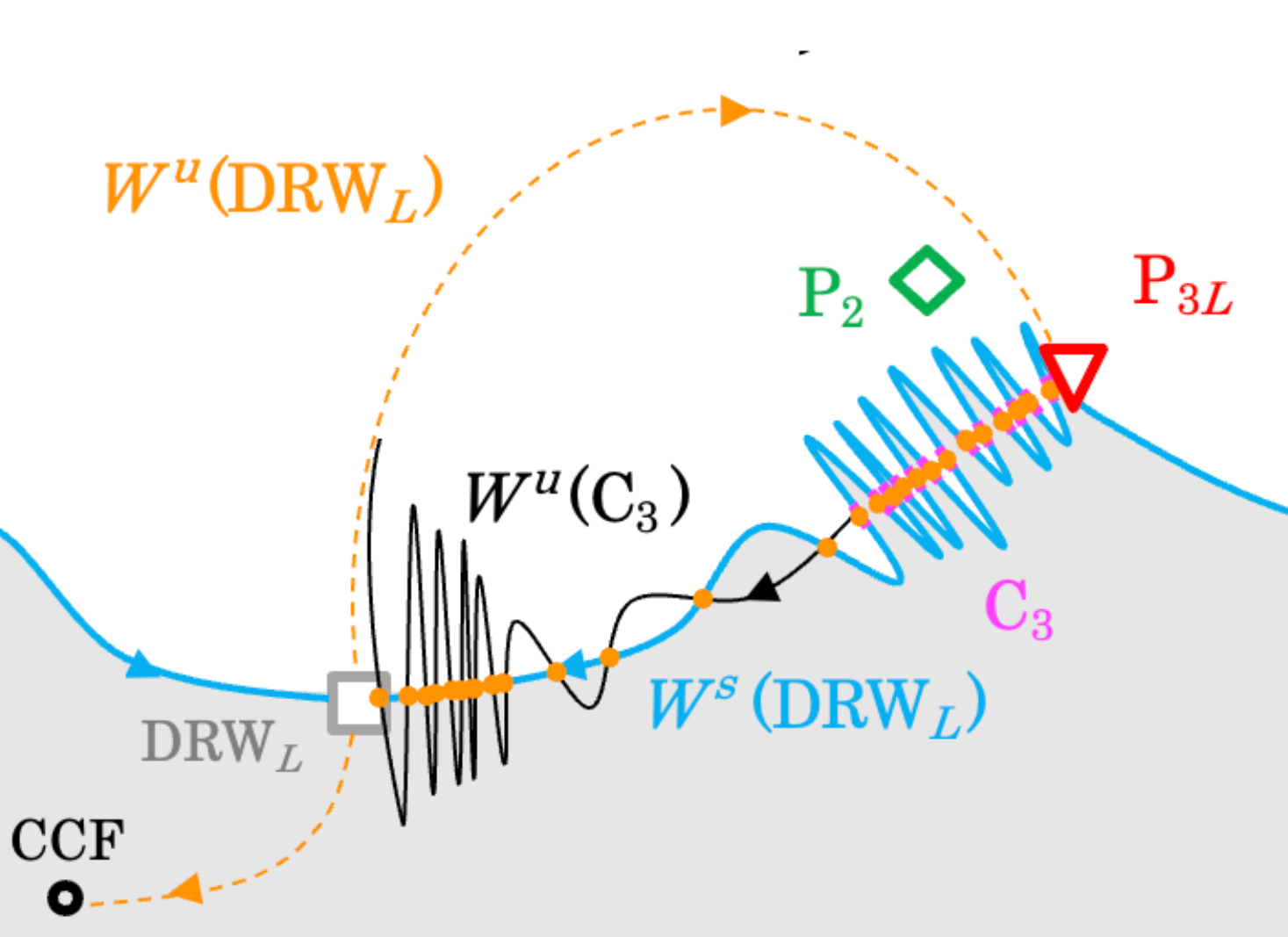} \\
  \end{tabular}
 \end{center}   
  \caption{{The heteroclinic bifurcation at $R_{het}$. Two trajectories started from a close vicinity of DRW$_L$ and leaving in the two different directions of $W_u$(DRW$_L$) are shown at (a) $R=395.646841\lesssim R_{het}$ and (b) $R=395.646875\gtrsim R_{het}$. The triangles indicate the 3rd, 6th and 9th Poincar\'e crossings, all three approaching P$_{3L}$ {(red dashed)} at very close quarters. (c) Phase map representation at $R=395.646875\gtrsim R_{het}$. {The orange trajectories are $W_u$(DRW$_L$), while the red orbit is $P_{3L}$.} (d) Sketch of the phase space at exactly $R_{het}$.}
  }
\label{fig:WDRWL}          
\end{figure}

\subsection{Heteroclinic cycle at $R=R_{\text{het}}$}\label{subsec:5.1}


Figures~\ref{fig:WDRWL}a and \ref{fig:WDRWL}b {track} $W^u$(DRW$_L$) {in its two directions, employing DNS at} {$R=395.646841 \lesssim R_{{het}}$ and $R=395.646875 \gtrsim R_{{het}}$}, respectively. As $R$ approaches $R_{{het}}$ from below, the time series {that eventually leads to} P$_2$ stays for an increasingly long {lapse shadowing} P$_{3L}$ ({{of which} one period is {shown with} a red dashed curve {for reference}}). 
The same shadowing occurs when decreasing $R$ towards $R_{het}$ from above for one of the trajectories, even if its final destination is now CCF.
As a result, the 3rd, 6th, and 9th $\Sigma$ crossings in figures~\ref{fig:WDRWL}a,b are graphically indistinguishable from the corresponding values for P$_{3L}$. This is a sign that a (non-robust) heteroclinic connection 
from DRW$_L$ to P$_{3L}$ occurs at the critical point, $R=R_{het}$. 
The yellow curve in figure~\ref{fig:WDRWL}c shows a three-dimensional phase map projection of the same time series {that visits P$_{3L}$ in} figure~\ref{fig:WDRWL}b, up to $t=2$. 
This trajectory provides a good approximation of the heteroclinic orbit, i.e., the 
intersection 
of $W^u$(DRW$_L$) and $W^s$(P$_{3L}$), which should actually happen at $R=R_{het}$.
%

Recall that the robust heteroclinic connection from P$_{3L}$ back to DRW$_L$ exists for $R>R_{c3}'$ (blue line in figure \ref{fig:edgeR395p63}b), since P$_{3L}$ lies on the basin boundary. 
The emergence of the non-robust  heteroclinic orbit {from DRW$_L$ to P$_{3L}$} at $R=R_{\text{het}}$ (yellow line in figure~\ref{fig:WDRWL}c) {generates} a heteroclinic cycle. This heteroclinic cycle {has been} sketched in figure~\ref{fig:WDRWL}d using a format similar to that of figure \ref{fig:skbasin}. 

\begin{figure}                                  
  \begin{center}
   \begin{tabular}{ll}
   (a) & (b) \\
        \includegraphics[width=.42\linewidth]{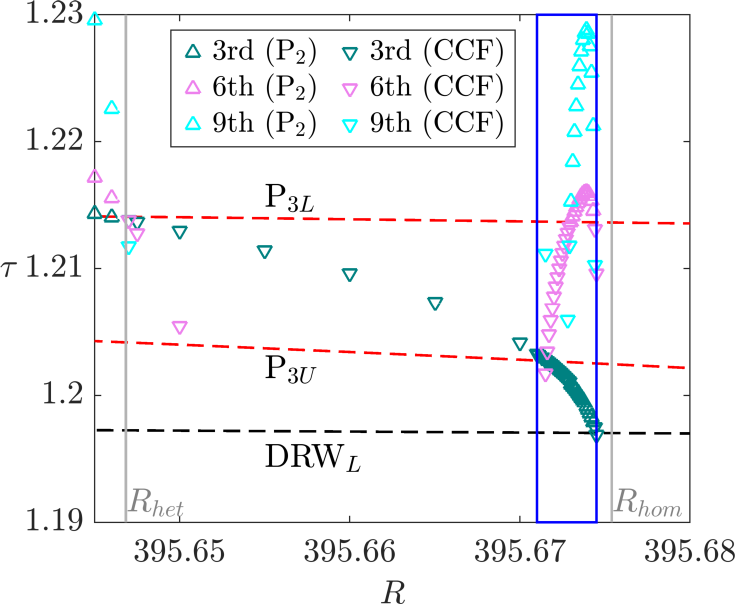} &
         \includegraphics[width=.45\linewidth]{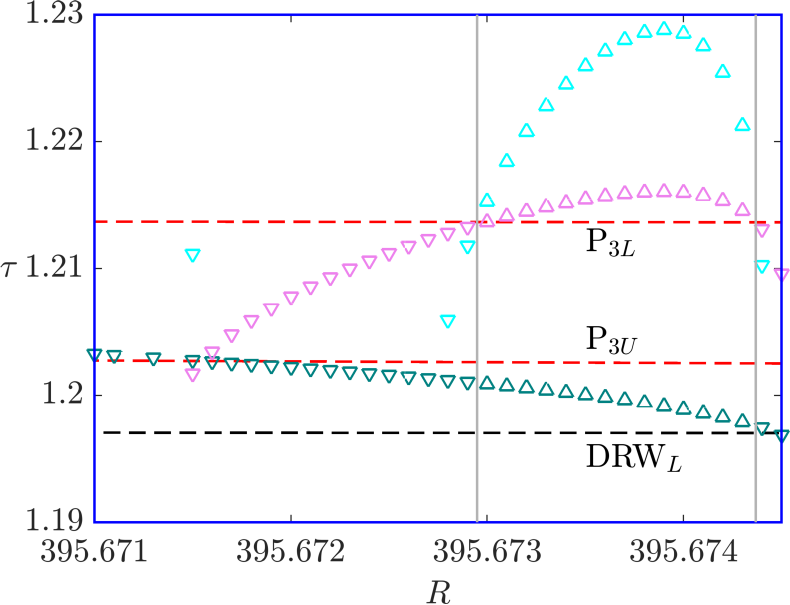} \\
   \end{tabular}
 \end{center}    
  \caption{
  Parameter dependence of $W^u$(DRW$_L$). {The 3rd, 6th and 9th} $\Sigma$ {crossings} are indicated with triangles {(see {the legend and} figure \ref{fig:WDRWL}ab)}. {Up-pointing} triangles indicate that the final state is P$_2$, while {down-pointing triangles} mark {trajectories} ending in CCF.  The red dashed curve is P$_3$ (the upper one is P$_{3L}$). The black dashed curve is DRW$_L$.
  }
\label{fig:bifaroundhet}          
\end{figure}

We have analysed in figure~\ref{fig:bifaroundhet}a the dependence of $W^u$(DRW$_L$) on $R$ by recording the 3rd, 6th and 9th $\Sigma$-crossings (see the legend) of trajectories departing from a vicinity of DRW$_L$ and transiently approaching P$_{3L}$ before heading towards either P$_2$ (up-pointing triangles) or CCF (down-pointing triangles). From this analysis, we estimate $R_{{het}}\approx 395.64686\pm2\times10^{-5}$ as the value of the parameter for which the sequence of $\Sigma$-crossings asymptotically approaches P$_{3L}$ (top red dashed line) and P$_2$ ceases to be accessible from DRW$_L$.
{Incidentally, additional heteroclinic bifurcation points seem to occur in succession {with increasing $R$} and, possibly, accumulate as $R_{hom}$ is approached (see the region bounded by the blue rectangle and the magnification in figure~\ref{fig:bifaroundhet}b; {two such bifurcations are shown by the vertical grey lines}). At these points, the connection from DRW$_L$ to P$_2$ is alternatively restored and broken. A detailed study of the apparent cascade of global bifurcations leading up to $R_{hom}$ would require specifically designed numerical methods and is, therefore, beyond the scope of the paper.}

\subsection{Edge tracking {immediately above $R_{het}$}}\label{subsec:5.2} 

\begin{figure}                                                           
  \begin{center}
   \begin{tabular}{l}
   (a) \\
        \includegraphics[width=.9\linewidth]{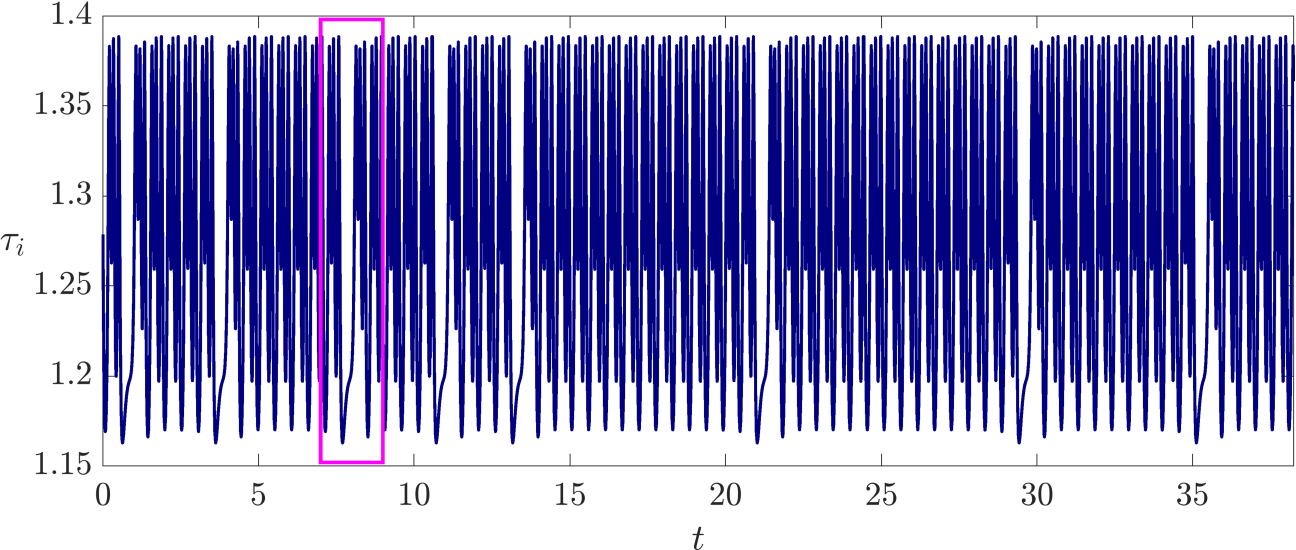} \\
        \begin{tabular}{ll}
        (b) & (c) \\
           \includegraphics[height=.43\linewidth]{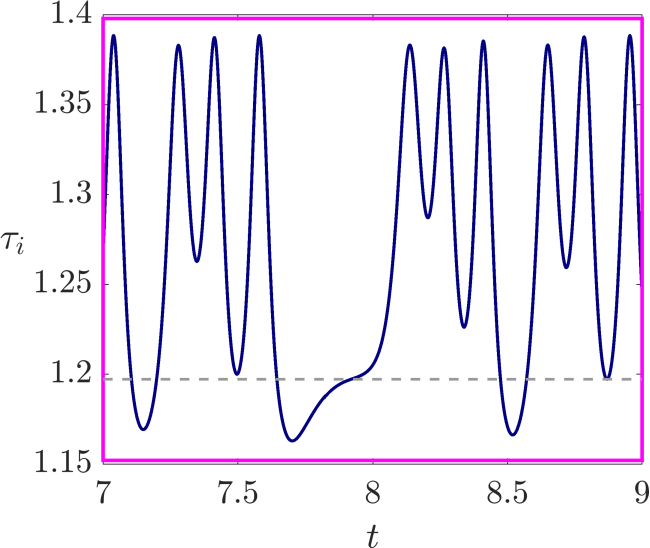} &  \includegraphics[height=.43\linewidth]{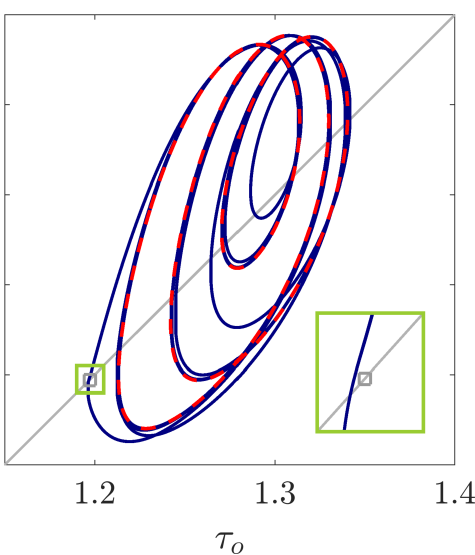} 
        \end{tabular} \\
   \end{tabular}
 \end{center}  
  \caption{{Edge state at $R=395.6468459\gtrsim R_{het}$. (a) Time series of the statistically converged edge state. (b) Magnification of the portion enclosed in the magenta rectangular region. The horizontal dashed line indicates DRW$_L$. (c) Phase map representation of the edge state. P$_{3L}$ is overlaid with a red dashed line. The green square region is magnified in the inset. DRW$_L$ is indicated with a gray square.}
}
\label{fig:edgechaos}          
\end{figure} 

\begin{figure}
  \begin{center}

  \includegraphics[width=.6\linewidth]{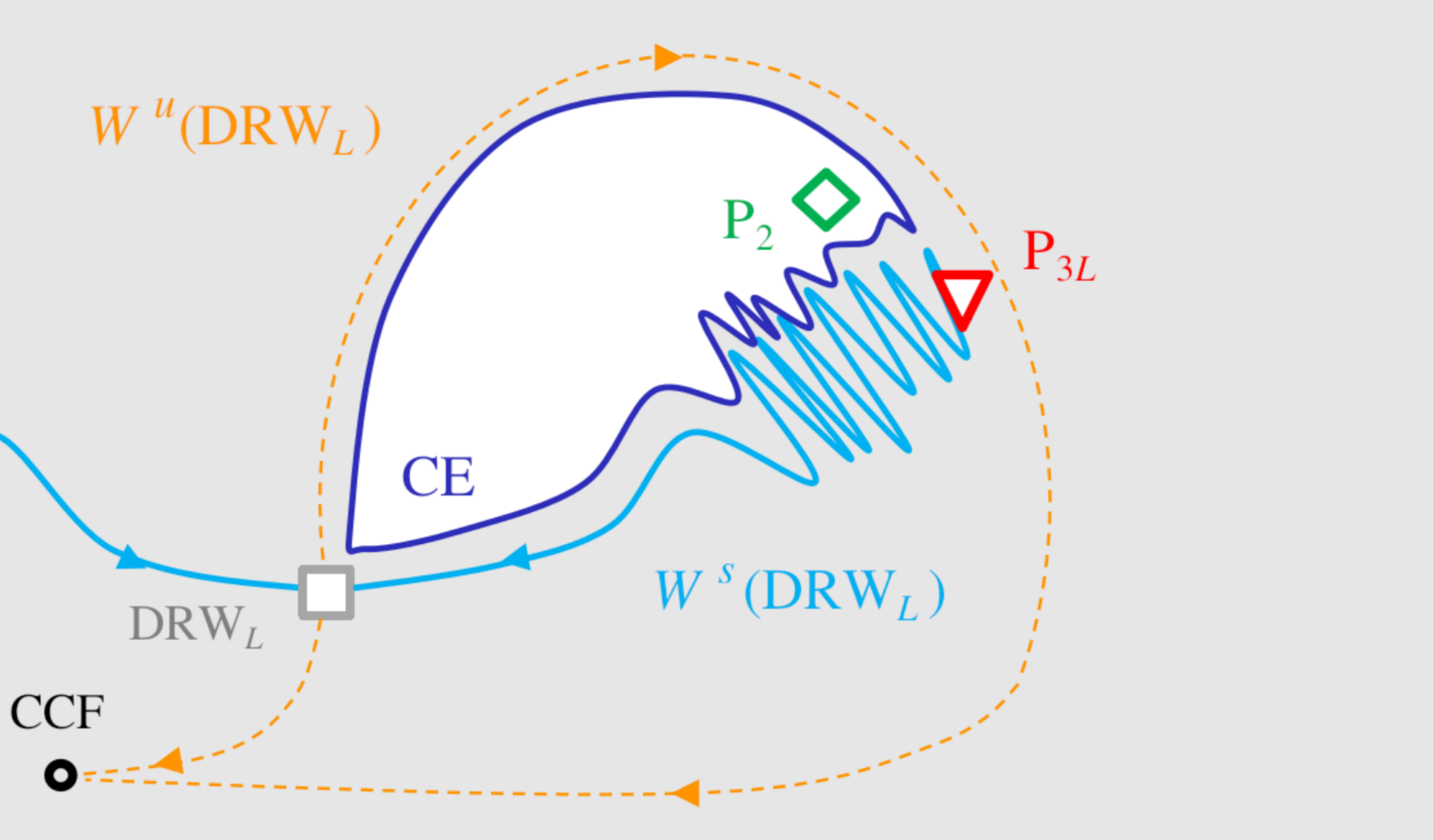}
 \end{center}  
  \caption{Sketch of the phase space for $R$ slightly larger than $R_{het}$. CE represents the chaotic edge state.
  }
\label{fig:edgesketch}          
\end{figure} 

{Trajectories originating in a neighbourhood of DRW$_L$ at $R\gtrsim R_{{het}}$ always relaminarise, so the solution is no longer an edge state. Since there are still two attractors, CCF and P$_2$, the question arises as to which solution replaces DRW$_L$ in the role of the edge state.}

Figure \ref{fig:edgechaos} shows the result of edge tracking at {$R=395.6468459\gtrsim R_{{het}}$}.
{Only the statistically converged edge state is shown, having discarded all preliminary transient evolution on the critical threshold. The long time series of figure~\ref{fig:edgechaos}a reveals that the edge state is chaotic, the dynamics transiently slowing down at irregular intervals. A close inspection (see magnification in figure~\ref{fig:edgechaos}b) reveals that the long lapses of fast dynamics orbit in close proximity of P$_{3L}$, while the intermittent slowdowns correspond to transient approaches to DRW$_L$ (horizontal grey dashed line). The phase map representation of figure~\ref{fig:edgechaos}c further substantiates the claim that the chaotic edge state looks like P$_{3L}$ (overlaid in dashed red) most of the time but occasionally visits DRW$_L$ (see the inset).}

{
The peculiar {dynamics 
in figure \ref{fig:edgechaos}}
strongly suggest that the chaotic edge state must have originated from the heteroclinic cycle identified in \S\ref{subsec:5.1}. Figure~\ref{fig:edgesketch} provides a sketch of {how} the phase space {topology could have evolved for $R\gtrsim R_{{het}}$}. In this figure, the CE (i.e. the {dark blue} loop) is the chaotic edge state originating from the heteroclinic cycle (see figure~\ref{fig:WDRWL}d). The stable manifold $W^s$(CE) now forms the basin boundary; although it is not shown in the figure, one may imagine it {as approaching CE 
along a third coordinate axis.}
{Figure~\ref{fig:edgesketch} is drawn in a format similar to the previous sketches (figures~\ref{fig:skbasin}, \ref{fig:WDRWL}d). 
However, unlike earlier sketches, it does not merely show the discrete map on the Poincare section.
This is {evinced by} the fact that{, while} periodic orbits P$_2$ and P$_3$ are represented as points, CE is shown as a simple closed loop. This abstraction is a compromise made to illustrate the {complications of an} inherently high-dimensional phase space topology {as we have in} the {problem at hand}. It should also be noted that the basin boundary is a hypersurface 
{that contains}
the chaotic trajectory CE, and hence has a structure too complex to be visualised even in a three dimensional representation.}

\section{{Summary and conclusions}}\label{sec:conclusions}

Subcritical transition in minimal box shear flows {can often be} explained {in the light of} simple ECS arising at saddle-node bifurcations, with the {saddle} branch, together with its stable manifold, serving as the {threshold} for turbulent transition. However, as {we show here}, chaotic solutions {emerged from} global bifurcations may {sometimes} play a consequential part in the transition, even at relatively low Reynolds numbers. 

{Of the three possible types of chaotic saddles, those dwelling on the basin boundary, \emph{i.e.}, types (ii) and (iii), are instrumental in elucidating the turbulent transition of subcritical flows.}
{However}, only the mechanisms for the formation of type (i) saddles were known.
{By investigating the origin of type (ii) and type (iii) saddles, this paper contributes} 
a theoretical framework that {addresses} the sensitive dependence on initial conditions observed in subcritical shear flow transition scenarios. {The morphing of a chaotic saddle from one type to another happens through global bifurcations.}

To summarise our findings, it is {convenient to revisit the bifurcation diagram in} 
figure~\ref{fig:biffig}. 
{The chaotic attractor C$_3$, emerged from a period doubling cascade, becomes a saddle in a boundary crisis at $R_{c3}$.}
{The saddle} is type (i) because all trajectories {starting from its neighbourhood} eventually leak to the non-trivial attractor (P$_2$), none relaminarising (see \S 3). 
{In other words,} {C$_3$ does not belong in} the boundary {that separates} the basins of attraction of P$_2$ and the laminar circular Couette flow. 
The edge state is the simple solution DRW$_L$, which has acted as such since its emergence in a saddle-node bifurcation {together with the non-trivial attractor that later evolves into P$_2$}.
The type (i) saddle C$_3$ is then subsumed into the basin boundary and becomes type (ii) in a heteroclinic bifurcation (another crisis) at $R'_{c3}$ involving the edge state, which at this point remains DRW$_L$ (see \S 4). Finally, the occurrence of a second heteroclinic bifurcation at $R_{het}$ {generates} a heteroclinic loop between DRW$_L$ and a solution (P$_{3L}$) at the boundary of the chaotic set C$_3$ {that} transforms the saddle into type (iii) (see \S 5). From this point on, the chaotic saddle {replaces DRW$_L$ in the role of} 
edge state.
}


Our theoretical { framework,} 
supported by detailed numerical simulations of Taylor–Couette flow{,} 
explains some {unaccounted for phenomena previously reported in the context of} plane Couette and pipe flows. 
{Notably, our findings in \S\ref{sec:emergence} provide a precise rationale for the}
Poisson processes 
observed in various 
subcritical shear flows in the form of memoryless decay of turbulence \citep{EcSchHoWe07, AvBaHo23}. 
{Our results show how} 
Poisson processes in type (i) saddles are closely related to the fractal structure in phase space of the chaotic set. 
To the best of our knowledge, this is the first study that accounts directly for the origin of the fractal structure of a chaotic saddle in a fluid mechanics system. 

Complexity in the transition to sustained turbulence {requires the presence of type (ii) or type (iii) chaotic saddles. We show that one way this may come about is through the transformation of a type (i) saddle.} 
The underlying fractal structure is {preserved} during {saddle type conversion}. 
Assuming that this is a {generic} phase-space evolution scenario, our findings
support the conjecture by \citet{ScEcYo07,KrEc12,KrEcSch14}, that the basin boundary of subcritical shear flows has a fractal structure.

Also, the fact that the emergence of type (iii) chaotic saddles requires two heteroclinic bifurcations appears to possess a certain degree of universality.
{The importance of heteroclinic bifurcations to the onset of transient turbulence was emphasised by \citet{LuKaVa19} in the context of plane Couette flow.
In our system, the final step producing the type (iii) saddle {involves} 
the replacement of the edge state via a global bifurcation.
It is noteworthy that \cite{LuShKa23} independently reported a similar edge switching phenomenon, although their case {did not entail} chaos. If the edge state is a simple ECS, transition control may be {feasible} \citep[as suggested by][]{WaGiWa07}. However, once a type (iii) saddle appears, sensitivity to initial conditions leads to significant difficulties in {devising effective} control.





Finally, we note that global bifurcations that qualitatively change the nature of the edge state as the Reynolds number is increased {seem to} occur far more frequently than anticipated. We hypothesise that this is caused by the tangle of manifolds near the edge state. 
{If global bifurcations that change the character of edge states or attractors are ubiquitous, this may explain why identifying a critical Reynolds number is notoriously difficult in subcritical shear flow transition. In long pipe flows, a critical Reynolds number has been identified statistically based on the {competition of puff decay and splitting rates, turbulence remaining locally transient}  \citep{AvMoLoAvBaHo11}. However, whether it is possible to determine a critical Reynolds number beyond which turbulence becomes sustained and the basin boundary can be rigorously defined remains an open question, even for short periodic pipes.}

\backsection[Acknowledgements]{
This research is supported by the Australian Research Council Discovery Project DP230102188 and the Ministerio de Ciencia, Innovación y Universidades, Agencia Estatal de Investigación, project PID2023-150029NB-I00 (MCIU/AEI/10.13039/501100011033/FEDER, UE). BW's and RA's research has been funded by the European Union’s Horizon 2020 research and innovation programme (Marie Sk\!\l{}odowska-Curie Grant Agreement No. 101034413). RA has also been funded by the Austrian Science Fund (FWF) 10.55776/ESP1481224. FM is a Serra-Húnter fellow.}

\backsection[Author ORCIDs]{B. Wang, https://orcid.org/0000-0002-6229-0336; R. Ayats, https://orcid.org/0000-0001-6572-0621; K. Deguchi, https://orcid.org/0000-0002-3709-3242; A. Meseguer, https://orcid.org/0000-0002-2022-2001; F. Mellibovsky, https://orcid.org/0000-0003-0497-9052}

\backsection[Declaration of Interests]{
The authors report no conflicts of interest.
}

\bibliography{local}
\bibliographystyle{jfm}

\end{document}